\documentclass{aa}

\usepackage{graphicx}
\usepackage[colorlinks=true,linkcolor=red]{hyperref} 
\usepackage{txfonts}
\usepackage{caption}
\usepackage{amssymb}
\usepackage{amsfonts}
\usepackage{amsbsy}
\usepackage{amsmath}
\usepackage{pdflscape}
\usepackage{mathrsfs}

\bibpunct{(}{)}{;}{a}{}{,}

\newcommand{\SPFA}{{\rm SPFA}}
\newcommand{\PFA}{P_{\rm FA}}
\newcommand{\PD}{P_{\rm D}}

\newcommand{\oneb}{\boldsymbol{1}}
\newcommand{\ab}{\boldsymbol{a}}
\newcommand{\bb}{\boldsymbol{b}}
\newcommand{\cb}{\boldsymbol{c}}
\newcommand{\Ab}{\boldsymbol{A}}
\newcommand{\Bb}{\boldsymbol{B}}
\newcommand{\fb}{\boldsymbol{f}}

\newcommand{\Gb}{\boldsymbol{G}}
\newcommand{\db}{\boldsymbol{d}}

\newcommand{\gb}{\boldsymbol{g}}
\newcommand{\pb}{\boldsymbol{p}}
\newcommand{\Cb}{\boldsymbol{C}}

\newcommand{\Db}{\boldsymbol{D}}
\newcommand{\hb}{\boldsymbol{h}}
\newcommand{\Hb}{\boldsymbol{H}}
\newcommand{\Ib}{\boldsymbol{I}}

\newcommand{\nb}{\boldsymbol{n}}

\newcommand{\Pb}{\boldsymbol{P}}

\newcommand{\ssb}{\boldsymbol{s}}
\newcommand{\rb}{\boldsymbol{r}}

\newcommand{\Sb}{\boldsymbol{S}}

\newcommand{\xb}{\boldsymbol{x}}

\newcommand{\yb}{\boldsymbol{y}}

\newcommand{\taub}{\boldsymbol{\tau}}
\newcommand{\zerob}{\boldsymbol{0}}
\newcommand{\ooneb}{\boldsymbol{1}}
\newcommand{\lambdab}{\boldsymbol{\lambda}}

\newcommand{\sigmab}{\boldsymbol{\sigma}}

\newcommand{\ctb}{\boldsymbol{\widetilde c}}
\newcommand{\htb}{\boldsymbol{\widetilde h}}

\newcommand{\gtb}{\boldsymbol{\widetilde g}}
\newcommand{\xtb}{\boldsymbol{\widetilde x}}

\newcommand{\Ctb}{\boldsymbol{\widetilde C}}

\newcommand{\Hc}{\mathcal{H}}

\newcommand{\Xmatc}{\mathcal{X}}
\newcommand{\Smatc}{\mathcal{S}}
\newcommand{\Gmatc}{\mathcal{G}}
\newcommand{\Tmatc}{\mathcal{T}}
\newcommand{\Tc}{T}

\newcommand{\at}{\widehat a}
\newcommand{\ct}{\widetilde c}
\newcommand{\ft}{\widetilde f}
\newcommand{\gt}{\widetilde g}
\newcommand{\st}{\widetilde s}

\newcommand{\xhb}{\boldsymbol{\check x}}
\newcommand{\xhc}{\check x}

\newcommand{\xbl}{\underline{\boldsymbol{x}}}
\newcommand{\fbl}{\underline{\boldsymbol{f}}}
\newcommand{\gbl}{\underline{\boldsymbol{g}}}
\newcommand{\Gbl}{\underline{\boldsymbol{G}}}

\newcommand{\Ssbl}{\underline{\boldsymbol{S}}}
\newcommand{\ssbl}{\underline{\boldsymbol{s}}}
\newcommand{\nbl}{\underline{\boldsymbol{n}}}

\newcommand{\ftbl}{\underline{\boldsymbol{\widetilde f}}}
\newcommand{\gtbl}{\underline{\boldsymbol{\widetilde g}}}
\newcommand{\Gtbl}{\underline{\boldsymbol{\widetilde G}}}

\newcommand{\sbl}{\underline{\boldsymbol{s}}}

\newcommand{\Cbl}{\underline{\boldsymbol{C}}}

\newcommand{\Sigmatbl}{\underline{\boldsymbol{\widetilde \Sigma}}}
\newcommand{\Xmattb}{\boldsymbol{\widetilde{\mathcal X}}}

\newcommand{\Cmatt}{\widetilde{\mathcal C}}
\newcommand{\Gmatt}{\widetilde{\mathcal G}}
\newcommand{\Xmatt}{\widetilde{\mathcal X}}

\newcommand{\Smatb}{\boldsymbol{{\mathcal S}}}
\newcommand{\Xmatb}{\boldsymbol{{\mathcal X}}}

\newcommand{\Cmatb}{\boldsymbol{{\mathcal C}}}

\newcommand{\Fmatfb}{\boldsymbol{{\mathfrak F}}}

\newcommand{\Gmatb}{\boldsymbol{{\mathcal G}}}
\newcommand{\Nmatb}{\boldsymbol{{\mathcal N}}}

\newcommand{\Tmatb}{\boldsymbol{{\mathcal T}}}

\newcommand{\Smat}{\mathcal S}
\newcommand{\Gmat}{\mathcal G}

\newcommand{\Fmatf}{{\mathfrak F}}
\newcommand{\gc}{\mathfrak{g}}
\newcommand{\ssc}{\mathfrak{s}}
\newcommand{\Nc}{\mathcal{N}}

\newcommand{\flo}{f_{\rm LO}}

\begin{document}

\title{Everything you always wanted to know about matched filters \\  (but were afraid to ask)}
 
   \author{R. Vio
 	 \inst{1}
          \and
          P. Andreani 
         \inst{2}
         }

   \institute{Chip Computers Consulting s.r.l., Viale Don L.~Sturzo 82,
              S.Liberale di Marcon, 30020 Venice, Italy\\
              \email{robertovio@tin.it}
          \and
           ESO, Karl Schwarzschild strasse 2, 85748 Garching, Germany \\
             \email{pandrean@eso.org}             
 				}
   \date{Received....; accepted....}

 \abstract{In this paper we review the application of the matched filter (MF) technique and its application to detect weak, deterministic, smooth signals in a stationary, random, Gaussian noise. This is particular suitable in astronomy to detect emission lines in spectra and point-sources in two-dimensional maps.
A detailed theoretical development is already available in many books
\citep[e.g.][]{kay98, poo94, mcn05, hip02, mac05, wic02, bar05, tuz01, lev08}. Our aim is to examine some practical issues that are typically ignored in textbooks or even in specialized literature as, for example, the effects of the discretization of the signals and the non-Gaussian nature of the noise.
To this goal we present each item in the form of answers to specific questions. The relative mathematics and its demonstration are kept to a bare simplest minimum, in the hope
of a better understanding of the real performances of the MF in practical applications. For the ease of formalism, arguments will be developed for one-dimensional signals. The extension to the two-dimensional signals is trivial and will be highlighted in dedicated sections.}

   \titlerunning{Practical issues for the matched filter applications}
   \authorrunning{Vio \& Andreani}
   \maketitle
 
\section{Notation and Formalism}

In the following a bold lowercase letter indicates a  {\it column} array (vector), e.g.
\begin{equation}
\hb= \left(
\begin{array}{c}
h_1 \\
h_2 \\
\vdots \\
h_{N-1}\\
h_N
\end{array}
\right).
\end{equation}
with 
\begin{equation}
\hb^T= (h_1, h_2, \ldots, h_{N-1}, h_N)
\end{equation}
a {\it row} array, whereas a bold uppercase letter indicates a matrix, e.g.
\begin{equation}
\Hb= \left(
\begin{array}{cccc}
h_{1,1} & h_{1,2} & \cdots & h_{1,M} \\
h_{2,1} & h_{2,2} & \cdots & h_{2,M} \\
\vdots & \vdots & \vdots & \vdots \\
h_{N-1,1} & h_{N-1, 2}& \cdots & h_{N-1,M} \\
h_{N,1} & h_{N, 2}& \cdots & h_{N,M}
\end{array}
\right),
\end{equation}
with
\begin{equation}
\Hb^T= \left(
\begin{array}{cccc}
h_{1,1} & h_{2,1} & \cdots & h_{N,1} \\
h_{1,2} & h_{2,2} & \cdots & h_{N,2} \\
\vdots & \vdots & \vdots & \vdots \\
h_{1,M-1} & h_{2,M-1}& \cdots & h_{N,M-1} \\
h_{1,M} & h_{2,M}& \cdots & h_{N,M}
\end{array}
\right).
\end{equation}
Here, symbol $^T$ denotes vector or matrix transpose operator. In the case of complex arrays or matrices, symbol $^*$ denotes the complex conjugate operator whereas $^{\dagger}$ denotes the complex conjugate transpose operator. This last works similarly to the matrix transpose operator when each entry is replaced by its complex conjugate (i.e. $c=a + \imath b \rightarrow c^* = a - \imath b$, where $\imath = \sqrt{-1}$).  The notation $| \hb |$ expresses the norm  $\sqrt{\hb^T \hb}$ of a real vector as well  the norm $\sqrt{\hb^{\dagger} \hb}$ of a complex one. Two operations between the elements of two arrays will be used, in particular, 
the element-wise (or Hadamard) product $\odot$ and division $\oslash$. According to these operations, the $i$th entry of the arrays $\rb = \hb \odot \pb$ and $\rb = \hb \oslash \pb$ is given by $r_i = h_i \times p_i$ 
and $r_i = h_i / p_i$, respectively. Something similar holds for matrices.

There are two results of linear algebra that will be useful in the following. The first result is that, if a square $N \times N$ matrix $\Hb$ is diagonalizable, then it can be factorized in the form
 $\Hb=\Pb \Db \Pb^{-1}$, with $\Pb$ the matrix having columns the eigenvectors of $\Hb$ and  $\Db={\rm DIAG[\db]}$ a diagonal matrix \footnote{Given a $N \times 1$ array $\bb$, ${\rm DIAG} [\bb]$ is a 
$N \times N$ diagonal matrix whose diagonal contains the array $\bb$.} having as diagonal entries the corresponding eigenvalues $d_1, d_2, \ldots, d_N$. 
When one or more of these eigenvalues are close to zero, matrix $\Hb$ is said {\it bad-conditioned} or {\it ill-conditioned}. The resulting $\Hb^{-1}$ is either imprecise or unfeasible.
The second result is that the function $f(\Hb$) of a diagonalizable matrix $\Hb$ is given by $f(\Hb)=\Pb  \Db \Pb^{-1}$ with $\Db = {\rm DIAG[f(d_1), f(d_2), \ldots, f(d_N)]}$ \citep[][page 3]{hig08}.

Finally, if $\Ab$ is an $N \times M$ matrix and $\Bb$ is a $P \times Q$ matrix, then the Kronecker product $\Ab \otimes \Bb$ is the $NP \times MQ$ block matrix
\begin{equation}
\Ab \otimes \Bb =  \left(
\begin{array}{ccc}
a_{1,1} \Bb & \cdots & a_{1,M} \Bb \\
\vdots & \vdots & \vdots \\
a_{N,1} \Bb  & \cdots & a_{N,M} \Bb
\end{array}
\right),
\end{equation}
and ${\rm VEC}[\Hb]$ is the operator that transforms a matrix $\Hb$ into a column array by stacking its columns one below the other,
\begin{equation}
{\rm VEC}[\Hb]= \left(
\begin{array}{c}
h_{1,1}  \\
\vdots \\
h_{N,1} \\
h_{1,2}  \\
\vdots \\
h_{N,2} \\ 
h_{1,M}  \\
\vdots \\
h_{N,M} \\ 
\end{array}
\right).
\end{equation}

\section{What is a Matched Filter?} 

The MF is a linear filter used in problems of detection of weak signal embedded in a stationary Gaussian noise. It can be derived in various ways but the most common are the Neyman-Pearson approach and the maximization of the signal to noise ratio  ({\rm SNR}). The following assumptions are common to all methods:
\begin{enumerate}
\item The signal of interest is discrete and has the form $\ssb = a \gb$, with $a$ a positive scalar quantity (amplitude) and $\gb$ a template typically given by a smooth function somehow normalized, e.g. $\max{\left[ g[0], g[1], \ldots, g[N-1] \right] } = 1$; \\
\item Signal $\ssb$ is embedded within an additive noise $\nb$, i.e. the observed signal $\xb$ is given by $\xb = \ssb + \nb$. Without loss of generality, it is assumed that ${\rm E}[\nb] = 0$, where ${\rm E}[.]$ denotes the expectation operator; \\
\item The noise $\nb$ is the realization of a stationary, random, Gaussian process. This means that, for the autocovariance function $c[i, i+ d]={\rm E}[n[i] n[i+d]]$, it is $c[i,i + d] = c[|d|]$ with $d$ any integer value. 
In this case,  a $N \times N$ covariance matrix
\begin{equation} \label{eq:C}
\Cb = {\rm E}[\nb \nb^T]
\end{equation}
can be associated to $\nb$. This is a symmetric matrix of Toeplitz type, i.e. a matrix in which each descending diagonal from left to right and right to left is constant,
\begin{equation} \label{eq:Ctoe}
\Cb= \left(
\begin{array}{ccccc}
c[0] & c[1] & \cdots & c[N-2] & c[N-1] \\
c[1] & c[0] & \cdots & c[N-3] & c[N-2] \\
\vdots & \vdots & \ddots & \vdots \\
c[N-2] & c[N-3]& \cdots & c[0] & c[1] \\
c[N-1] & c[N-2]& \cdots & c[1] & c[0]
\end{array}
\right).
\end{equation}
\end{enumerate}   

\section{What is the Neyman-Pearson approach?} \label{sec:neyman}

According to the Neyman-Pearson approach, the detection problem consists of deciding whether $\xb$ is pure noise $\nb$ (hypothesis $\Hc_0$) or it contains a contribution from a signal $\ssb$ (hypothesis $\Hc_1$). 
In these terms, it is equivalent to a decision problem between the two hypotheses
\begin{equation} \label{eq:decision}
\left\{
\begin{array}{ll}
\Hc_0: & \quad \xb = \nb; \\
\Hc_1: & \quad \xb = \nb + \ssb.
\end{array}
\right.
\end{equation}

\begin{figure}
        \resizebox{\hsize}{!}{\includegraphics{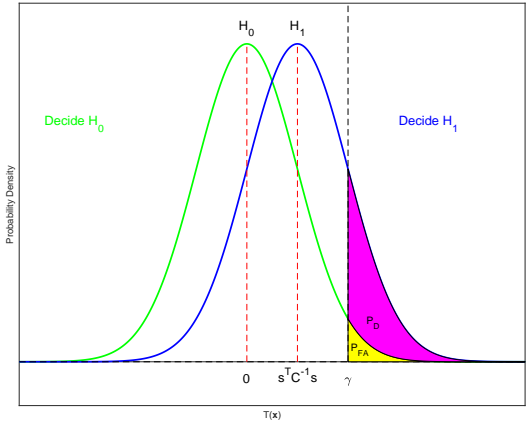}}
        \vskip -1cm
        \caption{Probability density function of the statistics $T(\xb)$ under the hypothesis $\Hc_0$ (noise-only hypothesis) and $\Hc_1$ (signal-present hypothesis). 
The detection-threshold is given by $\gamma$. The probability of false alarm ($\PFA$), called also probability of false detection,
and the probability of detection ($\PD$) are shown in red and magenta colors, respectively.}
        \label{fig:fig_gauss}
    \end{figure}

Any decision requires the definition of a criterion and, in this case, the Neyman-Pearson criterion is an effective choice. It consists of the maximization of the probability of detection $\PD$ under the constraint that the probability of false alarm $\PFA$ (i.e. the probability of a false detection) does not exceed a fixed value $\alpha$. According to the Neyman-Pearson theorem \citep[e.g. see ][]{kay98}, $\Hc_1$ has to be chosen when the statistic $\Tc(\xb)$ satisfies the inequality
\begin{equation} \label{eq:test1}
\Tc(\xb) =   \xb^T \Cb^{-1} \ssb = \xb^T \fb_s > \gamma,
\end{equation}
with
\begin{equation} \label{eq:mf}
\fb_s = \Cb^{-1} \ssb
\end{equation}
representing the matched filter \footnote{Strictly speaking, $\fb_s$ is not a filter since the term $\xb^T \fb_s$ in Eq.~\eqref{eq:test1} corresponds to a correlation and not to a convolution as required by the filtering operation.}. The detection threshold $\gamma$ for a fixed $\PFA = \alpha$ is given by
\begin{equation} \label{eq:gammap}
\gamma = \Phi_c^{-1}(\alpha) \sqrt{ \ssb^T \Cb^{-1} \ssb },
\end{equation}
where $\Phi_c^{-1}(.)$ is the inverse of the Gaussian complementary cumulative distribution function
\begin{equation}
\Phi_c(x) = 1- \Phi(x),
\end{equation}
with
\begin{equation}
\Phi(x) = \int_{t=-\infty}^{x} \phi(t) dt,
\end{equation}
and 
\begin{equation}
\phi(t) =\frac{1}{\sqrt{2 \pi}} \exp{\left( - \frac{1}{2} t^2 \right)}.
\end{equation}
The function $\Phi_c^{-1}(\alpha)$ provides the value of $x$ such that $\Phi_c(x) = \alpha$.
Equation~\eqref{eq:gammap} results from the fact that the statistic $\Tc(\xb)$ is a Gaussian random variable with variance $\ssb^T \Cb^{-1} \ssb$ and expected value equal to zero under the hypothesis $\Hc_0$ and $\ssb^T \Cb^{-1} \ssb$ under the hypothesis $\Hc_1$ (see Fig.~\ref{fig:fig_gauss}).

When the threshold $\gamma$ is fixed, the probability of false detection $\alpha$ can be computed by means of
\begin{equation} \label{eq:pfa}
\alpha = \Phi_c \left( \frac{\gamma}{\left[ \ssb^T \Cb^{-1} \ssb \right]^{1/2}} \right).
\end{equation}
For $\PFA=\alpha$, the probability of detection $\PD$ is 
\begin{equation} \label{eq:pd}
\PD= \Phi_c \left( \Phi_c^{-1} \left( \alpha \right) - \sqrt{ \ssb^T \Cb^{-1} \ssb} \right)
\end{equation}
 (again, see Fig.~\ref{fig:fig_gauss}).

\section{What is the {\rm SNR} maximization approach?} \label{sec:SNR}

A MF can be also derived as  the filter that maximizes the signal-to-noise ratio ({\rm SNR})
\footnote{Here, the quantity {\rm SNR} is defined as the ratio between the squared amplitude of the filtered signal
with the variance of the filtered noise.} or, in other words,
the filter which provides the greatest amplification of the signal with respect to the noise.  This can be obtained through the minimization of 
the variance of the filtered noise $\fb^T \nb$
with the constraint that $\fb^T \ssb = a$ (i.e. the filter $\fb$ does not modify the amplitude of the signal). Since the variance of the filtered noise is given by 
\begin{equation}
{\rm E}[(\fb^T \nb)(\fb^T \nb)]=\fb^T {\rm E}[\nb \nb^T] \fb=\fb^T \Cb \fb,
\end{equation}
the optimisation model is  \footnote{We recall that the functions $\arg\min F(x)$ and $\arg\max F(x)$
provide the values of $x$ of for which the function $F(x)$ has the smallest, respectively, the greatest value.}
\begin{equation}
\fb_{\rm SNR} = \underset{ \fb }{\arg\min}[\fb^T \Cb \fb - \lambda (\fb^T \ssb - a)] \label{eq:snr}
\end{equation}
with $\lambda$ a Lagrange multiplier. It can be shown that
\begin{equation}
\fb_{\rm SNR} = \frac{a}{\ssb^T \Cb^{-1} \ssb} \fb_s.
\end{equation}
This means that, apart from a normalizing factor, $\fb_{\rm SNR}$ is equivalent to $\fb_s$.

\section{What about if the amplitude ''a'' of ''s'' is unknown?} \label{sec:unknowna}

The results provided by the MF are independent of the amplitude $a$. Indeed, the test~\eqref{eq:test1} can be rewritten in the form
\begin{equation} \label{eq:test2}
\Tc(\xb) = \xb^T  \Cb^{-1} \gb = \xb^T \fb_g > \gamma',
\end{equation}
where $\gamma'=\gamma/a =Q^{-1}(\alpha) \sqrt{ \gb^T \Cb^{-1} \gb }$, and
\begin{equation} \label{eq:mf2}
\fb_g = \Cb^{-1} \gb.
\end{equation}
In this way, a statistic independent of $a$ is obtained. This means that, for a fixed $\PFA$, $\PD$ is maximized
also when the amplitude of the source is unknown. The only consequence is that $\PD$ cannot be evaluated in advance.

This holds also for the {\rm SNR} maximization approach since
\begin{equation} \label{eq:test2c}
\Tc_{\rm SNR}(\xb) =  \frac{\xb^T  \Cb^{-1} \gb}{\gb^T \Cb^{-1} \gb}= \xb^T \fb_{\rm SNR},
\end{equation}
with
\begin{equation}  \label{eq:test2b}
\fb_{\rm SNR} = \frac{\Cb^{-1} \gb}{\gb^T \Cb^{-1} \gb}.
\end{equation}
Since, under the hypothesis $\Hc_0$, the PDF of $\Tc_{\rm SNR}(\xb)$ is a zero-mean Gaussian with standard deviation equal to $1/\sqrt{\gb^T \Cb^{-1} \gb}$, 
the test~\eqref{eq:test2} becomes
\begin{equation} \label{eq:test3}
\Tc_{\rm SNR}(\xb) > \gamma'',
\end{equation}
with $ \gamma'' = Q^{-1}(\alpha) /  \sqrt{ \gb^T \Cb^{-1} \gb}$.

\section{Is there any relationship between the MF and the  least-squares fit of ''s'' to ''x''?}

In the framework of the least-squares (LS) approach, a detection is claimed when the estimate  $\hat{a}$ of the amplitude $a$  is statistically significant with respect to the noise level.
This quantity can be obtained from the optimization model
\begin{equation} \label{eq:ls}
\hat{a}=\underset{a }{\arg\min}[ (\xb - a \gb)^T \Cb^{-1} (\xb - a \gb)].
\end{equation}
The result is
\begin{equation}
\hat{a} = \frac{\xb^T \Cb^{-1} \gb}{\gb^T \Cb^{-1} \gb}.
\end{equation}
A comparison with Eq.~\eqref{eq:test2c} shows that $\hat{a}=\Tc_{\rm SNR}(\xb)$. Moreover,
since the optimization model~\eqref{eq:ls} is linear,  the PDF of $\hat{a} $ is a Gaussian with expected value $a$ and standard deviation $\sigma_{\hat{a}}=1/\sqrt{\gb^T \Cb^{-1} \gb}$.
In the context of the  LS approach, the estimate $\hat{a}$ is statistically different from zero, with a confidence level $\alpha$, when $\hat{a} > \eta$ with $\eta$ a threshold such that  the probability that this inequality is true when $a=0$ is $\alpha$. This is the same test as the test~\eqref{eq:test3} with $\eta = \gamma''$. This means that the MF and the least-squares approach are equivalent.

\section{Is there any efficient approach for the computation of the statistics ''T(x)''? }

In the case of long signals the computation of the statistic $\Tc(\xb)$ by means of Eqs.~\eqref{eq:test1} and \eqref{eq:mf} can be a very CPU-expensive operation since the size of matrix $\Cb$  can
become so huge to the point of not fitting into the RAM. For these reasons,
it is preferable to work in the Fourier domain. This requires the approximation of $\Cb$ with a circulant matrix,
\begin{equation} \label{eq:Ccirc}
\Cb \approx \left(
\begin{array}{ccccc}
c[0] & c[1] & \cdots & c[N-2] & c[N-1] \\
c[N-1] & c[0] & \cdots & c[N-3] & c[N-2] \\
\vdots & \vdots & \ddots & \vdots & \vdots \\
c[2] & c[3]& \cdots & c[0] & c[1] \\
c[N-1] & c[N-2]& \cdots & c[1] & c[0]
\end{array}
\right).
\end{equation}
For large matrices such approach has only secondary effects since a Toeplitz matrix may be treated as asymptotically equivalent to a circulant matrix \citep{dav79}. If $\Fmatfb_N$ denotes the discrete Fourier matrix
with entries $\Fmatf_N[j,k]= \omega^{jk}/\sqrt{N}$, where $\omega = \exp{(-2 \pi \imath /N)}$ and $j,k=0,1, \ldots, N-1$,
\begin{equation}
\Fmatfb_N = \frac{1}{\sqrt{N}} \left(
\begin{array}{cccccc}
1 & 1 & 1 & 1 & \ldots & 1 \\
1 & \omega & \omega^2 & \omega^3 & \cdots & \omega^{N-1} \\
1 & \omega^2 &\omega^4 & \omega^6 &\cdots &  \omega^{2(N-1)} \\
1 & \omega^3 &\omega^6 & \omega^9 &\cdots &  \omega^{3(N-1)} \\
\vdots & \vdots & \vdots & \vdots & \ddots & \vdots \\
1 & \omega^{N-1} &\omega^{2(N-1)} &\omega^{3(N-1)} & \cdots &  \omega^{(N-1)(N-1)}
\end{array}
\right),
\end{equation}
it happens that $\Fmatfb_N \hb = \htb$, where $\htb_N= {\rm DFT[\hb]}$ with ${\rm DFT[.]}$ the discrete Fourier transform (DFT) operator. In this case, the statistic $\Tc(\xb)$ in  Eq.~\eqref{eq:test2} can be written as
\begin{align}
\Tc(\xb) &= \xb^T \Fmatfb_N^{\dagger} \Fmatfb_N \Cb^{-1} \Fmatfb_N^{\dagger} \Fmatfb_N \gb; \\
&= \xb^T \Fmatfb_N^{\dagger} \widetilde{\Db}^{-1} \Fmatfb_N \gb.
\end{align}
This is because $\Fmatfb_N^{-1} =  \Fmatfb_N^{\dagger}$, i.e. $\Fmatfb_N^{\dagger} \Fmatfb_N=\Ib$, with $\Ib$ the identity matrix. Now, since the columns of $\Fmatfb_N$ constitute the eigenvectors of the positive definite matrix $\Cb$, this last can be diagonalized in the matrix $\Db$,
\begin{equation}
\Db= {\rm DIAG} \left[\ct[0], \ct[1], \ldots, \ct[N-1] \right],
\end{equation}
whose real positive entries are the elements of  $\ctb={\rm DFT[\cb]}$.
Because of this $\widetilde{\Db}^{-1}=\Fmatfb_N \Cb^{-1} \Fmatfb_N^{\dagger}$ is a diagonal positive definite matrix
$\widetilde{\Db}^{-1} = {\rm DIAG} \left[\ct^{-1}[0], \ct^{-1}[1], \ldots, \ct^{-1}[N-1] \right]$.
Hence,
\begin{equation} \label{eq:dft}
\Tc(\xb) = \xtb^{\dagger} (\gtb  \oslash \ctb).
\end{equation}

\section{Is the computation of the MF a numerically stable operation?}

\begin{figure}
        \resizebox{\hsize}{!}{\includegraphics{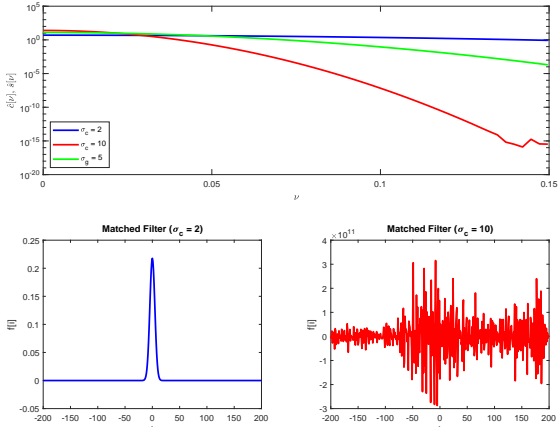}}
        \vskip-1.cm
        \caption{Top panel: Spectrum $|\ct[\nu]|$ (frequencies in Nyquist units $\nu / N$) of the Gaussian-shaped autocovariance function $c[d] = \exp(-0.5 d^2/\sigma_c^2)$ with, respectively,  $\sigma_c = 2$ (blue line) and $\sigma_c=10$ (red line) vs the spectrum 
        $|\st[\nu]|$ of a Gaussian-shaped signal $s[i] = a \exp(-0.5 i^2/\sigma_g^2)$ with $a =1$ and $\sigma_g=5$ (green line). Bottom panels: the corresponding MF.}
        \label{fig:fig_mf}
    \end{figure}

Despite its operational simplicity, there are situations where the computation of the MF is numerically unstable. This could be due to two different, although not necessarily independent, issues. 
The first issue is linked to the spectral characteristics of $\gb$ and $\nb$. To understand this point  it is advantageous to work in the Fourier domain. In particular, it is useful to write Eq.~\eqref{eq:dft} in  the explicit form
\begin{equation}
\Tc(\xb) =\sum_{i=0}^{N-1} \widetilde{x}^{*}[\nu_i] \ft[\nu_i],
\end{equation}
where
\begin{equation}
 \ft[\nu_i] = \frac{\gt[\nu_i]}{\ct[\nu_i]}
\end{equation}
is the DFT of the MF 
and $\nu_i \in  [-N/2 , -N/2 + 1 , \ldots, N/2 - 2, N/2 -1]$, for $N$ even, or $\nu_i \in [-N/2 , -N/2 + 1 , \ldots, N/2 - 1, N/2 ]$,  for $N$ odd, the discrete Fourier frequencies.
Now, if for $\nu_i \rightarrow N/2-1  \text{ (N even) or } (N+1)/2 \text { (N odd)}$ \footnote{Recall that ${\rm DFT}[\hb]$ provides an array $\htb$ symmetric with respect to its midpoint.} $\ctb$ goes to zero faster than $\gtb$,
the ratios $\gt(\nu) / \ct(\nu)$ increase without bound. This typically happens when the correlation length of $\gb$ is shorter than the correlation length of $\nb$ as, for example, a compact source on a slowly
changing background. As a
consequence the resulting MF will show a spurious heavy oscillating behavior due to the floating point representation of the real numbers (see Fig.~\ref{fig:fig_mf}).

The second issue occurs when the matrix $\Cb$ is numerically ill-conditioned. Typically, this can happen in the case of long signals when the support of $\gb$ (i.e. the interval where it is appreciably different from zero) and the correlation length of $\nb$ are much shorter than the length of $\xb$. This is because for $\nu_i \rightarrow N/2-1  \text{ (N even) or } (N+1)/2 \text { (N odd)}$ both $\gt[\nu_i]$ and $\ct[\nu_i]$ go very close to zero resulting numerically in a ratio $0/0$.

\section{What to do in the case of numerically unstable outcomes?} \label{sec:C_ill}

When $\ct[\nu]$ goes to zero faster than $\gt[\nu]$, the entries of $\nb$ change significantly only on sequences much longer than the support of $\gb$ (i.e. $n[i] \approx n[i + \Delta i_g]$ with $\Delta i_g$ the support of $\gb$). In this case, a way out to the ill-conditioning of $\Cb$ is a least-squares approach where the quantity 
\begin{equation}
	S^2 = (\xb - a \gb-\text{Pol}[\xb])^T \Cb^{-1} (\xb - a \gb - \text{Pol}[\xb]),
\end{equation}
with $\text{Pol}[\xb]$ a low-degree polynomial function, has to be minimized with respect to $a$ and the coefficients of $\text{Pol}[\xb]$. After that, the statistical significance of  $\at$ has to be tested against the variance of the residuals.

Conversely, if $\ctb$ goes to zero slower than $\gtb$, a solution is viable in the Fourier domain. It consists in forcing the ratio $\gt[\nu]/\ct[\nu]=0$ if $\ct[\nu] < {\rm tol}$ with ''${\rm tol}$'' a small value (e.g. $10^{-8}$). This corresponds to compute the Moore-Penrose pseudo-inverse of $\Cb$. An alternative, particularly useful in certain situations (e.g. isotropic noise and template $\gb$ with symmetric shapes) consists in the discretization of the MF obtained under the assumption that $\ssb$ and $\nb$ are continuous signals \citep[for details, see ][]{vio02}.

\section{What about if the position of ''s'' in ''x'' is unknown?} \label{sec:unknown}

In the previous section it was implicitly assumed that $\xb$ and $\gb$ have the same length $N$. In this way, the position of $\ssb$ in $\xb$ is {\it de facto} fixed. In practical application, however, the length $N_s$ of the support of $\ssb$ is shorter than $N$ and its position in $\xb$ unknown. Strictly speaking, under this condition, the MF could not be applied. This problem can be circumvented if the position of $\ssb$ is estimated before applying the MF.
This results in the standard procedure according to which a detection is claimed when
\begin{equation} \label{eq:testp}
\Tc_{i_p}[\xb] > u \hat{\sigma}_{\taub},
\end{equation}
where
\begin{equation} \label{eq:dectx}
\Tc_{i_p}[\xb] = \tau_{i_p},
\end{equation}
with
\begin{equation} \label{eq:dectt}
\tau_{i_p} = \max[\taub].
\end{equation}
Here, $\taub = [\tau[0], \tau[1], \ldots, \tau[N-N_s]$ with
\begin{equation} \label{eq:corrx}
\tau[i] = \sum_{j=i}^{i + N_s - 1} x[j] f_g[j-i]; \quad i=0, 1, \ldots, N - N_s,
\end{equation}
$i_p$ is the position of the greatest peak in $\taub$, $u$ a value typically in the range $[3, 5]$ and $\hat{\sigma}_{\taub}$ the standard deviation of the sequence $\taub$. In words,
the observed  signal $\xb$ is cross-correlated with the the template $\gb$,
the position $i_p$ of the greatest peak determined in the resulting sequence $\taub$ and finally the amplitude of this peak is tested whether it exceeds a threshold set to $u$ times $\hat{\sigma}_{\taub}$.
In the affirmative case the peak corresponds to a detection, otherwise it is considered noise. The rationale behind this procedure is that, if a signal $\ssb$ is really present then its most probable position corresponds
to the greatest peak in $\taub$ \citep{kay98}. Obviously this is not a certainty. However, if such peak does not pass the detection test, then 
no other peak as well any other entry in $\taub$ will do it.

The sequence $\taub$ can be obtained also working in the Fourier domain. In this case, two preliminary operations are necessary \citep[see chapter 13 in][]{pre07}. The first is to zero-pad $\gb$ in such a way to obtain the same length of $\xb$,
i.e. $\gb^T \rightarrow [0, \ldots, 0, \gb^T, 0, \ldots, 0] $. The second is that
$\gb$ must be arranged in wrap-around order, i.e. $[ g[0], g[1],  \dots, g[N-1], g[N] ]$ $\rightarrow$  $[ g[N^{\star}/2], g[N^{\star}/2 +1], \ldots g[N], g[1], g[2], \ldots, g[N^{\star}/2-1]]$, where $N^{\star} = N$ for $N$ even,
and $N^{\star} = N+1$ for $N$ odd with $N^{\star}/2$ the center of $\gb$. After that,
\begin{equation}
\taub = {\rm IDFT}[\xtb^* \odot \gtb \oslash \ctb],
\end{equation}
with $ {\rm IDFT}[.]$ the {\it inverse discrete Fourier transform} operator. 

\section{Why the computation of the Probability of False Alarm by means of the standard approach requires caution if the position of ''s'' in ''x'' is unknown?}

It is common practice that the $\PFA$ of a peak in $\taub$ is given by
\begin{equation} \label{eq:fd1}
\alpha = \Phi_c(u).
\end{equation}
However, such a practice is not correct since it can lead to severely underestimate $\alpha$ \citep{vio16, vio17, vio18}  \footnote{NB. in \citet{vio16} $\Phi_c (.)$ is erroneously denoted as  $\Phi(.)$.}.
Indeed, with the test~\eqref{eq:test2}, we check if at the true position of the hypothetical signal $\ssb$, the statistic $\Tc(\xb)$ exceeds the detection threshold. Under the hypothesis $\Hc_0$ (i.e. no signal is present in $\xb$), there is no reason why such a position must coincide with a peak. In fact, it corresponds to a generic point of the Gaussian noise process. This is the reason why the PDF of $\Tc(\xb)$ is a Gaussian. On the other hand, with the test~\eqref{eq:testp}, we check whether the highest peak of the sequence $\taub$ exceeds the detection threshold. Now, contrary to the previous case, under the hypothesis $\Hc_0$, the position $i_p$ does not correspond to a generic point of the Gaussian noise process, but rather to the subset of its local maxima. Since the PDF of the local maxima of a Gaussian random process is not a Gaussian, the PDF of $\tau[i_p]$ cannot be a Gaussian. In other words, the tests~\eqref{eq:test2} and  \eqref{eq:testp} are not equivalent. 

Once the sequence $\taub$ is standardised to zero-mean and unit-variance, it can be shown  \citep{vio16, vio17, vio19} that the correct $\PFA$ for a peak of amplitude $z$ is given by 
\begin{equation} \label{eq:corra}
 \alpha = \Psi_c(u),
\end{equation}
where \footnote{NB. in \citet{vio16} $\Psi_c(.)$ is denoted as  $\Psi(.)$, and in Eqs.~(24)-(25) the function $\Phi(.)$ has to be intended as the Gaussian cumulative distribution function and not its complement as it erroneously appears.}
\begin{equation}
\Psi_c(u)= 1 - \Psi(u)
\end{equation} 
with
\begin{equation}
\Psi(u)=\int_{-\infty}^{u} \psi(z) dz
\end{equation} 
and
\begin{equation} \label{eq:pdf_z1}
\psi(z) = \frac{\sqrt{3 - \kappa^2}}{\sqrt{6 \pi}} \exp{\left( -\frac{3 z^2}{2(3 - \kappa^2)} \right)} + \frac{2 \kappa z \sqrt{\pi}}{\sqrt{6}} \phi(z) \Phi\left(\frac{\kappa z}{\sqrt{3 - \kappa^2}} \right)
\end{equation}
providing the PDF of the local maxima of a zero-mean unit-variance smooth stationary one-dimensional continuous Gaussian random field  \citep{che15a, che15b}. Here,
\begin{equation} \label{eq:kd}
\kappa = - \frac{\rho'(0)}{\sqrt{\rho''(0)}},
\end{equation}
where $\rho'(0)$ and $\rho''(0)$ are, respectively, the first and second derivative with respect to $d^2$ of the autocorrelation function $\rho(d)$ at $d=0$. To notice that $\kappa=1$
corresponds to a Gaussian shaped $\rho(d)$. The condition of smoothness for the random
field requires that $\rho(d)$ be differentiable at least six times \footnote{This condition is required for a rigorous proof of the arguments by \citet{che15a, che15b}. However,  in real applications it is expected that
a function $\rho(r)$ differentiable four times is sufficient \citep{che16}.} with respect to $d$.
The left panel in Fig.~\ref{fig:fig_pdf_peaks1} shows that the PDF $\psi(z)$ peaks at greater values of $z$ than the Gaussian $\phi(z)$. Hence, as shown in the right panel of the same figure, the value 
of the $\PFA$ computed by means of Eq.~\eqref{eq:fd1} is systematically smaller than the correct one provided by Eq.~\eqref{eq:corra}.

\section{How to compute the correct $\PFA$?}

The correct computation of $\PFA$ requires that two problems have to be addressed. The first problem is that, strictly speaking, the equations above apply only to continuous signals. However, it is reasonable to expect that they can also be applied with good results to the discrete random fields if the support of $\rho(d)$ is not too ''narrow'' with respect to the pixel size, or too ''wide'' with respect to the area spanned by the data. In other words, the correlation length of the random field must be greater than the pixel size and smaller than the data length. 
The second problem is the estimation of parameter $\kappa$. If, as happens often, $\rho(d)$ is not available, one possibility, suggested by \citet{vio16}, is to estimate such function by fitting the discrete sample autocorrelation function of  $\taub$ with an appropriate analytical function. The reason is that the estimation of the correlation function of the noise is required also by the MF and, therefore, it is not an additional condition of the procedure.
However, a reliable estimate of $\rho'(0)$ and $\rho''(0)$ is a delicate issue. A robust alternative is to estimate $\kappa$ through a maximum likelihood approach
\begin{equation} \label{eq:ml}
\hat{\kappa} = \underset{\kappa }{\arg\max} \sum_{i=1}^{N_p} \log{\left(\psi(z_i; \kappa)\right)},
\end{equation}
where $\{ z_i \}$, $i=1,2,\ldots, N_p$, are the local maxima of $\taub$. 

\section{What about if also the number of signals ''s'' in ''x'' is unknown?}

If the number of signals $\ssb$ is unknown, the only consequence is that the detection test has to be applied to the most prominent peaks in $\taub$. Here, however, it is necessary to stress that the number of 
the signals $\ssb$ must be such to
not appreciably modify the statistical characteristics of the noise $\nb$ otherwise the estimation of parameter $\kappa$ via the ML approach~\eqref{eq:ml} will fail.

   \begin{figure}
        \resizebox{\hsize}{!}{\includegraphics{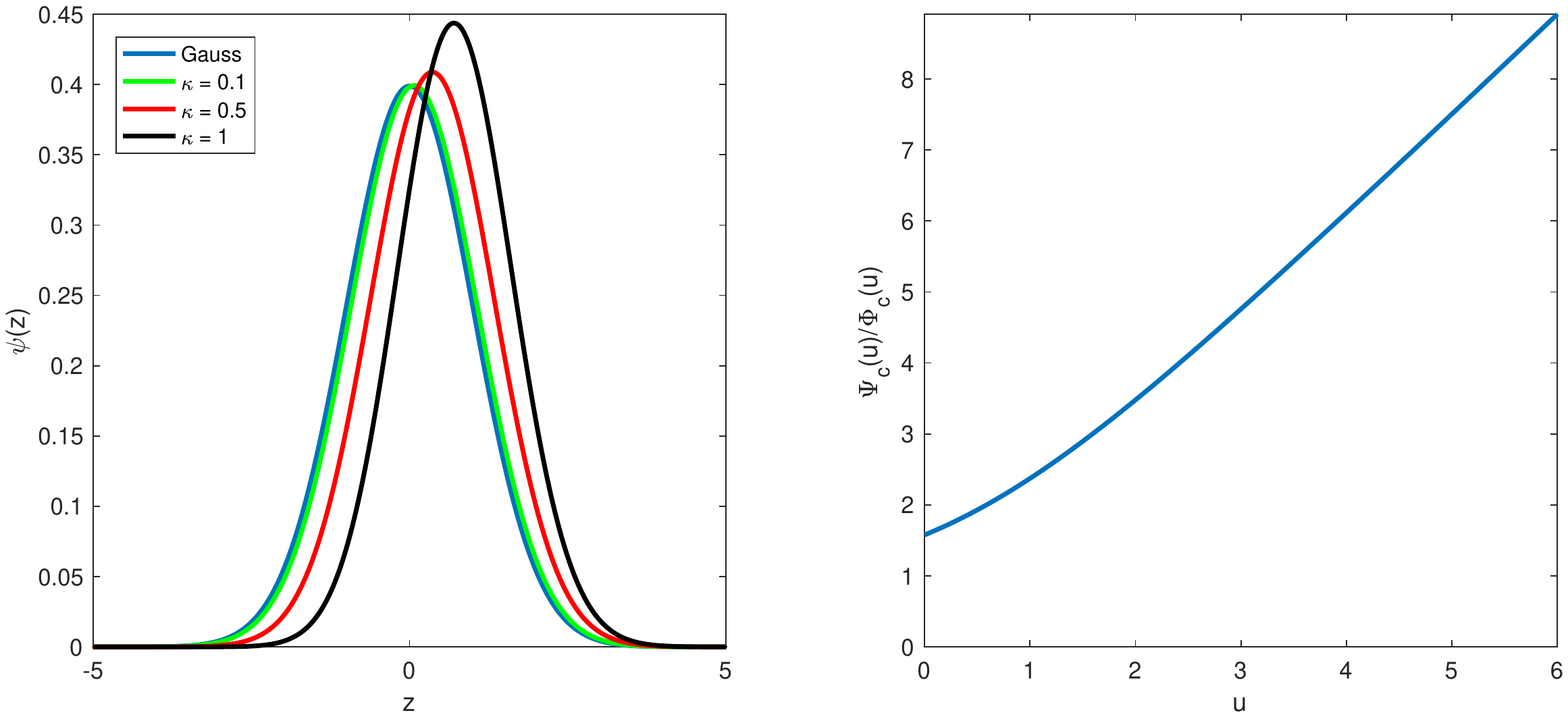}}
        \caption{Left panel: Comparison of the correct PDF $\Psi(z)$ of the amplitude $z$ of a peak in a stationary one-dimensional, zero-mean, unit-variance,  Gaussian random process with the standard Gaussian $\Phi(z)$ adopted in the classical detection procedure. Here, $\Psi(z)$ is computed for $\kappa = 0.1$ , $0.5$, and $1$.
Right panel: Ratio $\Psi_c(u) / \Phi_c(u)$ for the case $\kappa=1$ as function of the threshold $u$. As explained in the text, $\Psi_c(u) $ and $\Phi_c(u)$ provide the $\PFA$ as function of $u$ for the correct and the standard method, respectively.}
        \label{fig:fig_pdf_peaks1}
    \end{figure}

\section{Why does not the $\PFA$ provide the correct probability of false detection if the position of signal ''s'' in ''x'' is unknown?} \label{sec:spurious}

Contrary to what one could believe at first glance, the $\PFA$ given by Eq.~\eqref{eq:corra} does not provide the probability $\alpha$ that a specific detection is spurious but the probability that a generic peak due to the noise in $\taub$  
can exceed by chance the threshold $u$.
If $N_p$ peaks due to the noise are present in $\taub$, then a number  $\alpha \times N_p$  among them  is expected to exceed the prefixed detection threshold. 
For example, if in $\taub$ there are $1000$ peaks, then there is a high probability that a single detection with a $\PFA$ equal to $10^{-3}$ is spurious.
As a consequence, in spite of the low $\PFA$, the reliability of the detection is actually small.
A possible strategy  to avoid this problem is to fix a threshold $u$ such as $\alpha \times N_p \ll 1$. 
However, in this way there is the concrete risk to be too conservative and miss some true detections. The solution, after a preselection based on the $\PFA$, is to compute the probability of false detection for each 
specific detection.  We call it {\it specific probability of false alarm} ($\SPFA$).
This quantity can be computed by means of the order statistics, in particular by exploiting  the statistical characteristics of the greatest value of a finite sample of {\it identical and independently distributed} (iid) random variable from a given 
PDF  \citep{hog13}. Under the iid condition, the PDF $\upsilon(z_{\max})$ of the largest value among a set of $N_p$ peaks $\{ z_i \}$ is given by
\begin{equation} \label{eq:gz}
\upsilon(z_{\max}) = N_p \left[ \Psi(z_{\max}) \right]^{N_p-1} \psi(z_{\max}).
\end{equation}
Hence, the $\SPFA$ can be evaluated by means of
\begin{equation} \label{eq:intz}
\alpha = \int_{z_{\max}}^{\infty} \upsilon(z') dz'.
\end{equation}

The importance of the $\SPFA$ is demonstrated by Fig.~\ref{fig:fig_max_test1} where the PDF $\upsilon(z_{\max})$,
corresponding to the PDF $\psi(z)$ of the peaks of a stationary, zero-mean, unit-variance, Gaussian random process with $\kappa=1$, is plotted for three different values of the sample size $N_p$,
say $10^2$, $10^3$ and $10^4$.  The color filled areas provide the respective $\SPFA$  for a detection threshold $u$ corresponding to a $\PFA$ equal to $10^{-4}$.
It is evident that a detection threshold independent of $N_p$ is not able to quantify the risk of a false detection. 

\subsection{Is the iid condition for the amplitudes of the peaks of a random signal always satisfied?}

\begin{figure}
        \resizebox{\hsize}{!}{\includegraphics{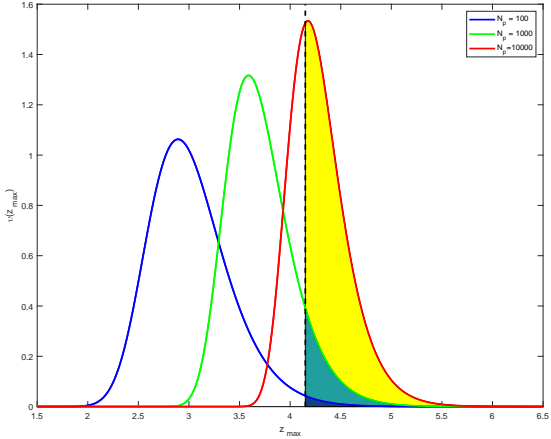}}
		\vskip -0.7cm
        \caption{PDF $\upsilon(z_{\rm max})$ of the greatest value of a finite sample of $N_p=10^2, 10^3$ and $10^4$ identical and independently distributed random variables from the PDF $\psi(z)$ of the peaks of a one-dimensional, zero-mean, unit-variance, stationary, random Gaussian process with $\kappa=1$. The color filled areas provide the respective $\SPFA$  for a detection threshold $u$ corresponding to a $\PFA$ given by Eq.~\eqref{eq:corra} equal to $10^{-4}$. It is evident that a detection threshold independent of $N_p$ is not able to quantify the risk of a false detection.}
        \label{fig:fig_max_test1}
    \end{figure}

In principle, the amplitudes of the peaks of a random signal are not necessarily iid. Therefore, before applying the previous procedure, such condition has to be checked.
A way to measure the degree of dependence of the peak amplitudes is the two-point correlation function $\rho_p[d]$. This discrete function is computed on a 
set of non-overlapping and contiguous distance bins of size $\Delta d$,
\begin{equation}
\rho_p[d]= \sum_{\substack{i,j = 1  \\ d - \Delta d/ 2 < i - j \le  d + \Delta d/ 2 }}^{N_d} \frac{z[i] z[j]}{N_d} / \sum_{i=1}^{N_p} \frac{z[i] z[i]}{N_p},
\end{equation}
with $N_d$ the number of peak couples with a distance in the range $(d - \Delta d/ 2, d + \Delta d/ 2]$. It measures the tendency of two peaks with similar value to be next to each other.
Hence,  if $\rho_p[d]$ is ''narrow'' with respect the
area spanned by the data (a necessary situation for the application of the MF), the iid condition can be expected to hold with good accuracy. The rationale is that two peaks with a distance $d$ such that $\rho_p[d] \approx 0$ 
are essentially independent. Therefore, most of the peaks in a signal $\xb$ of length $N \gg d_*$, with $d_*$ the distance for which  $\rho_p[d_*]$ is appreciably greater than zero,
can be expected to be approximately  iid. In this case,  Eq.~\eqref{eq:gz} is applicable but possibly with an effective number $N^+_p < N_p$ 
\citep[see Sect. 6 in][]{may05}. This last point is due to the dependence among a set of random variable which lowers its number of degrees of freedom \footnote{The term degrees of freedom refers to the number of
items that can be freely varied in calculating a statistic without violating any constraints.}.
 However, Fig.~\ref{fig:fig_max_test1} shows  that the exact value of
$N^+_p$ is not a critical quantity since $\upsilon(z_{\max})$ is a slow changing function of $N_p$. Therefore, for weakly dependent peaks it can be assumed that $N_p \approx N_p^+$.

\subsection{How to compute the $\SPFA$?} \label{sec:efficient}

The numerical evaluation of the integral~\eqref{eq:intz} does not present particular difficulties since
\begin{align}
\alpha & = N_p \int_{z_{\max}}^{\infty} \left[\Psi(z) \right]^{N_p-1} d\Psi(z); \\
          & = \left[ \Psi(z) \right]^{N_p} \Big|_{z_{\rm max}}^{\infty}; \\
          & = 1-\left[ \Psi(z_{\rm max}) \right]^{N_p}.
\end{align}
If the number of signals $\ssb$ present in  $\taub$ is unknown, the above procedure can be applied, in an order of decreasing amplitude, to all the peaks with a $\PFA$ smaller 
than a prefixed $\alpha$, and reducing $N_p$ of one unit after any confirmed detection. The last step is based on the rationale that if a peak can be assigned to a signal $\ssb$ in $\xb$, it can be removed from the set of the peaks related to the noise.

       \begin{figure}
        \resizebox{\hsize}{!}{\includegraphics{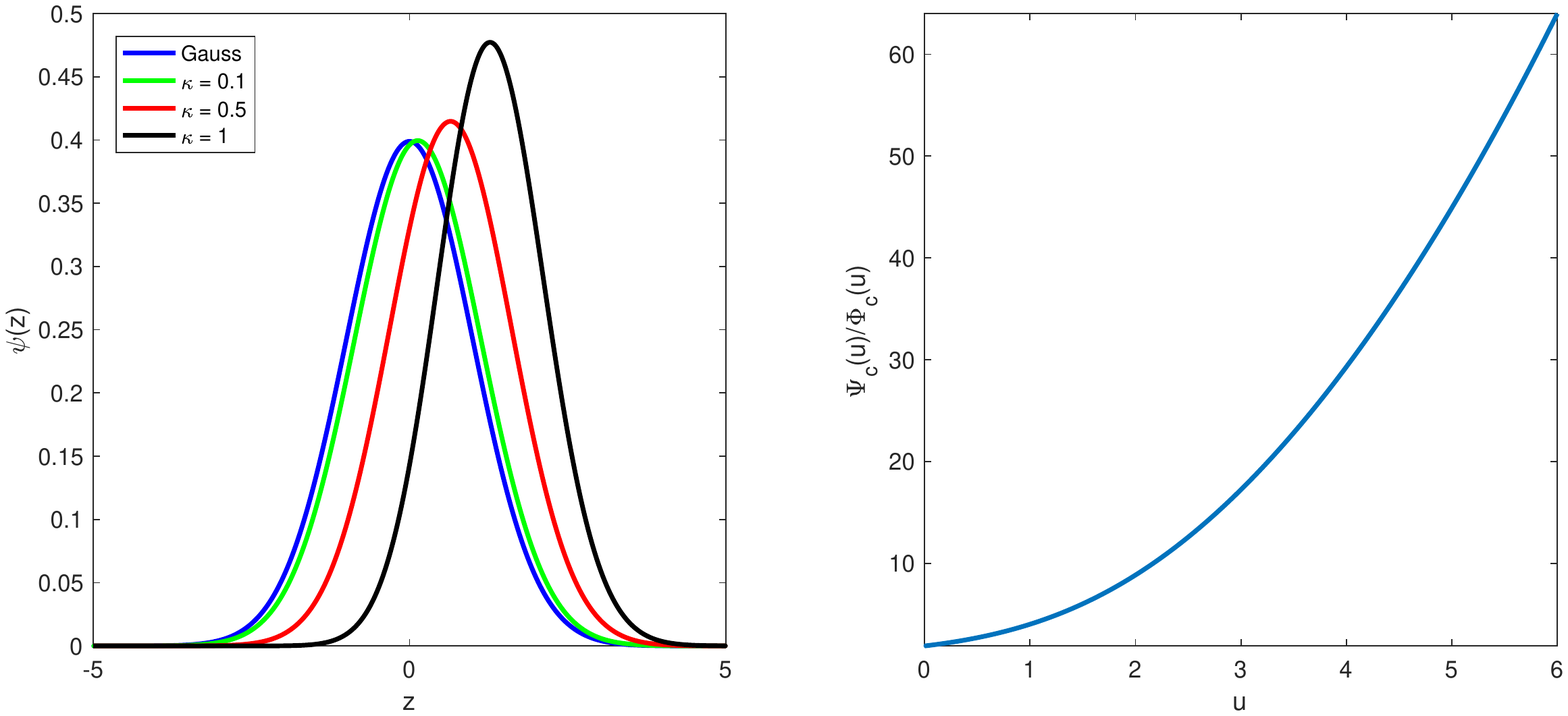}}
        \vskip -1cm
        \caption{Left panel: Comparison of the correct PDF $\Psi(z)$ of the amplitude $z$ of a peak in a stationary two-dimensional, zero-mean, unit-variance, isotropic, Gaussian random process with the standard Gaussian $\Phi(z)$ adopted in the classical detection procedure. Here, $\Psi(z)$ is computed for $\kappa = 0.1$ , $0.5$, and $1$.
        Right panel: Ratio $\Psi_c(u) / \Phi_c(u)$ for the case $\kappa=1$ as function of the threshold $u$. As explained in the text, $\Psi_c(u) $ and $\Phi_c(u)$ provide the $\PFA$ as function of $u$ for the correct and the standard method, respectively.}
        \label{fig:fig_pdf_peaks2}
    \end{figure}

\section{Is it possible to improve the rejection of spurious detections by means of a morphological analysis of the shape of the peaks?}

It is believed that it is possible to improve the rejection of spurious detections by means of a morphological analysis of the shape of the peaks in the sequence $\taub$. Such conviction is based on the assumption  that a peak produced by the signal $\ssb$ looks different from that produced by the noise. Unfortunately, the situation is more complex. 

Let assume, for the moment, that $\xb=\nb$ with $\nb$ a standard Gaussian white-noise. In this case, $\taub$ is a random Gaussian process with autocorrelation function $\rho[d]$ given by 
\begin{equation}
\rho[d] = \gb \star \gb, \quad r=0,1,2,\ldots, N-N_s
\end{equation}
where $\star$ denotes the correlation operator. Here, the point is that  $\ssb \star \gb \propto \gb \star \gb$ provides the shape of the signals $\ssb$ after the matched filtering. Moreover, the conditional expectation of  $\tau[i_p + d]$ given $i_p$ is
\begin{equation}
{\rm E}[\tau[i_p+d | i_p]] = a \rho[d],
\end{equation}
where $a = \tau[i_p]$ is the amplitude of the peak. This means that the expected shape of a peak in $\taub$ due to the noise is identical to that of the signal $\ssb$ after the matched filtering. 

Something similar holds when the noise is of non-white type. Indeed, under the hypothesis $\Hc_0$,  the quantity $\Tc(\xb) = \xb^T  \Cb^{-1} \gb$ in  Eq.~\eqref{eq:test2} can be written in the form $\Tc(\xb) = \yb^T \hb$ where $\yb^T = \xb^T \Cb^{-1/2}$ and
$\hb = \Cb^{-1/2} \gb$. Now, ${\rm E}[\yb \yb^T] = {\rm E}[\Cb^{-1/2} \xb \xb^T \Cb^{-1/2}] = \Cb^{-1/2} {\rm E}[\xb \xb^T] \Cb^{-1/2} = \Cb^{-1/2} \Cb \Cb^{-1/2} = \Ib$. This means that the case of a noise of non-white type can be brought back to that of a white type noise,
where now the matched filter takes the form $\fb = \hb$.

\section{Is it possible improve the detection performance of the MF?}

According to the Neyman-Pearson theorem, when the noise is of Gaussian type, the MF is the filter which provides the highest $\PD$ for a fixed $\PFA$. This means that no other filter can outperform it. Despite this, in the past some modified versions of the MF were proposed which, according to their authors, are able to improve the detection capability \citep{san01, bar03, lop05}. However, as shown in a series of papers \citep{vio02, vio04, vio17} such claim is not correct because they are based on wrong assumptions and incorrect numerical simulations.

\section{Are the previous arguments applicable to two-dimensional signals?}

The extension of MF to the two-dimensional $N \times M$ signals $\Xmatb$, $\Smatb$, $\Gmatb$ and $\Nmatb$ is conceptually trivial. Indeed, if one sets
\begin{align}
\xb & = {\rm VEC}[\Xmatb],\label{eq:stack1} \\
\ssb & = {\rm VEC}[\Smatb], \label{eq:stack2} \\
\gb & = {\rm VEC}[\Gmatb], \label{eq:stack3} \\
\nb & = {\rm VEC}[\Nmatb], \label{eq:stack4}
\end{align}
a problem similar to that of the one-dimensional signals is formally obtained. There are only two differences. The first is that
$\Cb$ becomes a $(N M) \times (N M)$ block-Toeplitz with Toeplitz-block (BTTB) matrix, i.e. a matrix which contains blocks that are repeated down the diagonals of the matrix, as a Toeplitz matrix has elements repeated down the diagonal. The individual block matrix elements are also Toeplitz matrices. The second concerns the PDF of local maxima that, in the case of a zero-mean, unit-variance, isotropic, Gaussian noise, is given by \citep{che15a, che15b}
\begin{multline} \label{eq:pdf_z2}
\psi(z) = \sqrt{3} \kappa^2 (z^2-1) \phi(z) \Phi \left( \frac{\kappa z}{\sqrt{2 - \kappa^2}} \right) + \frac{\kappa z \sqrt{3 ( 2 - \kappa^2)}}{2 \pi} \exp{\left( -\frac{z^2}{2 - \kappa^2} \right)}\\
+\frac{\sqrt{6}}{\sqrt{\pi (3 - \kappa^2)}} \exp{\left( -\frac{3 z^2}{2 (3-\kappa^2)} \right) } \Phi\left( \frac{\kappa z}{\sqrt{(3 - \kappa^2) (2 - \kappa^2)}} \right).
\end{multline}

Figure~\ref{fig:fig_pdf_peaks2}, to compare with Fig.~\ref{fig:fig_pdf_peaks1}, shows that for the two-dimensional signals the use of the correct $\PFA$ is even a more critical issue than for the one-dimensional signals.

\begin{figure}
        \resizebox{\hsize}{!}{\includegraphics{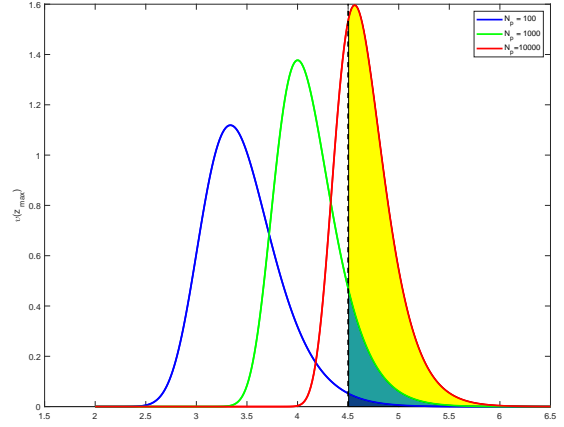}}
        \vskip -1cm
        \caption{PDF $\upsilon(z_{\rm max})$ of the greatest value of a finite sample of $N_p=10^2, 10^3$ and $10^4$ identical and independently distributed random variable from the PDF $\psi(z)$ of the peaks of a two-dimensional zero-mean unit-variance isotropic Gaussian random field with $\kappa=1$. The color filled areas provide the respective $\SPFA$  for a detection threshold $u$ corresponding to a $\PFA$~\eqref{eq:corra} equal to $10^{-4}$. It is evident that a detection threshold independent of $N_p$ is not able to quantify the risk of a false detection.}
        \label{fig:fig_max_test2}
    \end{figure}

Similarly,  Fig.~\ref{fig:fig_max_test2} shows the importance of the $\SPFA$. 
Here, the PDF $\upsilon(z_{\max})$, corresponding to the PDF $\psi(z)$ of the peaks of a stationary two-dimensional zero-mean unit-variance Gaussian random field with $\kappa=1$, is plotted for three different values of the sample size $N_p$, $10^2$, $10^3$ and $10^4$.  Again, the color filled areas provide the respective $\SPFA$  for a detection threshold $u$ corresponding to a $\PFA$ equal to $10^{-4}$.

In the two-dimensional case, a computational problem arises due to the fact that, even for maps of moderate size, the covariance matrix $\Cb$
becomes rapidly huge.  Hence, some efficient numerical methods based on a Fourier approach have to be used as in
\citet[][Chap.~5]{vog02},  \citet[][page $145$]{jai89}, \citet[][Appendix B]{els13}, and \citet[][Appendix A]{lag91}. This is because, similarly to the one-dimensional case the Fourier domain offers a simple solution 
when the Toeplitz blocks are approximated with circulant ones. 
In this way $\Cb$ becomes a block-circulant with circulant-blocks (BCCB) matrix which can be easily diagonalized. 
Indeed, the two-dimensional Fourier transform of a $N \times M$ image $\Xmatb$ is given by
\begin{equation} 
\Xmattb= \Fmatfb_N \Xmatb  \Fmatfb_M.
\end{equation}
Because a property of the $ {\rm VEC}[.]$ operator, this equation can be rewritten in the vectorized form
\begin{equation}
{\rm VEC}[\Xmattb] = (\Fmatfb_M^T \otimes \Fmatfb_N) {\rm VEC}[\Xmatb].
\end{equation}
Now, similarly to the one-dimensional case, with the BCCB approximation it is 
\begin{equation}
(\Fmatfb_M^T \otimes \Fmatfb_N ) \Cmatb^{\rm -1}  (\Fmatfb_M^T \otimes \Fmatfb_N)^* = \widetilde{\Db}^{-1}
\end{equation}
with $\widetilde{\Db}^{-1} = {\rm DIAG[{\rm VEC}[\oneb \oslash \Ctb]]}$, with $\oneb$ an $(N \times M) \times (N \times M)$ matrix of only ones. Hence,
\begin{equation}
T(\Xmatb) =\sum_{i=0}^{N-1} \sum_{j=0}^{M-1} \Xmatt^*[\nu_i, \nu_j] \frac{\Gmatt[\nu_i, \nu_j]}{\Cmatt[\nu_i, \nu_j]}.
\end{equation}

\subsection{An example of application of the MF to a two-dimensional signal}

\begin{figure}
        \resizebox{\hsize}{!}{\includegraphics{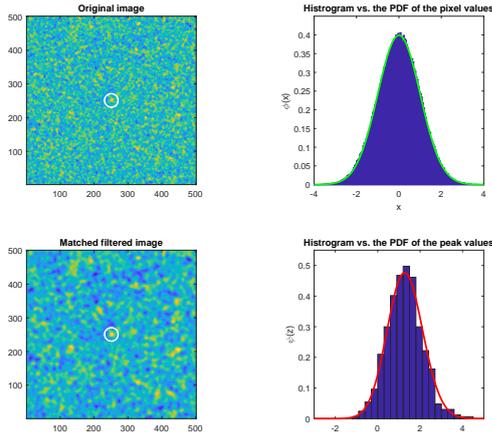}}
         \vskip -1cm
        \caption{Top-left panel: original map $\Xmatb$ (see text). The white circle marks the position of the source; Top-right panel: histogram of the pixel values of the map $\Xmatb$; Bottom-left panel: 
map $\Tmatb$ standardized to zero-mean and unit-variance. Again, the white circle marks the position of the source; Bottom-right panel: histogram vs the estimated PDF of the
amplitude of the peaks of the map $\Tmatb$.}
        \label{fig:fig_test}
    \end{figure}

As an example of application, we test the above methodology with a simulated two-dimensional, zero-mean, unit-variance, isotropic Gaussian random noise $\Nmatb$  with a circular Gaussian-shaped autocovariance function
\begin{equation}
c[d] = \exp{\left( -\frac{d^2}{2 \sigma_c^2} \right)},
\end{equation}
where $\sigma_c=3$, when a Gaussian-shaped source $\Smatc[i,j]= a \Gmatc[i,j]$ with $a=3.0$,
\begin{equation} \label{eq:tpsf}
\Gmatc[i,j]=  \exp{ \left( -\frac{(i-251)^2 + (j-251)^2}{2 \sigma_s^2} \right)}, \qquad i,j=0,1, \ldots, 500
\end{equation}
and $\sigma_s=5$, is placed in its center. The resulting map $\Xmatb$ is shown in the top-left panel of Fig.~\ref{fig:fig_test}. In the top-right panel of the same figure the histogramme of the pixel values is compared with the standard Gaussian PDF.

Working under the hypothesis that the position of the source is unknown, the procedure in Sect.~\ref{sec:unknown} requires that $\Xmatb$ be correlated with $\Cb^{-1} \Gmatb$. Given the size of the map, it is convenient to work in the Fourier domain. In this case, it is
\begin{equation} 
\Tmatc[i,j] = {\rm IDFT2}\left[\Xmatt^*[\nu_i, \nu_j] \frac{\Gmatt[\nu_i, \nu_j]}{\Cmatt[\nu_i, \nu_j]}\right],
\end{equation}
with $ {\rm IDFT2}[.]$ the {\it inverse two-dimensional Fourier transform}. Since matrix $\Cb$ is numerically ill-conditioned, as explained in Sec.~\ref{sec:C_ill}, the ratio $\Gmatt[\nu_i, \nu_j] / \Cmatt[\nu_i, \nu_j]$, and hence $\Tmatc[i,j]$, 
is set to zero whenever $\Cmatt[\nu_i, \nu_j] < 10^{-8}$.
As for the one-dimensional case, the computation of $\Tmatb$ requires that $\Gmatb$ is arranged in wraparound order. The bottom-left panel of Fig.~\ref{fig:fig_test} shows the
matched filtered map standardized to zero-mean and unit-variance. The bottom-right panel of the same figure shows the histogram of the values of the peaks of this map and compares it with  the maximum likelihood estimate of the PDF $\psi(z)$. This last provides $\kappa = 0.96$. The agreement is clearly good.

       \begin{figure}
        \resizebox{\hsize}{!}{\includegraphics{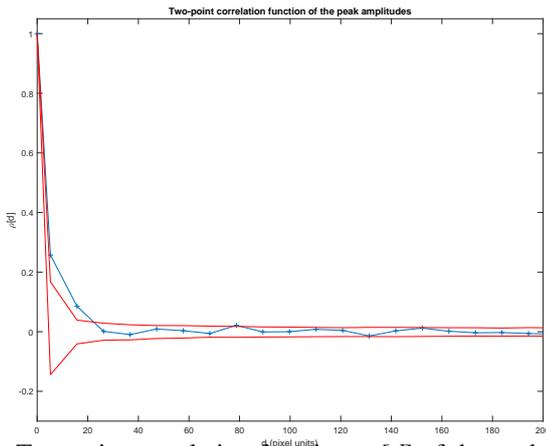}}
        \vskip -1cm
        \caption{Two-point correlation function $\rho_p[d]$ of the peak amplitudes in the bottom-left panel of Fig.~\ref{fig:fig_test}. The bin size is $10$ pixels. The two red lines define the $95\%$ confidence band. They are obtained by means of a bootstrap method based on the $95\%$ percentile envelopes of the two-point correlation functions obtained from $1000$ resampled sets of peaks with the same spatial coordinates as in the original signal but whose values are randomly permuted. }
        \label{fig:fig_corr_test}
    \end{figure}

\begin{figure}
        \resizebox{\hsize}{!}{\includegraphics{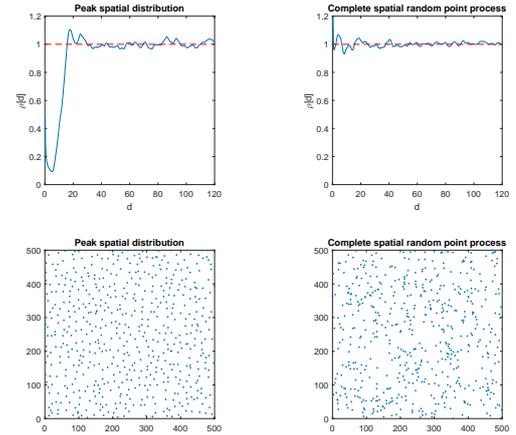}}
        \vskip -1cm
        \caption{Top-left panel: sample spatial pair-correlation function $\varrho[d]$ of the peaks in the Gaussian random field in Fig.~\ref{fig:fig_test}; Top-right panel: sample spatial pair-correlation function $\varrho[d]$ of a complete spatial random point process (CSRPP) with the same sizes and containing a number of points equal to number of peaks as in the Gaussian random field. The red lines provide the corresponding theoretical PDF's   due to a CSRPP; Bottom-right panel: spatial distribution of the peaks in the Gaussian random field; Bottom-left panel: realization of a CSRPP containing a number of points equal to number of peaks as in the previous panel.}
        \label{fig:fig_distance}
    \end{figure}

The $\PFA$ of the central source is $2.2 \times 10^{-4}$ and the  $\SPFA$ is $\approx 0.12$. Hence, the source is detected with a confidence level of about $88\%$.
However, the reliability of the estimated $\SPFA$ needs to verify the statistical independence of the peak amplitudes. Figure~\ref{fig:fig_corr_test} presents the corresponding two-point correlation function $\rho_p[d]$. Only the peaks closer than $15$ pixels show a correlation $\ge 0.1$. Hence, most of them can be considered iid. 
This is confirmed by the top-right panel of Fig.~\ref{fig:fig_distance}  where the sample pair-correlation function \footnote{The pair correlation function $\varrho(d)$
of the spatial distribution of a set of points is given by
$\varrho(d) = K'(d)/2 \pi r$ with $K'(d)$ the derivative of the Ripley's $K$-function with respect to $d$ \citep{bad16}.}  $\varrho(d)$
indicates that for $d \ge 15$ the spatial distribution of the peaks is compatible with a complete spatial random point process (CSRPP) for which, independently of $d$, it is
$\varrho(d)=1$. As it is visible also in the bottom panels of the same figure, the difference involves only the small scales where the lack of nearby points for the spatial distribution of the peaks with respect the CSRPP is apparent.  Since the autocorrelation function  of $\Tmatb$ goes to zero when approximately $d \ge 15$ pixels, this implies that most of the peak amplitudes can be considered iid.

\begin{figure}
	\resizebox{\hsize}{!}{\includegraphics{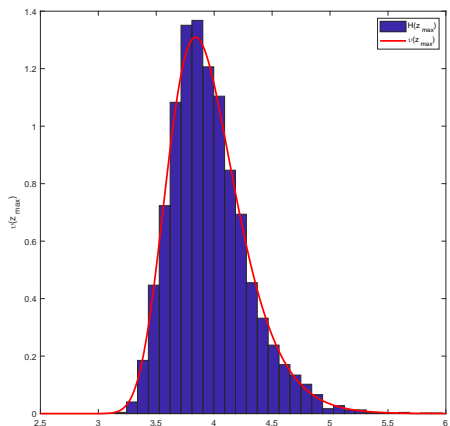}}
	\vskip -1cm
	\caption{Histogram $H(z_{\rm max})$ vs the expected PDF $\upsilon(z_{\rm max})$  of the largest peak value from $5000$ Gaussian random fields with characteristics similar to that shown in the bottom-left panel of Fig.~\ref{fig:fig_test}.
		Since each simulated map is characterized by a different number of peaks and of the value of $\hat{\kappa}$, the corresponding PDFs
		are slightly different one from another. For this reason, the displayed  $\upsilon(z_{\rm max})$ plotted in red corresponds to the mean number of peaks and of $\hat{\kappa}$.}
	\label{fig:fig_max}
\end{figure}

Finally, as countercheck of the the reliability of the iid condition for the peak amplitudes, Fig.~\ref{fig:fig_max} shows the good agreement of the histogram of the greatest peak value from a set of $5000$ Gaussian random fields with the same characteristics as the original $\Xmatb$ with the expected PDF $g({z_{\rm max}})$.

\section{Is the discretization of ''s'' and ''g'' a critical issue?}

\begin{figure}
        \resizebox{\hsize}{!}{\includegraphics{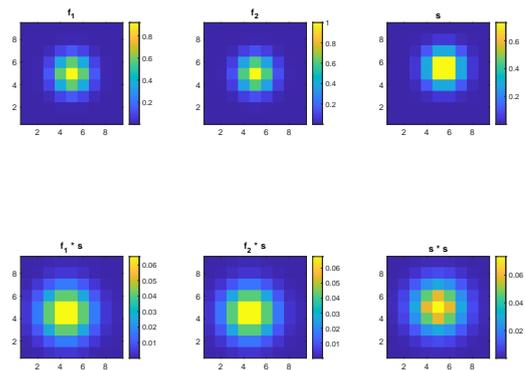}}
        \vskip -1cm
        \caption{Upper panels: MFs $\fb_1$ and $\fb_2$ and signal $\ssb$ as obtained by a circular bivariate Gaussian with standard deviation set to one pixel. For $\fb_1$ and $\fb_2$ the Gaussian is assumed to be placed at the middle of the central pixel of the map, whereas $\ssb$ is set midway between pixels in both dimensions. Moreover, $\fb_1$ and $\ssb$ are computed by integrating the Gaussian over the corresponding pixel, whereas $\fb_2$ is obtained by sampling the Gaussian at the central point of each pixel; Bottom panels: correlation of $\ssb$ with $\fb_1$, $\fb_2$ and itself, respectively. To remark is that only the latter case corresponds to the correct MF operation in the case of white-noise.}
        \label{fig:fig_discrete}
    \end{figure}

Up to now, we have worked under the implicit condition that $\ssb$ is a continuous signal sampled on a discrete grid of points and located at the center of a pixel (e.g. see Eq.~\eqref{eq:tpsf}). Actually, the position of $\ssb$ can vary within the pixel. Moreover, the value of a  specific pixel is given by the integral of $\ssb$ over the area of that pixel. However, unless the size of $\ssb$ is comparable to the dimension of the pixel, large effects on the results are not expected. This is visible in Fig.~\ref{fig:fig_discrete} where in the upper panels two versions, $\fb_1$ and $\fb_2$, of the MF are shown in addition to the signal $\ssb$. Both $\fb_1$ and $\fb_2$ as well $\ssb$ are a circular bivariate Gaussian with a standard deviation sets to one pixel. However, for $\fb_1$ and $\fb_2$ the Gaussian is assumed to be placed at the middle of the central pixel of the map, whereas $\ssb$
is set midway between pixels in both dimensions. Moreover, $\fb_1$ and $\ssb$ are computed by integrating the Gaussian over each pixel of the corresponding supports, whereas $\fb_2$ is obtained simply sampling the Gaussian at the central point of each pixel. The three bottom panels show the correlation of $\ssb$ with $\fb_1$, $\fb_2$ and itself, respectively. Under the hypothesis of white-noise, this operation corresponds to the MF filtering. The best results are obtained by the latter case which 
represents the correct application of the MF. In particular, $\fb_1$ and $\fb_2$ provide peak amplitudes that are, respectively, $11\%$ and $8\%$ smaller than the amplitude obtained with the correct procedure. This experiment reproduces an extreme situation, already with a standard deviation of the Gaussian set to $1.5$ pixels these percentages go down to $5\%$ and $3\%$. In conclusion, but in the case of signals $\ssb$ with a very small support with respect the pixel size, the effect of the discretization is negligible. 

\section{What about if the shape of template ''g'' is unknown?}

When the functional form of the template $\gb$ is not available there is no general procedure to obtain the MF. In astronomy, this kind of situations arise in source detection in digital images of the sky. Because of the optics of the telescopes,
an observed astronomical image is given by the convolution of the true sky with the point spread function (PSF). This last describes the response of an imaging system to a point-source. As a consequence, the template
$\gb$ is known only for this kind of objects. For an extended object, $\gb$ is given be the true shape of the object convolved with the PSF. Hence, in general this is unknown. However, in most cases the PSF  is
given by a smooth function. For this reason, there are situations where the unknown shape of $\gb$ is not a critical issue. 

\begin{figure}
		\resizebox{\hsize}{!}{\includegraphics{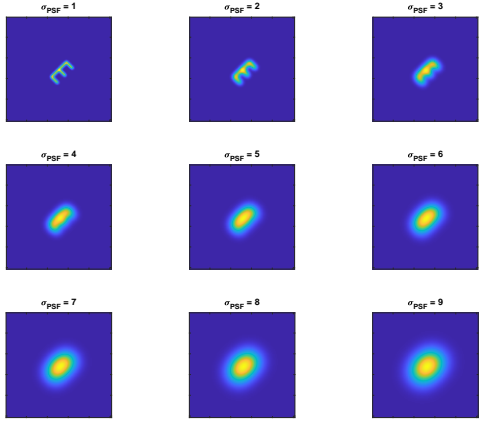}}
		\vskip -1cm
		\caption{Set of noise-free images used in the numerical experiment described in the text. Each map is obtained by convolving the original signal with a circular Gaussian PSF with dispersion $\sigma_{\rm PSF}$ and then normalized to have values in the range $[0, 1]$.}
		\label{fig:fig_E1}
	\end{figure}

\begin{figure}
        \resizebox{\hsize}{!}{\includegraphics{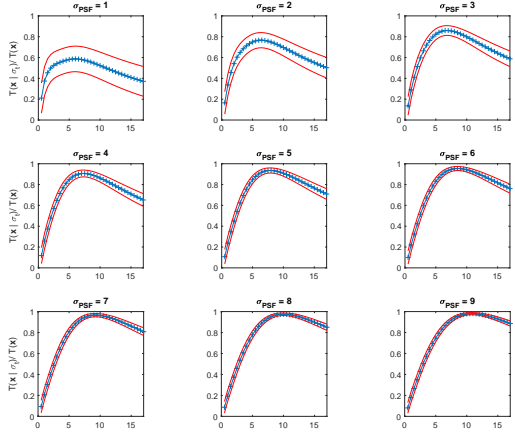}}
        \vskip -1cm
        \caption{Blue line: ratios $\Tc(\xb | \sigma_t)/\Tc(\xb)$ vs of the dispersion $\sigma_t$,  under the hypothesis that the position of the source is known. Here, $\Tc(\xb)$ is the statistic obtainable with the correct MF for the images in Fig.~\ref{fig:fig_E1} when added with a zero-mean, unit-variance, Gaussian white-noise, whereas $\Tc(\xb | \sigma_t)$ provides the same statistic when the MF is assumed to be a circular Gaussian PSF with dispersion  $\sigma_t=0.5, 1, 1.5 \ldots, 20$. For each $\sigma_t$
the value of the ratio are obtained as the mean of $1000$ numerical simulations of the added noise. Red lines: standard deviation interval of the estimated ratios.}
        \label{fig:fig_E2}
    \end{figure}

As an example, Fig.~\ref{fig:fig_E1} shows the effects on a E-shaped source of a circular Gaussian PSF 
\begin{equation}
{\rm PSF}[d]=\exp{\left( -\frac{d^2}{2 \sigma^2_{\rm PSF}}\right)}
\end{equation}
for different values of the dispersion $\sigma_{\rm PSF}$. All the images are standardized to have values in the range $[0, 1]$. Moreover, after the convolution with the PSF,  they are added to a zero-mean, unit-variance, Gaussian white-noise. The shape of the source is chosen because of its sharp edges which complicate the detection if the correct form of the MF is unknown.  
Under the hypothesis that the position of the source is known, Fig.~\ref{fig:fig_E2} shows the ratio $\Tc(\xb | \sigma_t)/\Tc(\xb)$,
where  $\Tc(\xb)$ represents the statistic obtainable with the correct MF,  here given by the source convolved with the PSF, whereas $\Tc(\xb | \sigma_t)$ provides the same statistic when the MF is assumed to be a circular Gaussian PSF with dispersion  $\sigma_t=0.5, 1, 1.5 \ldots, 20$. It is evident that, despite the clear asymmetry of the source, already for  $\sigma_{\rm PSF} = 3$ the circular Gaussian MF is able to provide reasonable good results for an opportune $\sigma_t > \sigma_{\rm PSF}$. This is confirmed in Fig.~\ref{fig:fig_PFA} which, for each $\sigma_{\rm PSF}$, shows the $\PFA$ corresponding to the MF, whose $\sigma_t$ provides the greatest $\Tc(\xb | \sigma_t)$, when the $\PFA$ corresponding to the correct MF
is $10^{-4}$. These results suggest that, in the case of extended objects, a way to produce a reasonable MF is to use a ''wider'' version of the PSF. The explanation for this fact is that, if the frequency characteristics of the noise is different from that of
the source, whatever filter which, more or less optimally, smooths out the Fourier frequencies of the noise will provide an improvement of the {\rm SNR}. 
The ''best'' width can be determined among a set of values, for instance,. those providing the greatest $\Tc(\xb)$.

\section{May the MF be extended to the multi-frequency signals?} \label{sec:multiple}

In the one-dimensional case, the arguments in Sects.{\ref{sec:neyman}-\ref{sec:unknowna} can be extended to the multi-frequency signals, i.e. when $M$ signals are available such that 
\begin{equation}
\xb_k = \ssb_k + \nb_k,  \quad k = 1, 2, \ldots, M;
\end{equation}
with
\begin{equation}
\ssb_k = a_k \gb_k.
\end{equation}
Here, $a_k$ is the amplitude of the signal in $k$th channel,
whereas the template $\gb_k$ represents the corresponding template. The array $\ab = [a_1, a_2, \ldots, a_M]^T$ is called the spectrum of $\ssb$. 
A typical example of this kind of signal is represented by multiple-frequency observations, i.e. when an astronomical object is observed at more frequencies.  

For ease of notation, all signals are assumed to have the same length $N$. In general, the amplitudes $\{ a_k \} $ as well as the
templates $\{ \gb_k \}$ are different for different $k$. However, if one sets
\begin{align}
\xb & = [\xb_1^T, \xb_2^T, \ldots, \xb_M^T]^T, \label{eq:mxb} \\
\ssb & = [\ssb_1^T, \ssb_2^T, \ldots, \ssb_M^T]^T, \label{eq:msb} \\
\nb & = [\nb_1^T, \nb_2^T, \ldots, \nb_M^T]^T, \label{eq:mnb} \\
\gb & = [\gb_1^T, \gb_2^T, \ldots, \gb_M^T]^T, \label{eq:mgb} \\
\fb & = [\fb_1^T, \fb_2^T, \ldots, \fb_M^T]^T, \label{eq:mfb}
\end{align}
it is possible to obtain a problem that is formally identical to that of the previous section. Similarly to Eq.~\eqref{eq:mf}, the MF is still given by $\fb_s = \Cb^{-1} \ssb$ and
is called {\it multi-frequency matched filter} (MMF). The only difference with the classic MF is that now the covariance matrix $\Cb$ is a $(N M) \times (N M)$ 
block matrix with Toeplitz blocks (BTB)
\begin{equation} \label{eq:covariance}
\Cb = 
\left( \begin{array}{ccc}
\Cb_{11} & \ldots & \Cb_{1M} \\
\vdots & \ddots & \vdots \\
\Cb_{M1} & \ldots & \Cb_{MM} \\
\end{array} \right),
\end{equation}
i.e. each of the $\Cb_{ij}$ blocks is constituted by a $N \times N$ Toeplitz matrix. In particular, 
$\Cb_{ii} = {\rm E}[\nb_i \nb_i^T]$ provides the autocovariance matrix of
the $i$th noise, whereas $\Cb_{ij} = {\rm E}[\nb_i \nb_j^T]$, $i \neq j$, the cross-covariance matrix between the $i$th and the $j$th channel.
When the noises $\{ \nb_k\}$ are uncorrelated, i.e. $\Cb_{ij} = \zerob$
for $i \neq j$, the test~\eqref{eq:test1} is equivalent to separately apply the MF to each $\xb_k$, normalized to unit variance by division with the quantity $\sigma^2_k={\rm E}[n_k^2]$ and then 
$\Tc(\xb)=\sum_{k=1}^M \Tc_k(\xb)$.

   \begin{figure}
    \resizebox{\hsize}{!}{\includegraphics{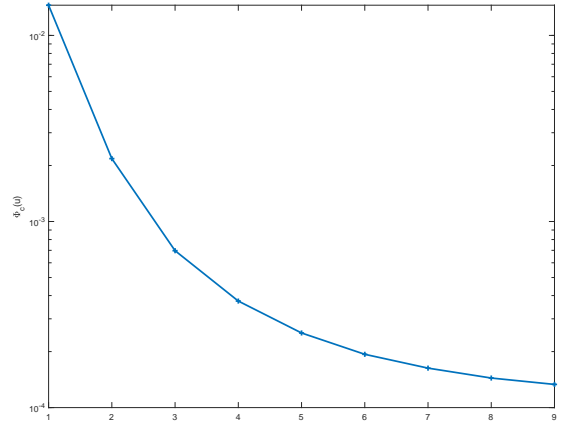}}
    \vskip -1cm
        \caption{$\PFA$ for different values of the dispersion $\sigma_{\rm PSF}$ of the PFS, corresponding to the MF whose $\sigma_t$ provides the greatest $T[\xb | \sigma_t]$ in Fig.~\ref{fig:fig_E2} when the $\PFA$ corresponding to the correct MF is $10^{-4}$.}
        \label{fig:fig_PFA}
    \end{figure}

\section{Is the MMF as optimal as the MF?}

MMF is an optimal filter only if the amplitudes $\{ a_k \}$ are known. Otherwise, when $M > 1$  the statistic $\Tc(\xb)$ cannot be written in a form equivalent to
Eq.~\eqref{eq:test2} and as a consequence it cannot be computed. This means that, if the spectral characteristics of the signal are unknown, the MMF is not applicable. 

\subsection{The need of a modified multi-frequency matched filter}

Since it is not possible to exploit the Neyman-Pearson theorem when the amplitudes $\{ a_k \}$ are unknown, an available alternative is to use the maximization of the {\rm SNR}.
Following this approach, the model~\eqref{eq:snr} can be modified as
\begin{equation} \label{eq:solu}
\fb_{\rm SNR} = \underset{ \fb }{\arg\min} [\fb^T \Cb \fb - \lambdab^T (\Sb^T \fb - \ab)],
\end{equation}
where $\fb = [\fb_1^T, \fb_2^T, \ldots, \fb_M^T]^T$ is an $(N M) \times 1$ array, 
$\lambdab = [\lambda_1, \lambda_2, \ldots, \lambda_M]^T$, $\ab = [a_1, a_2, \ldots, a_M]^T$ and $\Sb$ is a $(N M) \times M$ matrix
\begin{equation} \label{eq:S}
\Sb = \left(
\begin{array}{cccc}
\ssb_1 & \zerob & \ldots & \zerob \\
\zerob & \ssb_2 & \ddots & \zerob \\
\vdots & \ddots & \ddots & \zerob \\
\zerob & \zerob & \ldots & \ssb_M
\end{array} 
\right),
\end{equation}
with $\zerob = [0, 0, \ldots, 0]^T$ a $N \times 1$ array.
Now, since $\ab = {\rm DIAG}[\ab] \ooneb$, with $\ooneb = [1, 1, \ldots, 1]^T$, and $\Sb^T = {\rm DIAG}[\ab] \Gb^T$,
Eq.~\eqref{eq:solu} can be written in the form
\begin{equation} \label{eq:musnr}
\fb_{\rm SNR} = \underset{ \fb }{\arg\min} [\fb^T \Cb \fb - \lambdab_*^T (\Gb^T \fb - \ooneb) ],
\end{equation}
with $\lambdab_* = ({\rm DIAG}[\ab])^{-1} \lambdab$ and 
\begin{equation}
\Gb = \left(
\begin{array}{cccc}
\gb_1 & \zerob & \ldots & \zerob \\
\zerob & \gb_2 & \ddots & \zerob \\
\vdots & \ddots & \ddots & \zerob \\
\zerob & \zerob & \ldots & \gb_M
\end{array} 
\right).
\end{equation}
The solution is
\begin{equation} \label{eq:usnr}
\fb_{\rm SNR} = \Cb^{-1} \Gb (\Gb^T \Cb^{-1} \Gb)^{-1} \ooneb.
\end{equation}
It is evident that, contrary to the MMF,  with $\fb_{\rm SNR}$ 
it is possible to obtain a statistic $\Tc(\xb)$,
\begin{equation}
\Tc(\xb) = \xb^T \fb_{\rm SNR},
\end{equation}
independent of the unknown spectrum $\ab$. 
For a given threshold $\gamma$, the standard $\PFA$~\eqref{eq:fd1} is given by $\alpha$ with
\begin{equation}
\alpha = \Phi\left( \frac{\gamma}{[ \ooneb^T (\Gb^T \Cb^{-1} \Gb)^{-1} \ooneb]^{1/2} } \right),
\end{equation} 
that again is a quantity independent of the signal amplitude.
We call $\fb_{\rm SNR}$ the {\it modified multi-frequency matched filter} (MMMF).

\section{Is the MMMF as optimal as the MMF?}

It is necessary to stress that the MMMF does not share the optimal properties of MMF. For example, if the amplitude $a_k$ of the $k$th channel is zero (i.e. no signal is present), $(\fb_s)_k=\zerob$ but the same is not true for $(\fb_{\rm SNR})_k$. As a consequence, while the MMF does not use the $k$th channel in the computation of the statistics $\Tc(\xb)$, the same is not true for the 
MMMF. In the latter case, the effect of using a channel with no signal is to increase the variance of $\Tc(\xb)$.
Moreover, the MMMF is not invariant against the normalization of the templates $\{ \gb_k \}$. This implies that the MMMF is sensitive to the changes in the template amplitudes if these latter are different at the different frequencies. 
Hence, there must be an internal consistency of the normalization at the various frequencies. 
A useful normalization for each channel $k$ is $g_k[0] + g_k[1] + \ldots + g_k[N-1] = 1$.

A closer look at MMMF reveals that this filter corresponds to MMF when all the amplitudes $\{ a_k \}$ are identical. For this reason, the utility of MMMF could appear limited. However, the detection algorithms are useful only when looking for signals
with amplitudes at the level of the noise if not less (strong signals are detectable by naked eye). If there are channels where the corresponding amplitudes are much smaller than the other ones,
their contribution to the construction of the MMF is negligible. In practice, those channels are not used. This suggests that, when searching for signals with unknown spectra, a detection procedure should consist in applying MMMF to different
subsets of the available channels in such a way to determine those whose contribution is marginal.

\section{How to compute the $\PFA$ and the $\SPFA$ with the MMF and  the MMMF?} \label{sec:dan1}

The arguments presented in Sect.~\ref{sec:spurious}, concerning the computation of the $\PFA$ and the $\SPFA$ when the position of the signals $\ssb$ is unknown, can be extended to the MMF and  the MMMF with some adaptions since now $\fb_s$ and $\fb_{\rm SNR}$ are $(NM) \times 1$ arrays. In particular, each signal $\xb_k$ has to be filtered by the appropriate $\fb_k$ and summed up as shown in the following
\begin{equation}
 \tau[i] = \sum_{k=1}^M \sum_{j=i}^{i + N_s - 1} x_k[j] f_k[j-i]; \quad i = 0, 1, \ldots, N-N_s.
\end{equation}
In the case of the MMF each filter $ \fb_k $ is obtained from 
\begin{equation}
\fb_k=[[{\rm IVEC_{NM}}[ \fb_s]]_k  
\end{equation}
and in the case of the MMMF from
\begin{equation}
\fb_k=[ {\rm IVEC_{NM}}[ \fb_{\rm SNR}] ]_k,  
\end{equation}
where ${\rm IVEC}_{NM}[.]$ is an operator which for a given $(N M) \times 1$ array, provides a $N \times M$ matrix. It is the inverse of the ${\rm VEC[.]}$ operator. 

\subsection{Computational considerations} \label{sec:considerations}

In the case of  one-dimensional, multi-frequency, stationary signals with short auto-correlation and cross-correlation functions, it is useful to rearrange the elements of $\xb$, $\ssb$, $\nb$ and $\fb_s$ or $\fb_{\rm SNR}$ according to the so called {\it column rollout} order, i.e.,
\begin{equation} \label{eq:rollout}
\bar{\xb}= [\underline{\xb}^T[0],\underline{\xb}^T[1], \ldots, \underline{\xb}^T[N-1]]^T,
\end{equation}
with $\xbl[i] = [x_1[i], x_2[i], \ldots, x_M[i]]^T$, and similarly for $\bar{\ssb}$, $\bar{\nb}$, $\bar{\fb}_{\rm SNR}$ and $\bar{\fb}_s$. The MMMF $\bar{\fb}_{\rm SNR}$ is given by Eqs.~\eqref{eq:usnr},
with $\Cb$ and $\Gb$ replaced, respectively, by
\begin{equation} \label{eq:Cs}
\Cbl = \left(
\begin{array}{cccc}
\Cbl[0] & \Cbl[1] & \ldots & \Cbl[N-1] \\
\Cbl[-1] & \Cbl[0] & \ldots & \Cbl[N-2] \\
\vdots & \vdots & \ddots & \vdots \\
\Cbl[-(N-1)] & \Cbl[-(N-2)] & \ldots & \Cbl[0]
\end{array}
\right),
\end{equation}
where $\Cbl[d]={\rm E}[\nbl[i] \nbl^T[i+d]]$, and
\begin{equation} \label{eq:Ss}
\Gbl = [{\rm DCS}_0[\gbl_1], {\rm DCS}_1[\gbl_2], \ldots, {\rm DCS}_{M-1}[\gbl_M]],
\end{equation}
with
\begin{equation}
\gbl_k = [g_k[0], \zerob^T_{[M-1]}, g_k[1], \zerob^T_{[M-1]}, \ldots, g_k[N-1], \zerob^T_{[M-1]}]^T. 
\end{equation}
Here, ${\rm DCS}_l[.]$ denotes the {\it down circulant shifting operator} that circularly down shifts the elements of a column array 
by $l$ positions. Similarly, the MMF $\fb_s$ is given by Eq.~\eqref{eq:mf} with $\Cb$ and $\ssb$ replaced,respectively, by $\Cbl$ and
\begin{equation} \label{eq:Sss}
	\Ssbl = [{\rm DCS}_0[\ssbl_1], {\rm DCS}_1[\ssbl_2], \ldots, {\rm DCS}_{M-1}[\ssbl_M]],
\end{equation}
with
\begin{equation}
	\sbl_k = [s_k[0], \zerob^T_{[M-1]}, s_k[1], \zerob^T_{[M-1]}, \ldots, s_k[N-1], \zerob^T_{[M-1]}]^T. 
\end{equation}

In case of noises ${\nb_k}$ of white type and correlated
each other only in correspondence to the same index} $i$ (i.e. $E[n_k[i] n_l[j]] = 0$ if $i \neq j$), the matrix $\Cbl$ becomes a block diagonal,
\begin{equation} \label{eq:Cs0}
\Cbl = \left(
\begin{array}{cccc}
\Cbl[0] & \zerob & \ldots & \zerob \\
\zerob & \Cbl[0] & \ldots & \zerob \\
\vdots & \vdots & \ddots & \vdots \\
\zerob & \zerob& \ldots & \Cbl[0]
\end{array}
\right),
\end{equation}
i.e., a form easy to deal with. 

The arrangement~\eqref{eq:rollout} is useful also if the MMF and MMMF are computed in the Fourier domain. Indeed, by imposing periodic boundary conditions on the covariance function of each signal  
$\xb_i$ (i.e. ${\rm E}[x_i[l] x_i[l+k]] = {\rm E}[x_i[l] x_i[l+N-1-k]]$), it happens that each Toeplitz matrix $\Cb_{ij}$ can be approximated by a circulant matrix. For example, under this approximation, it can be shown  \citep{kay98} 
that the discrete Fourier transform  $\ftbl_{\rm SNR}$ of $\fbl_{\rm SNR}$ is given by
\begin{equation} \label{eq:Utbl}
\ftbl_{\rm SNR} = \Sigmatbl^{-1} \Gtbl (\Gtbl^H \Sigmatbl^{-1} \Gtbl)^{-1}.
\end{equation}
Here, $\Sigmatbl$ is a block diagonal matrix
\begin{equation}
\Sigmatbl = \left( \begin{array}{ccl}
\Sigmatbl_0 & &  \Large{\zerob} \\
& \ddots & \\
\Large{\zerob} & & \Sigmatbl_{N-1}
\end{array} \right),
\end{equation}
with
\begin{equation}
\Sigmatbl_i = \left( \begin{array}{cccc}
\Pb_{11}(\nu_i) & \Pb_{12}(\nu_i) & \ldots & \Pb_{1M}(\nu_i) \\
\vdots & \vdots & \ddots & \vdots \\
\Pb_{M1}(\nu_i) & \Pb_{M2}(\nu_i) & \ldots & \Pb_{MM}(\nu_i)
\end{array} \right),
\end{equation}
$\nu_i = i/N$, $i=0, 1, \ldots, N-1$, and $\Pb_{kl}(\nu_i)$  representing the cross power-spectrum at frequency $\nu_i$ between $\nb_k$ and $\nb_l$. These quantities can be computed by means of the discrete Fourier transform
of the correlation and cross-correlation functions of the signals $\{ \xb_k \}$. Matrix $\Gtbl$ is given by
\begin{equation}
\Gtbl = [{\rm DCS}_0[\gtbl_1], {\rm DCS}_1[\gtbl_2], \ldots, {\rm DCS}_{M-1}[\gtbl_M]],
\end{equation}
with
\begin{equation}
\gtbl_k = [\gt_k[\nu_0], \zerob^T_{(M-1)}, \gt_k[\nu_1], \zerob^T_{(M-1)}, \ldots, \gt_k[\nu_{N-1}], \zerob^T_{(M-1)}]^T.
\end{equation}
where the entries $\{ \gt_k[\nu_i] \}$  are obtained by means of the discrete Fourier transform of the array $\gb_k$.
The advantage of such an approach is that both $\Sigmatbl$ and  $\Gtbl$ are highly sparse matrices. As a consequence, $\fb_{\rm SNR}$ can be computed very efficiently by means of block-matrix operations.

\subsection{An example of application of the MMMF}

\begin{figure}
        \resizebox{\hsize}{!}{\includegraphics{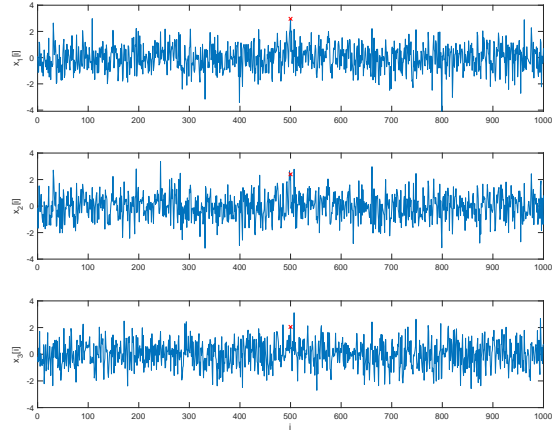}}
        \vskip -0.7cm
        \caption{Simulated multi-frequency signal $\{ \xb_k \}$. The red crosses indicate the position of the signals $\ssb_k$.}       
        \label{fig:original_1}
    \end{figure}

\begin{figure}
        \resizebox{\hsize}{!}{\includegraphics{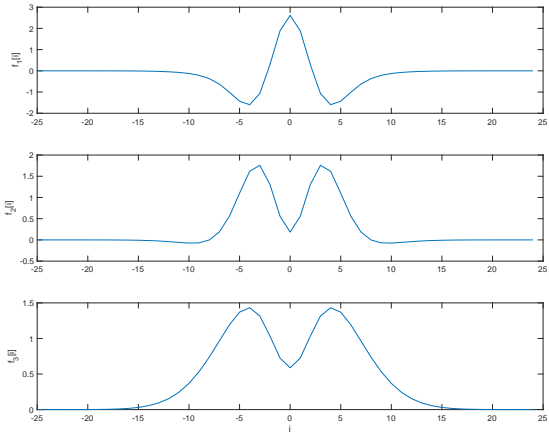}}
        \vskip -0.7cm
        \caption{MMMF filters obtained for the multi-frequency signal in Fig.~\ref{fig:original_1} . }
        \label{fig:MF_1}
    \end{figure}

\begin{figure}
        \resizebox{\hsize}{!}{\includegraphics{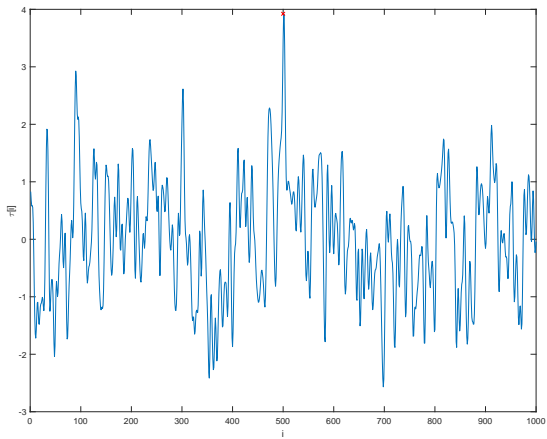}}
        \vskip -0.7cm
        \caption{Final data sequence  $\taub$ obtained after the MMMF filtering of the multi-frequency signal $\{\xb_k\}$ shown in Fig.~\ref{fig:original_1}. The red cross indicates the position of the searched signal.}
        \label{fig:result_1}
    \end{figure}

\begin{figure}
    \vskip -2cm
        \resizebox{\hsize}{!}{\includegraphics{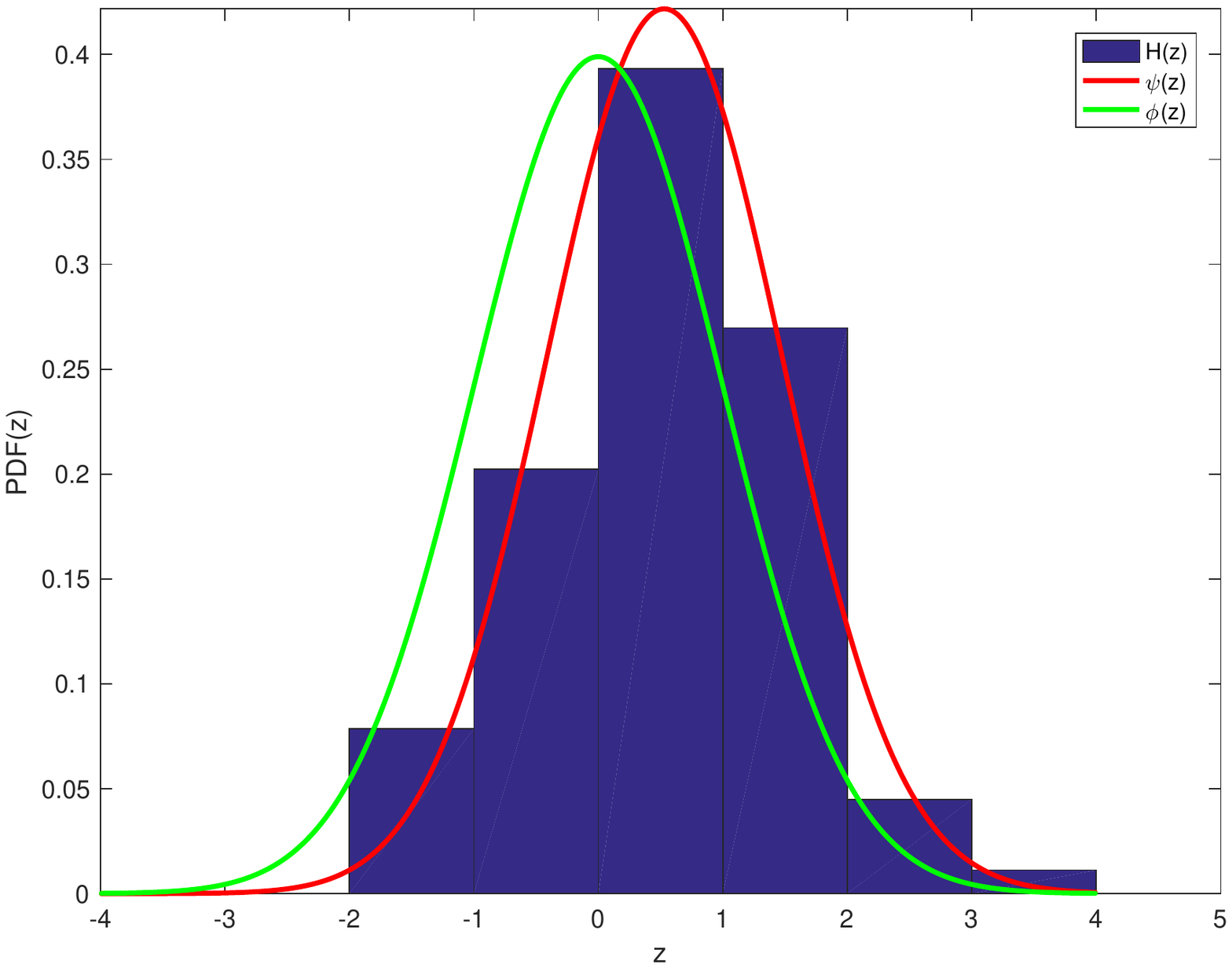}}
        \vskip -3cm
        \caption{Histogram $H(z)$ of the values of the peaks in the data sequence $\taub$ of Fig.~\ref{fig:result_1}. For comparison, the PDF $\psi(z)$ (red curve), corresponding to 
        	the maximum likelihood estimate $\kappa = 0.75$, and the standard Gaussian PDF $\phi(z)$ (green curve) are shown.} 
        \label{fig:pdf_1}
    \end{figure}

\begin{figure}
	\resizebox{\hsize}{!}{\includegraphics{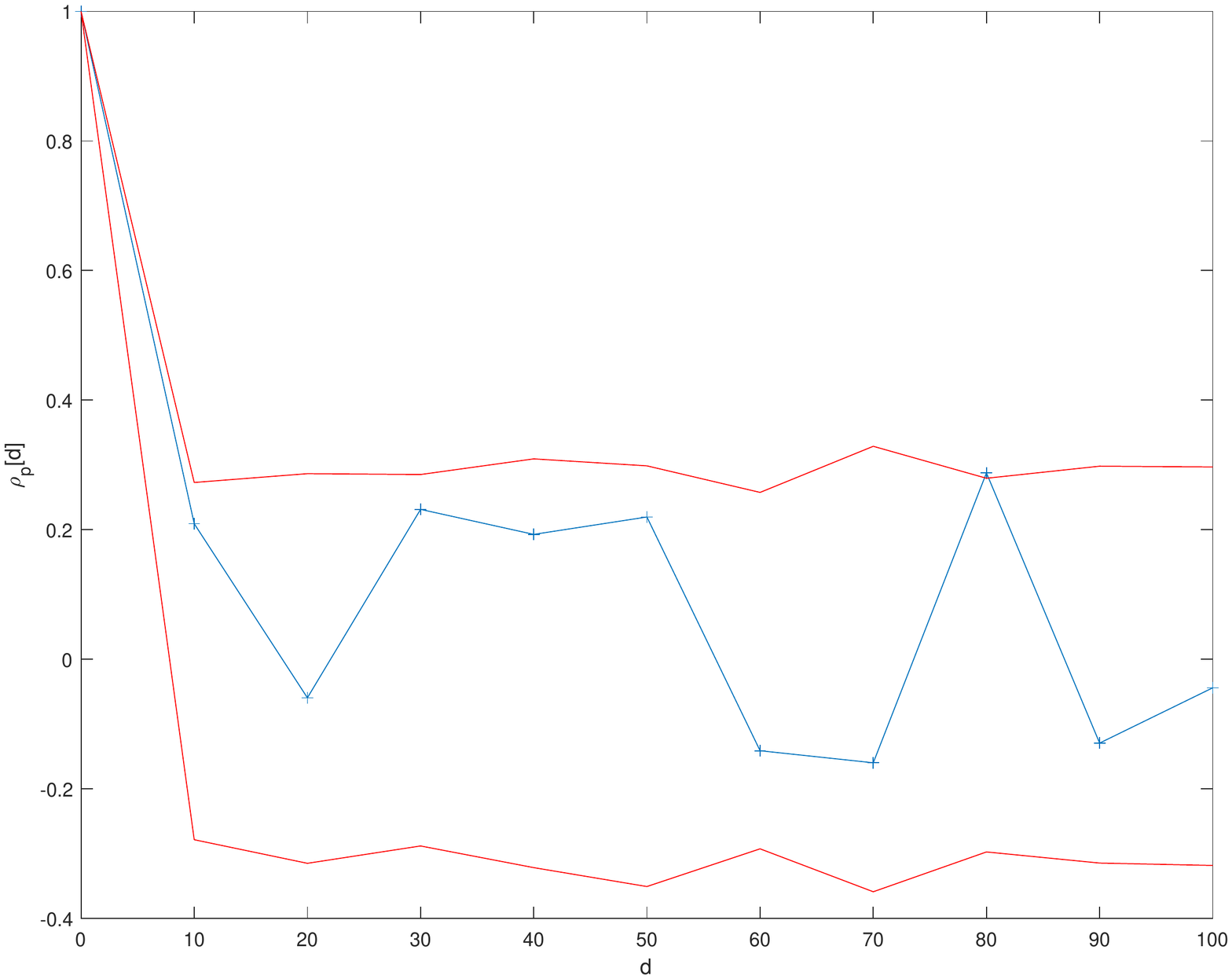}}
	\vskip -1cm
	\caption{Two-point correlation function $\rho_p[d]$ for the peaks in the signal $\taub$ in Fig.~\ref{fig:result_1}. The two red lines define the $90\%$ confidence band. They are  obtained by means of a bootstrap method based on the $90\%$ percentile envelopes of the two-point correlation functions obtained from $1000$ resampled sets of peaks with the same spatial coordinates as in the original signal but whose 
		values are randomly permuted.}
	\label{fig:fig_corr_test1}
\end{figure}

As an example of application of the MMMF, Fig.~\ref{fig:original_1} shows a multi-frequency signal $\xb_k = \ssb_k + \nb_k$, $k=1,2,3$, of length $N=1000$, where $s_k$ 
\begin{equation}
s_k[i] = a_k \exp{\left(- \frac{(i-N/2)^2}{ 2 \sigma_k^2}\right)},
\end{equation}
with $\ab = [2.0, 1.0, 0.5]$, $\sigmab = [2, 3 , 5]$, and $\{ n_k \}$ cross-correlated white-noise processes with a cross-correlation $\Cbl[0] = {\rm E}[\nbl[0] \nbl^T[0]]$, $\nbl(0) = [n_1[0], n_2[0], \ldots, n_M[0]]^T$
given by
\begin{equation} \label{eq:matrix0}
\Cbl[0] = \left(
\begin{array}{ccc}
1.0 & 0.8 & 0.5 \\ 
0.8 & 1.0 & 0.5 \\
0.5 & 0.5 & 1.0
\end{array}
\right).
\end{equation} 
This numerical experiment simulates an unfavorable situation where the noises are strong and highly correlated. The MMMF corresponding to this signal
is shown in Fig.~\ref{fig:MF_1}. It produces
the sequence $\taub$ in Fig.~\ref{fig:result_1} which contains $89$ peaks whose histogram $H(z)$ is shown in Fig.~\ref{fig:pdf_1}. In the same figure the PDF $\psi(z)$, corresponding to 
the maximum likelihood estimate $\kappa = 0.75$, and the standard Gaussian $\phi(z)$ are shown. 
The iid condition of the peaks, necessary for the computation of the $\SPFA$, is supported by the $\rho_p[d]$ in  Fig.~\ref{fig:fig_corr_test1} which is almost completely contained in its $90\%$ confidence band. This last is assigned as the $90\%$ percentile envelopes of the two-point correlation functions obtained from $1000$ resampled sets of peaks with the same spatial coordinates as in the original signal but whose values are randomly permuted.

The MMMF detects $\ssb$ with a $\PFA$
equal to $2.0 \times 10^{-4}$. This value is about five times greater than the value provided by the standard $\PFA$~\eqref{eq:fd1}, i.e. $4.4 \times 10^{-5}$. Given the small number of peaks and the small $\PFA$, the reliability of 
this detection is high. Indeed, the corresponding $\SPFA$ is $6.1 \times 10^{-2}$, i.e. the probability that this specific peak is due to the noise is only of $6\%$. 

\section{Is it possible to extend the MMF and MMMF to the two-dimensional case?} \label{sec:twodimensional}

Similar arguments concerning the two-dimensional MF hold for the MMF and the MMMF if one sets
\begin{align}
\ssb & = {\rm VEC}\left[ {\rm VEC}[ \Smatb_1 ], {\rm VEC}[ \Smatb_2 ], \ldots, {\rm VEC}[ \Smatb_M ] \right]; \label{eq:stack12} \\
\xb & = {\rm VEC}\left[ {\rm VEC}[ \Xmatb_1 ], {\rm VEC}[ \Xmatb_2 ], \ldots, {\rm VEC}[ \Xmatb_M ] \right]; \label{eq:stack22} \\
\nb & = {\rm VEC}\left[ {\rm VEC}[ \Nmatb_1 ], {\rm VEC}[ \Nmatb_2 ], \ldots, {\rm VEC}[ \Nmatb_M ] \right]. \label{eq:stack32}
\end{align}
The difference is that now $\Cb$
is a $(M N_p) \times (M N_p)$ block matrix where each of the $\Cb_{ij}$ blocks is constituted by a $N_p \times N_p$  block Toeplitz with Toeplitz blocks (BTTB) matrix. In particular, $\Cb_{ii}$ provides the autocovariance matrix of the $i$th noise, whereas $\Cb_{ij}$, $i \neq j$, the cross-covariance matrix between the $i$th and the $j$th noises. Also in this case it is possible to exploit an efficient numerical approach 
which works in the Fourier domain \citep{gal05}.
Likewise the one-frequency case, before carrying out the MF filtering, the arrays $\fb$, $\ssb$, $\xb$ have to be reshaped as matrices with the sizes of the 
corresponding original maps.

\subsection{Computational considerations for the two-dimensional MMMF} \label{sec:considerations2}

Also in the two-dimensional case the colum rollout arrangement of the entries of $\bar{\xb}$ in the form \eqref{eq:rollout} with
$\xbl(i) = {\rm VEC} [x_1[j,l], x_2[j,l], \ldots, x_M[j,l]]^T$, $i=1, 2, \ldots, N_p$, and similarly for  $\bar{\ssb}$, $\bar{\nb}$, $\bar{\gb}$, $\bar{\fb}_{\rm{SNR}}$ or $\bar{\fb}_{s}$,
provides a covariance matrix $\Cbl$ and a matrix $\Gbl$ or $\Ssbl$ as given in Eqs.~\eqref{eq:Cs} and \eqref{eq:Ss} or \eqref{eq:Sss}.
Matrix $\Cbl$ presents some computational advantages since in the Fourier domain can be block-diagonalized \citep{kat93}.

\subsection{An example of application of the two-dimensional MMMF}

\begin{figure}
        \resizebox{\hsize}{!}{\includegraphics{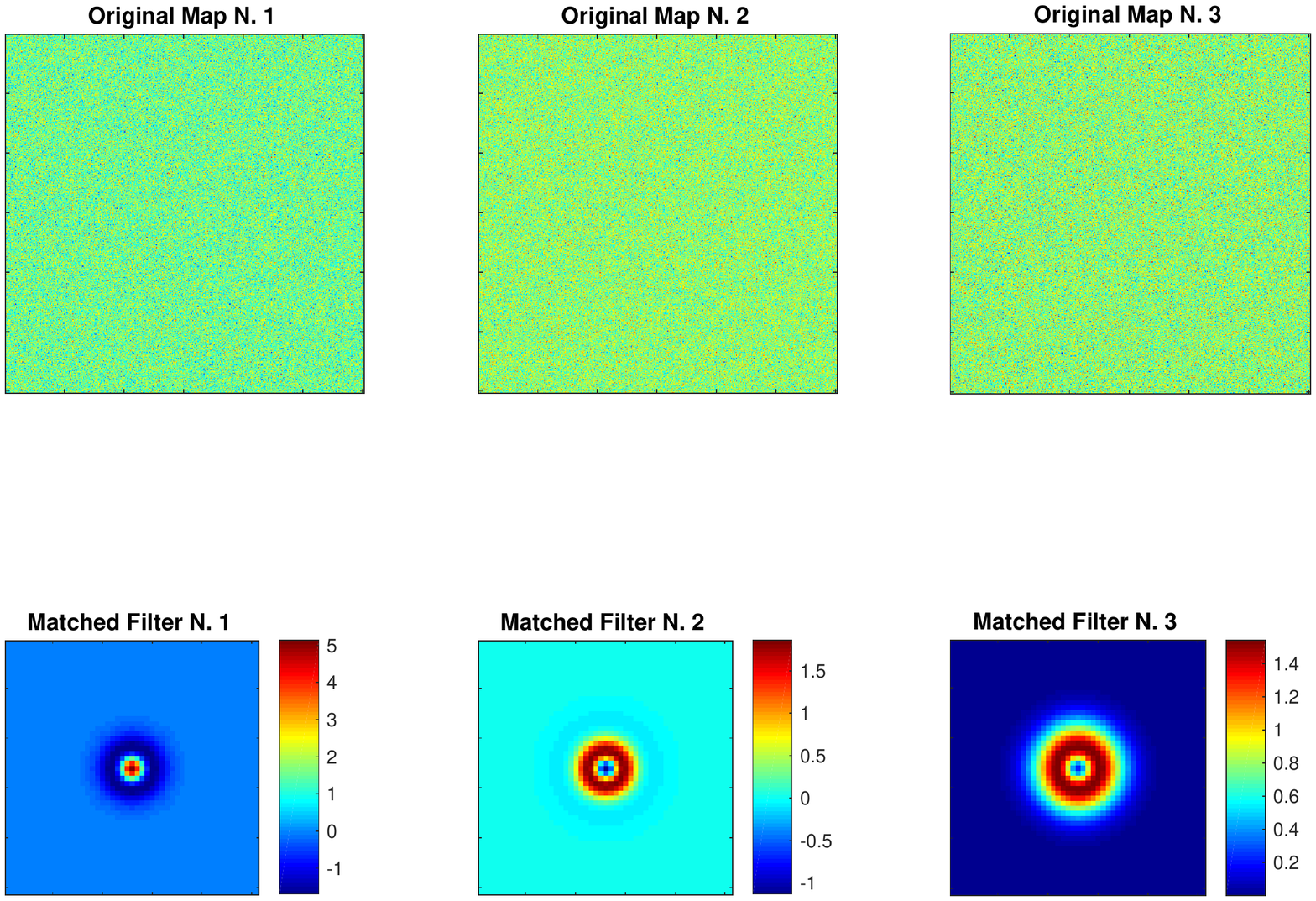}}
        \vskip -1cm
        \caption{Upper panels: simulated multi-frequency maps $\Xmatb$. Given the high level of noise, signal $\Smatb$, placed at the center of the map, is not visible.  Lower panels: the corresponding MMMFs.}
        \label{fig:original_2}
    \end{figure}

As an example of application of MMMF to the two-dimensional case, Fig.~\ref{fig:original_2} shows a multi-frequency map $\Xmatb_k = \Smatb_k + \Nmatb_k$, $k=1,2,3$, of sizes $N \times N$ with $N=500$, where $\Smat_k$ 
\begin{equation}
\Smat_k[i, j] = a_k \exp{\left(- \frac{(i-N/2)^2+(j-N/2)^2}{ 2 \sigma_k^2}\right)}
\end{equation}
with $\ab = [0.9, 0.6, 0.4]$, $\sigmab = [2, 3 , 5]$, and $\{ n_k \}$ cross-correlated white-noise processes with the same cross-correlation matrix~\eqref{eq:matrix0} used in the one-dimensional experiment.
 
\begin{figure}
           \vskip -2cm
        \resizebox{\hsize}{!}{\includegraphics{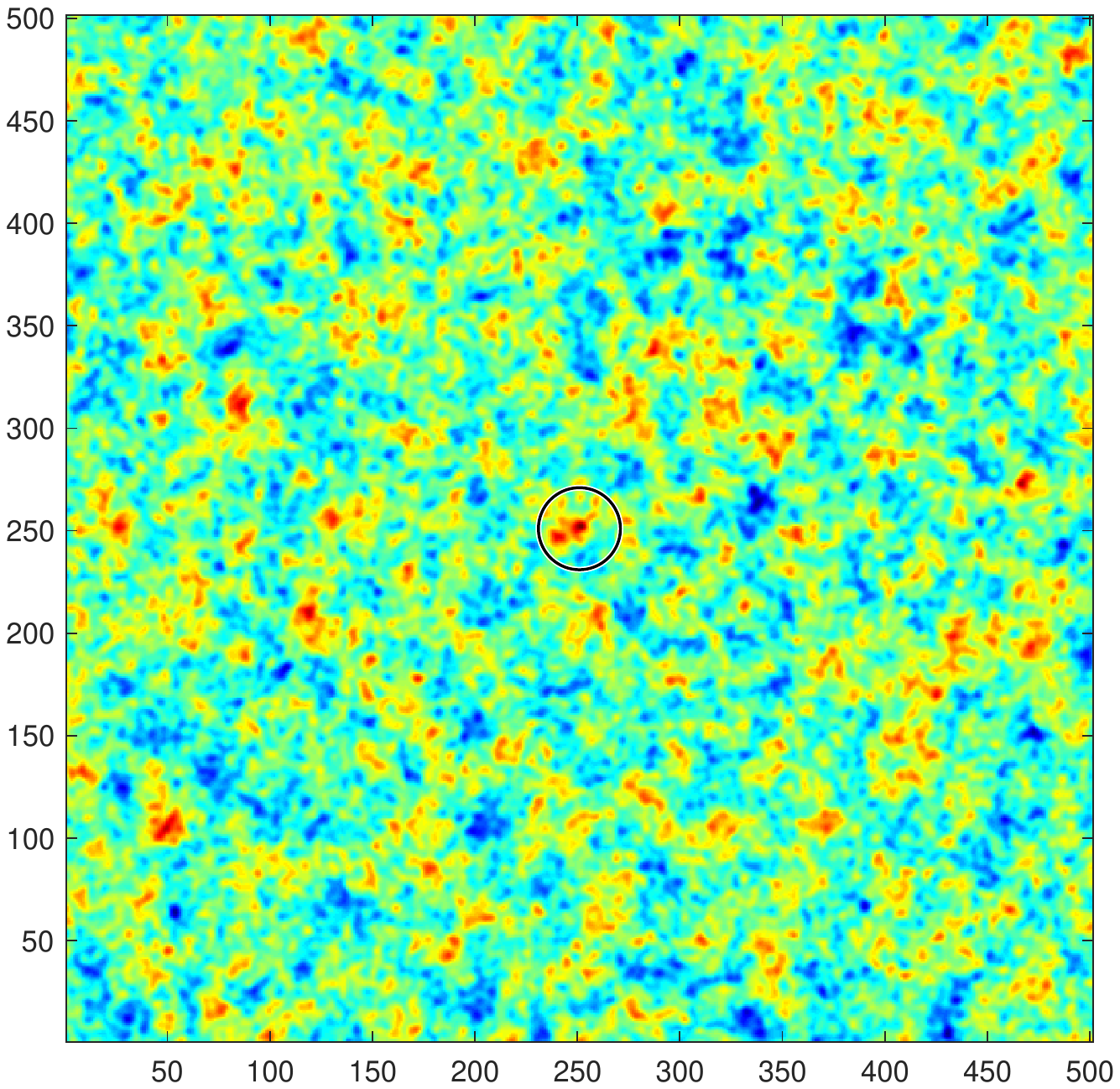}}
              \vskip -3cm
        \caption{Final map obtained after the MMMF filtering of the multi-frequency map in Fig.~\ref{fig:original_2}. The black circle in the center of the map highlights the detected signal.}
        \label{fig:result_2}
    \end{figure}

\begin{figure}
              \vskip -2cm
        \resizebox{\hsize}{!}{\includegraphics{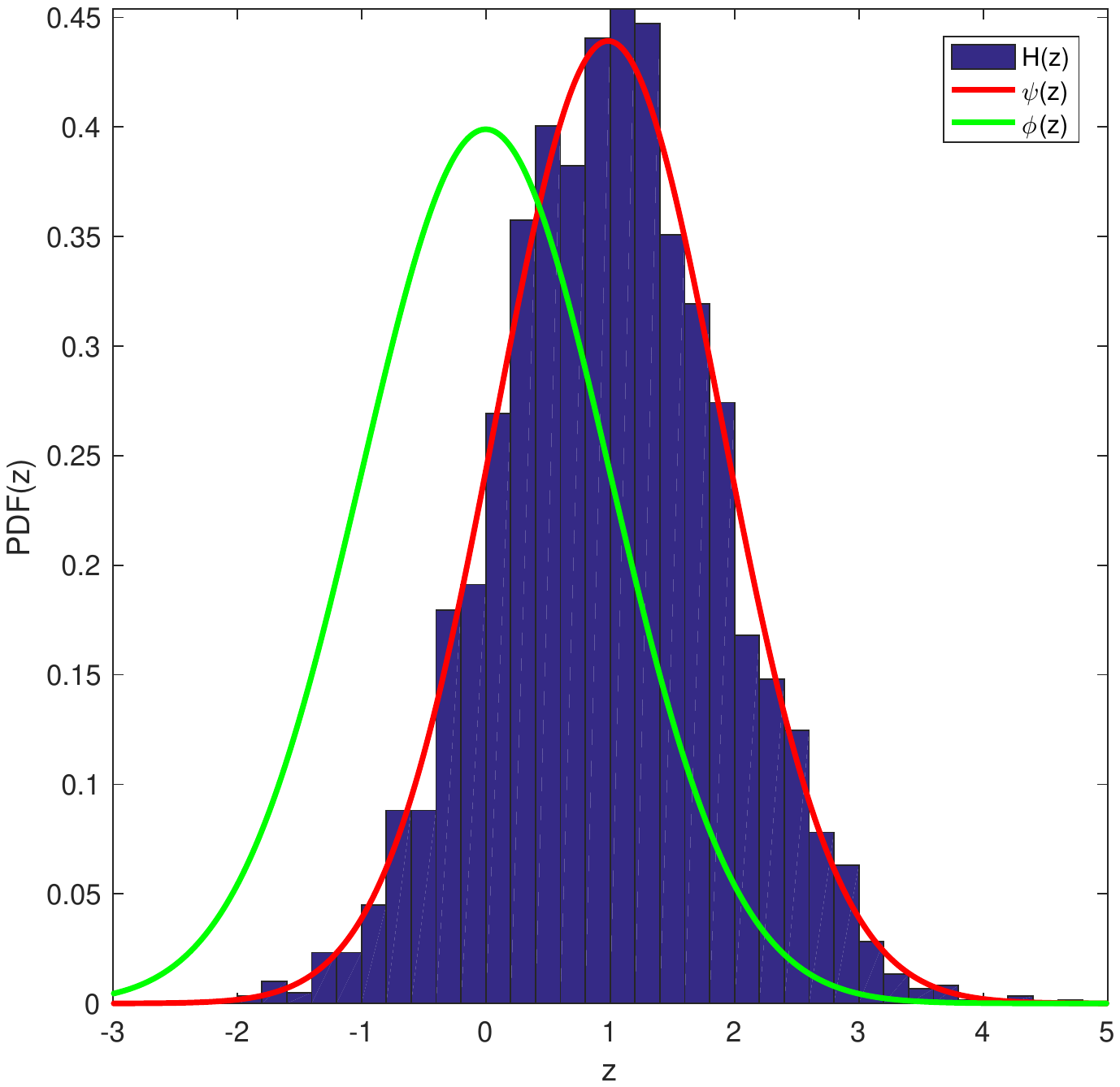}}
                \vskip -2cm
        \caption{Histogram $H(z)$ of the amplitudes of the peaks in the map in Fig.~\ref{fig:result_2}. For comparison, the PDF $\psi(z)$, corresponding to 
        	the maximum likelihood estimate $\kappa = 0.76$, and the standard Gaussian PDF $\phi(z)$ are shown.} 
        \label{fig:pdf_2}
    \end{figure}

Also this numerical experiment simulates an unfavorable situation where the noises are strong and highly correlated. The MMMF corresponding to this signal is shown in the lower panels in Fig.~\ref{fig:original_2} which then produces
the map in Fig.~\ref{fig:result_2}. In this map there are $3008$ peaks whose histogram $H(z)$ is shown in Fig.~\ref{fig:pdf_2}. In the same figure the PDF $\psi(z)$, corresponding to 
the maximum likelihood estimate $\kappa = 0.76$, is also shown as well the standard Gaussian $\phi(z)$.

\begin{figure}
        \resizebox{\hsize}{!}{\includegraphics{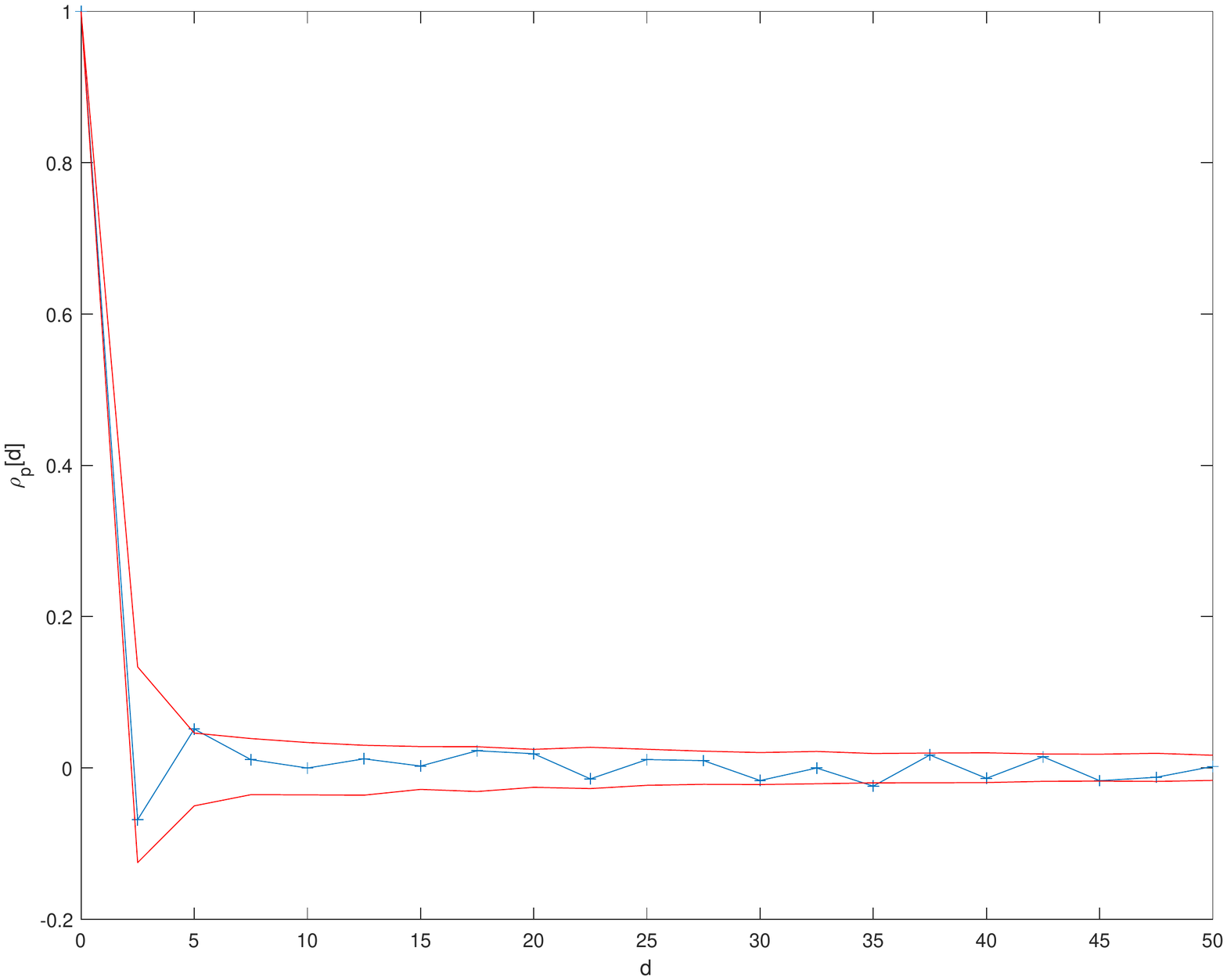}}
     	\vskip -1cm
        \caption{The two-point correlation function $\rho_p[d]$ of the peaks on the map in Fig.~\ref{fig:result_2}. The two red lines define the $90\%$ confidence band. They have
        been obtained by means of a bootstrap method based on the $90\%$ percentile envelopes of the two-point correlation functions obtained from $1000$ resampled sets of peaks with the same spatial coordinates as in the original map but whose values are randomly permuted.}
        \label{fig:fig_corr_test2}
    \end{figure}

Again, the iid condition for the peaks, necessary for the computation of the PSFA, is supported by the $\rho_p[d]$ in  Fig.~\ref{fig:fig_corr_test2} which is almost completely contained in its $90\%$ confidence band. Also in this
case the confidence band is obtained by means of a bootstrap method based on the $90\%$ percentile envelopes of the two-point correlation functions. These latter are obtained from $1000$ resampled sets of peaks with the same spatial coordinates as in the original map but whose amplitude values are randomly permuted.

The MMMF detects $\Smatb$ with a $\PFA$ equal to $2.4 \times 10^{-5}$.
This value is about twenty four times greater than the value provided by the standard $\PFA$~\eqref{eq:fd1}, i.e. $1.0 \times 10^{-6}$. Although the number of peaks is high, the $\PFA$ corresponding to
the greatest peak is very small, hence also in this case the reliability of the detection is high. 
This is confirmed by the corresponding $\SPFA$ which is $6.7 \times 10^{-2}$, i.e. the probability that this specific peak is due to the noise is less than $7\%$.

\section{What about if the noise is not of Gaussian type?} \label{sec:nongauss}

The property of the MF in Eq.~\eqref{eq:mf} to maximize the SNR is independent of the characteristics of the noise. However, in the case of non-Gaussian noise this fact does not guarantee the best detection performance. 
Unfortunately, in general the Neyman-Pearson approach is not able to provide 
a filter $\fb$ such that the statistic $\Tc(\xb)= \xb^T \fb$ can be computed. Even in the rare case where this is possible, the PDF of $T(\xb)$ cannot be obtained in analytical form. An example of the latter case is represented by the Poissonian noise for which the MF takes the form \citep{ofe18}
\begin{equation} \label{eq:mfp}
	\fb=\ln{\left( 1+ \frac{a \gb}{\lambda}\right)},
\end{equation}
with $\lambda$ the intensity of the Poisson noise. In this case $\Tc(\xb)$ is given by the weighted sum of Poisson random variables whose PDF cannot be obtained in analytical form. Moreover,
the MF~\eqref{eq:mfp} depends on the amplitude $a$ which often is unknown. Hence, the reliability of a detection cannot be computed. Two procedures were proposed to bypass this problem. The first is based on numerical simulations \citep{ofe18}, the second one on the approximation of the PDF of $\Tc(\xb)$ using
a saddle point approximation \citep{vio18}.

As shown below, in the non-Gaussian case a possible way out for weak signals is the {\it local optimal detector} (LOD) filter. This can be derived by expanding the so called {\it likelihood ratio test} (LRT) in a first-order Taylor expansion about $a=0$.  With this approach it is possible to obtain a detector with a form similar to that in Eq.~\eqref{eq:test1}
\begin{equation} \label{eq:test2a}
	\Tc(\xhb) = \xhb^T \gb  > \gamma,
\end{equation}
where $\xhb$ is a conveniently prefiltered version of $\xb$. 

	\begin{figure}
		\resizebox{\hsize}{!}{\includegraphics{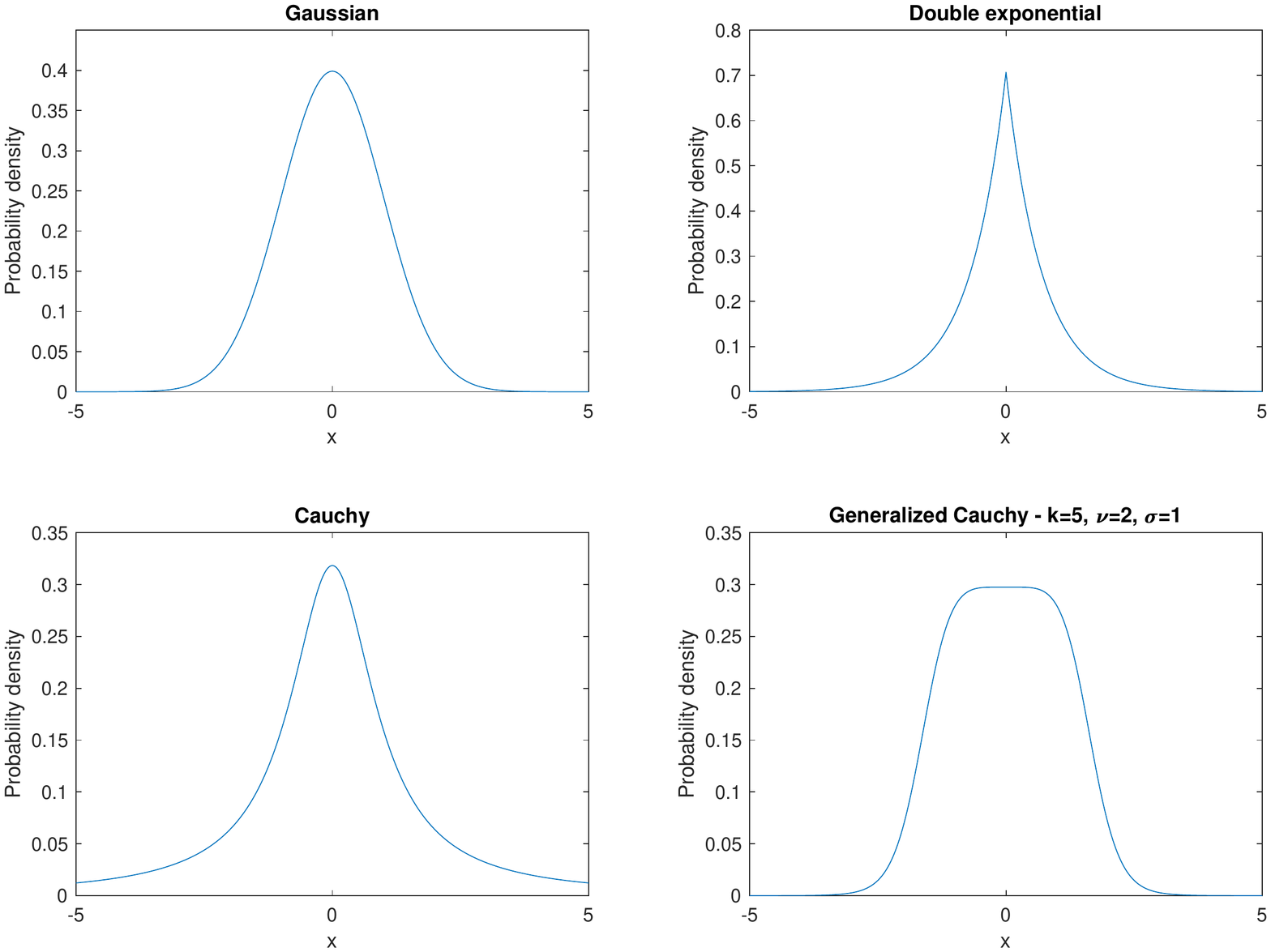}}
		\caption{Probability density functions whose corresponding local optimal filters $\flo(x)$ are shown in Fig.~\ref{fig:fig_LOs}.}
		\label{fig:fig_pdfs}
	\end{figure}

	\begin{figure}
		\resizebox{\hsize}{!}{\includegraphics{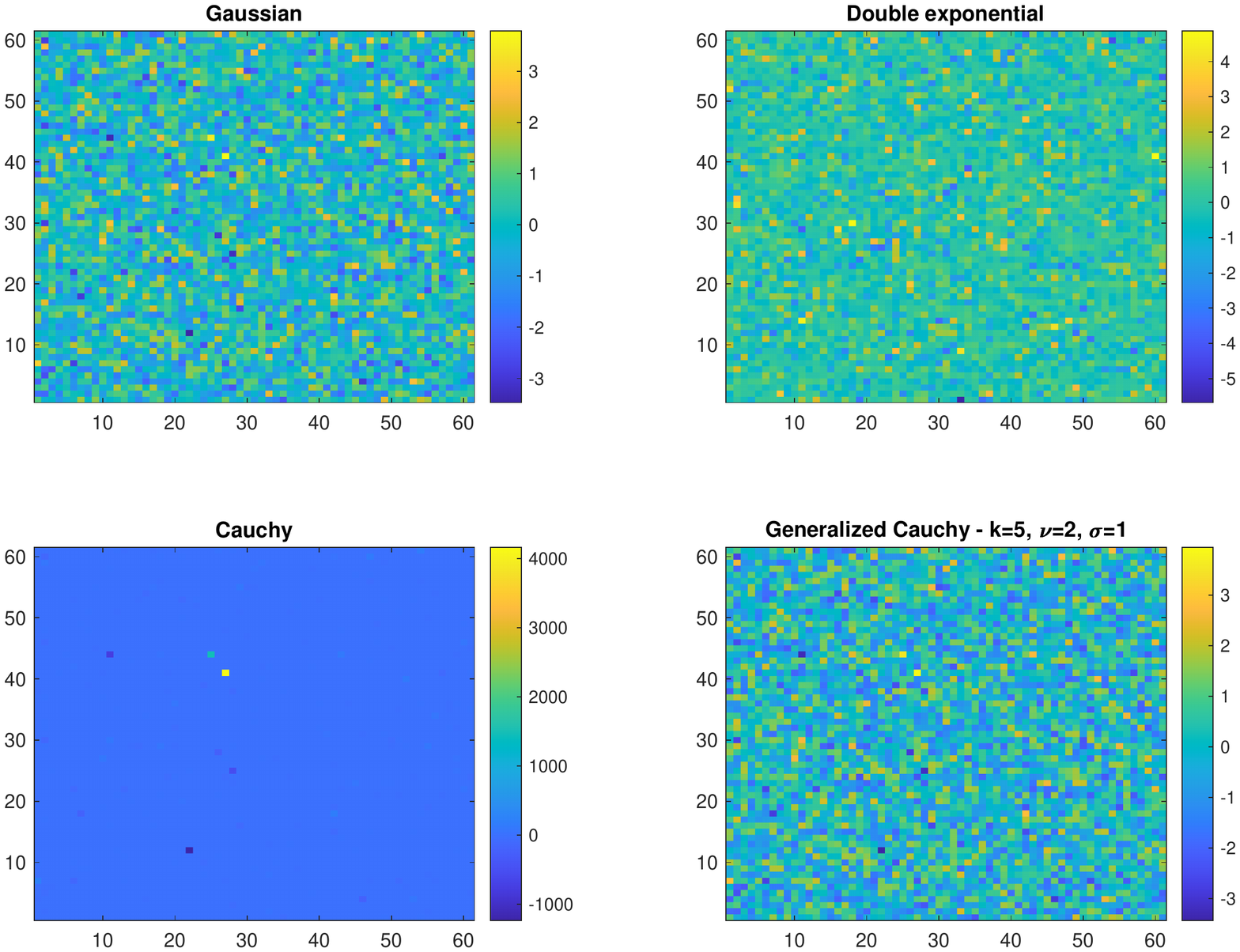}}
		\caption{White-noise maps corresponding to the PDFs given in Fig.~\ref{fig:fig_pdfs}.}
		\label{fig:fig_maps}
	\end{figure}

	\begin{figure}
		\resizebox{\hsize}{!}{\includegraphics{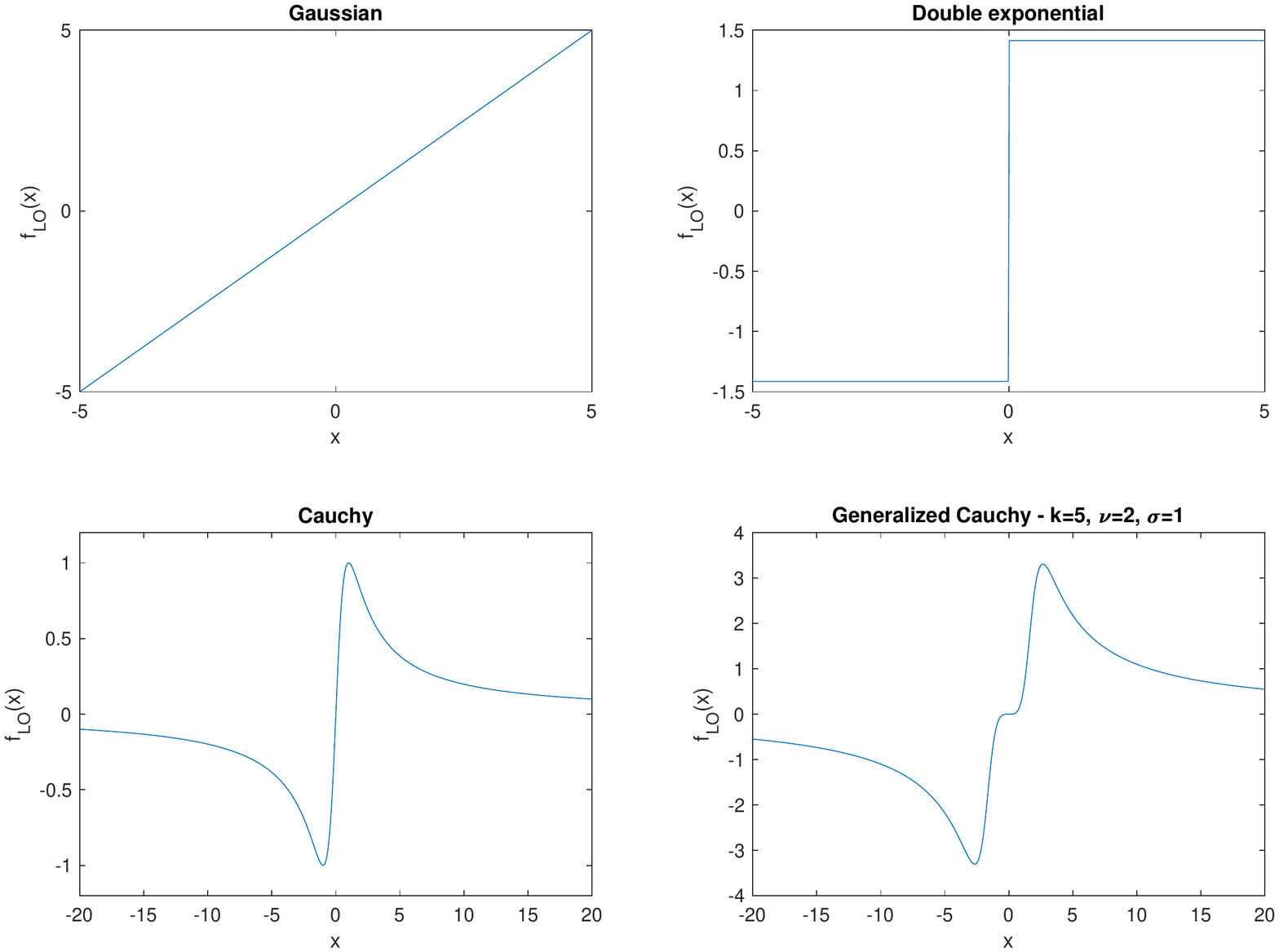}}
		\caption{Local optimal filters $\flo(x)$ corresponding to the probability density functions shown in  Fig.~\ref{fig:fig_pdfs}.}
		\label{fig:fig_LOs}
	\end{figure}

\subsection{Optimal local detectors: the theory}

As seen in Sect.~\ref{sec:neyman} the detection problem is a decision problem as given by Eq.~\eqref{eq:decision}
which consists of making a decision on whether $\xb$ is a pure
noise $\nb$ (hypothesis $\Hc_0$) or whether it contains the contribution of a signal $\ssb$ too
(hypothesis $\Hc_1$).
Under $\Hc_0$ the PDF of $\xb$ is given by $p(\xb| \Hc_0)$ whereas under $\Hc_1$ by $p(\xb| \Hc_1)$.
The most common criterion to claim a detection is that of Neyman-Pearson
which consists in the maximization of the probability of detection $\PD$ 
under the constraint that the probability of false alarm $\PFA$ does not exceed a fixed value $\alpha$.

The Neyman-Pearson theorem \citep[e.g., see ][]{kay98} is a powerful tool that allows to design
a decision process that pursues this aim:
to maximize $\PD$ for a given $\PFA=\alpha$, decide $\Hc_1$ 
if for the likelihood ratio (LR) $L(\xb)$ it is 
\begin{equation} \label{eq:ratio}
	L(\xb) = \frac{p(\xb| \Hc_1)}{p(\xb| \Hc_0)} > \gamma,
\end{equation}
where the threshold $\gamma$ is found from
\begin{equation} \label{eq:p1}
	\PFA = \int_{\{\xb: L(\xb) > \gamma\}} p(\xb| \Hc_0) d\xb = \alpha.
\end{equation}
Equation~\eqref{eq:ratio} represents the above mentioned LRT.

Following \citet{kay98}, the starting point to work with non-Gaussian noises is that the LRT can be written in the form
\begin{equation} \label{eq:ration}
	L(\xb) = \frac{\Pi_{i=0}^{N-1} p(x[i]| \Hc_1)}{\Pi_{i=0}^{N-1} p(x[i]| \Hc_0)} > \gamma,
\end{equation}
or
\begin{align} \label{eq:lration}
	\ln{L(\xb)} & = \sum_{i=0}^{N-1} \ln{\frac{p(x[i]| \Hc_1)}{p(x[i]| \Hc_0)}} > \gamma', \\
	& = \sum_{i=0}^{N-1} q_i(x[i]) >  \gamma'
\end{align}
where
\begin{align}
	q_i(x) &= \ln{\frac{p(x - a g[i])}{p(x)}} \\
	& \approx 0 + \left. \frac{\dfrac{d p(w)}{d w}}{p(w)} \right|_{w=x - a g[i], a=0} (-g[i]) a \\
	& = \left. - \frac{\dfrac{d p(x)}{d x}}{p(x)} \right|_{x=x[i]} g[i] a.
\end{align}
From Eq.~\eqref{eq:lration} one opts for $\Hc_1$ if
\begin{equation}
	\sum_{i=0}^{N-1} q_i(x[i]) \approx - \sum_{i=0}^{N-1}  \frac{\dfrac{d p(x)}{d x}}{p(x)} g[i] a > \gamma'
\end{equation}
or
\begin{equation}  \label{eq:test3a}
	\Tc(\xb) =  -  \sum_{i=0}^{N-1} \frac{\dfrac{d p(x)}{d x}}{p(x)} g[i] > \gamma''.
\end{equation}
We need to stress two points. $\Tc(\xb)$ is independent of the amplitude $a$ and it assumes the form given in Eq.~\eqref{eq:test2a} with 
\begin{equation}
	\xhc[i] =  \left.  - \frac{\dfrac{d p(x)}{d x}}{p(x)} \right|_{x=x[i]}.
\end{equation}
This last identity corresponds to prefiltering the signal $\xb$ by means of a nonlinear memoryless filter $\flo(x)$ such as $\xhc[i]= \flo(x[i])$. It can be shown that the asymptotic PDF of $\Tc(\xb)$ is
\begin{equation} \label{eq:cond3}
	\Tc(\xb) \sim
	\begin{cases}
		\Nc\left( 0, \sqrt{ I(a) \sum_{i=0}^{N-1} g^2[i]} \right) & \text{ under } \Hc_0, \\
		\Nc\left( a I(a) \sum_{i=0}^{N-1} g^2[i] , \sqrt{I(a) \sum_{i=0}^{N-1} g^2[i]} \right) & \text{ under } \Hc_1, \\
	\end{cases}
\end{equation} 
where, $\Nc(\mu, \sigma)$ is the Gaussian distribution of mean $\mu$ and standard deviation $\sigma$, whereas
\begin{equation}
	I(a) = \int_{-\infty}^{+\infty} \frac{\left(\dfrac{d p(x)}{d x}\right)^2}{p(x)} d x
\end{equation}
is the Fisher information.

\begin{figure}
		\resizebox{\hsize}{!}{\includegraphics{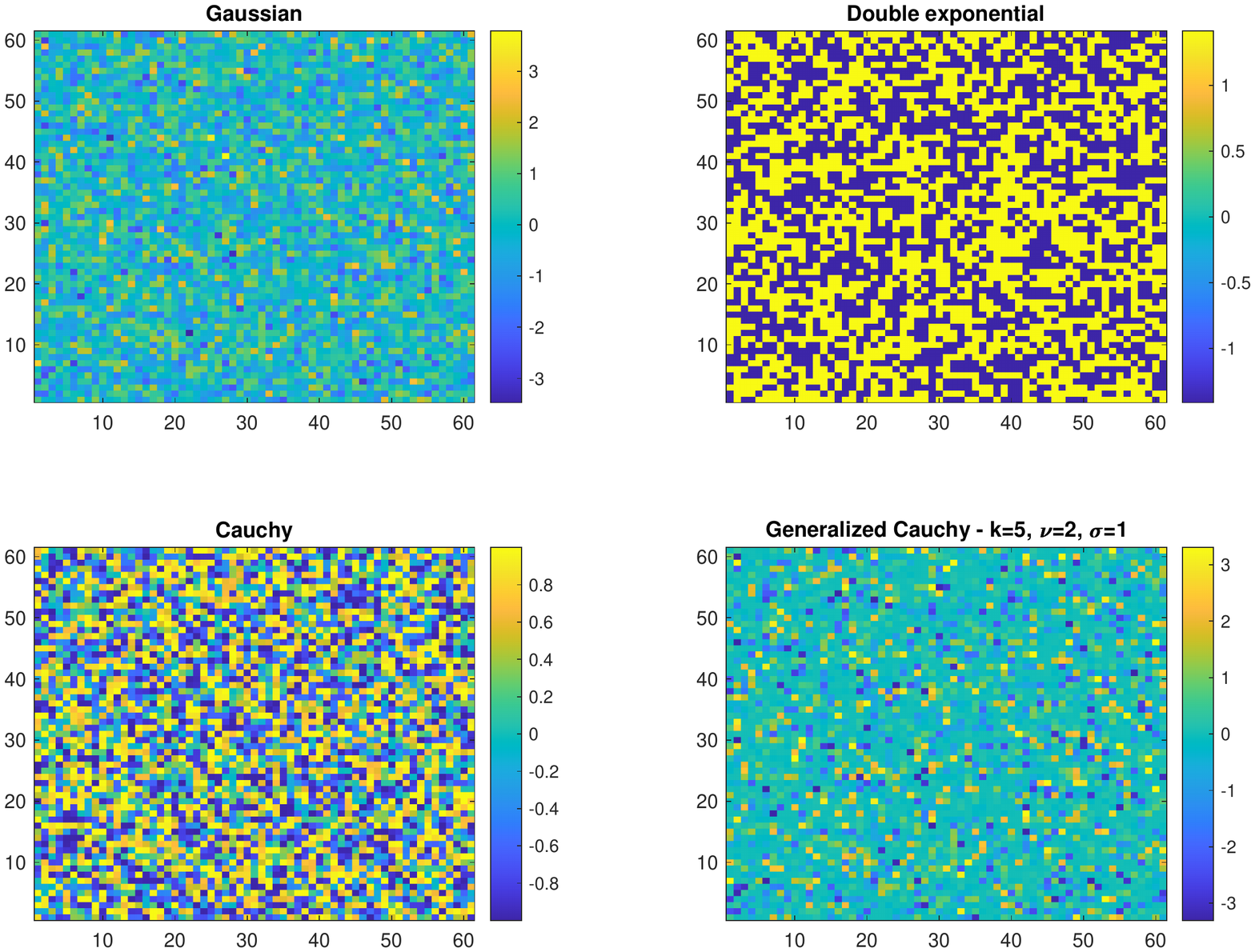}}
		\caption{LO filtered maps of Fig.~\ref{fig:fig_maps}.}
		\label{fig:fig_fmaps}
	\end{figure}

In order to get an idea of how the filter $\flo(x)$ works with different kind of non-Gaussianity, we examine four different types of white-noise whose PDFs consist of two generalized Gaussian and two generalized Cauchy distributions (see appendix~\ref{sec:appA}), respectively.
In detail, a standard Gaussian, a zero-mean unit-variance double exponential (or Laplacian), a Cauchy and a generalized Cauchy distribution with parameters $k=5$, $\nu=2$, and $\varsigma=1$ are considered and displayed in Fig.~\ref{fig:fig_pdfs}. Two-dimensional realizations of these processes are shown in Fig.~\ref{fig:fig_maps}.
From Fig.~\ref{fig:fig_pdfs} it is possible to see that the double exponential and the
Cauchy PDFs have more extended tails than the Gaussian whereas the reverse is true for the generalized Cauchy PDF. The corresponding filters $\flo(x)$ are visible in Fig.~\ref{fig:fig_LOs} (their analytical form is available in appendix~\ref{sec:appA}). 
From this figure it is evident that while for the Gaussian $\flo(x)=x$, i.e. no filtering is done, for the double exponential PDF $\flo(x) = \pm \sqrt{2}$, i.e. $\xhc$ is forced to assume only two values. This operation is necessary to limit the contribution to $\Tc(\xb)$ of the $x[i]$ with large magnitude. The Cauchy PDF represents an extreme situation since its tails are so extended that the corresponding variance is infinite.  Therefore, the filter $\flo(x)$ does not just limit but actually quickly zeroes the $x[i]$ with more and more large magnitude. Consequently, most of the entries in $\xb$ are not used in the computation of $\Tc(\xb)$. In the case of the PDF with shorter tails than the Gaussian, $\flo(x)$ filters out only the few
$x[i]$ with values very far from the central body of the PDF. The effects of these filters when applied to the corresponding maps of Fig.~\ref{fig:fig_maps} can be seen in Fig.~\ref{fig:fig_fmaps}.

	\begin{figure}
		\resizebox{\hsize}{!}{\includegraphics{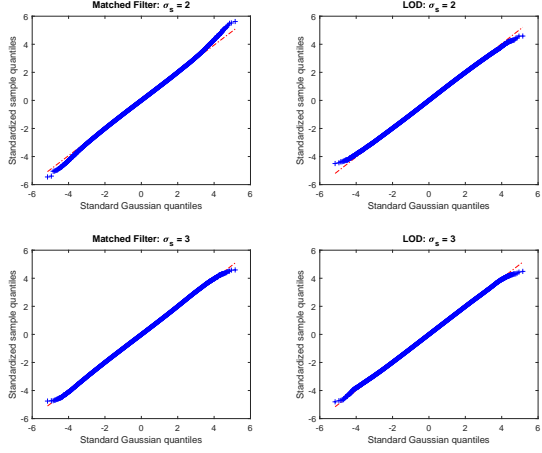}}
		\vskip -0.7cm
		\caption{Quantiles of the standard Gaussian PDF  vs.  the sample quantiles  of the statistic $\Tc(\xb)$ given by  Eq.~\eqref{eq:test1} (i.e. the classical MF) and Eq.~\eqref{eq:test2a} (i.e. the LO filter)
			computed for all the pixels of the 
			$2000 \times 2000$ pixels map described in the text. A good alignment of the blue crosses along the red line means a good compatibility of  the PDF of $\Tc(\xb)$ with the standard Gaussian. This condition do not appear to be satisfied at the extreme values of $\Tc(\xb)$.}
		\label{fig:fig_quantile_dexp_lo_mf}
	\end{figure}
\
	\begin{figure}
		\resizebox{\hsize}{!}{\includegraphics{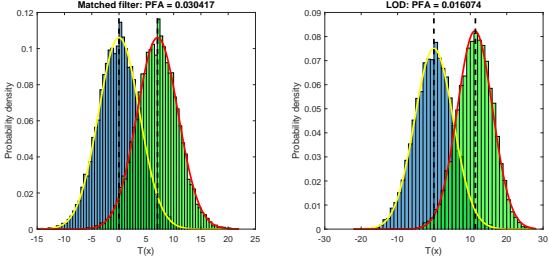}}
		\vskip -1cm
		\caption{Histogram of the  values of the test statistics $\Tc(\xb)$ for the MF and the LOD corresponding  to the central pixel of ten thousand $61 \times 61$ pixels maps containing only a white-noise with zero-mean, unit variance double exponential PDF (blues color) and when a circular Gaussian $\ssb$ with amplitude $a=0.5$ and dispersion
			$\sigma_s=3$ pixels is added to the noise at the central pixel (green color). The black dashed line marks the medians of the the distribution of the resulting $\Tc(\xb)$. The $\PFA$ indicated in the top of each panel, provides the probability of false alarm of the
			median of the computed $\Tc(\xb)$. The yellow line provides the expected Gaussian PDF of the $\Tc(\xb)$ under the hypothesis $\Hc_0$ of pure noise whereas the red line provides the expected Gaussian PDF under the alternative hypothesis $\Hc_1$.}
		\label{fig:fig_gen_gauss_detection_known}
	\end{figure}

	\begin{figure}
		\resizebox{\hsize}{!}{\includegraphics{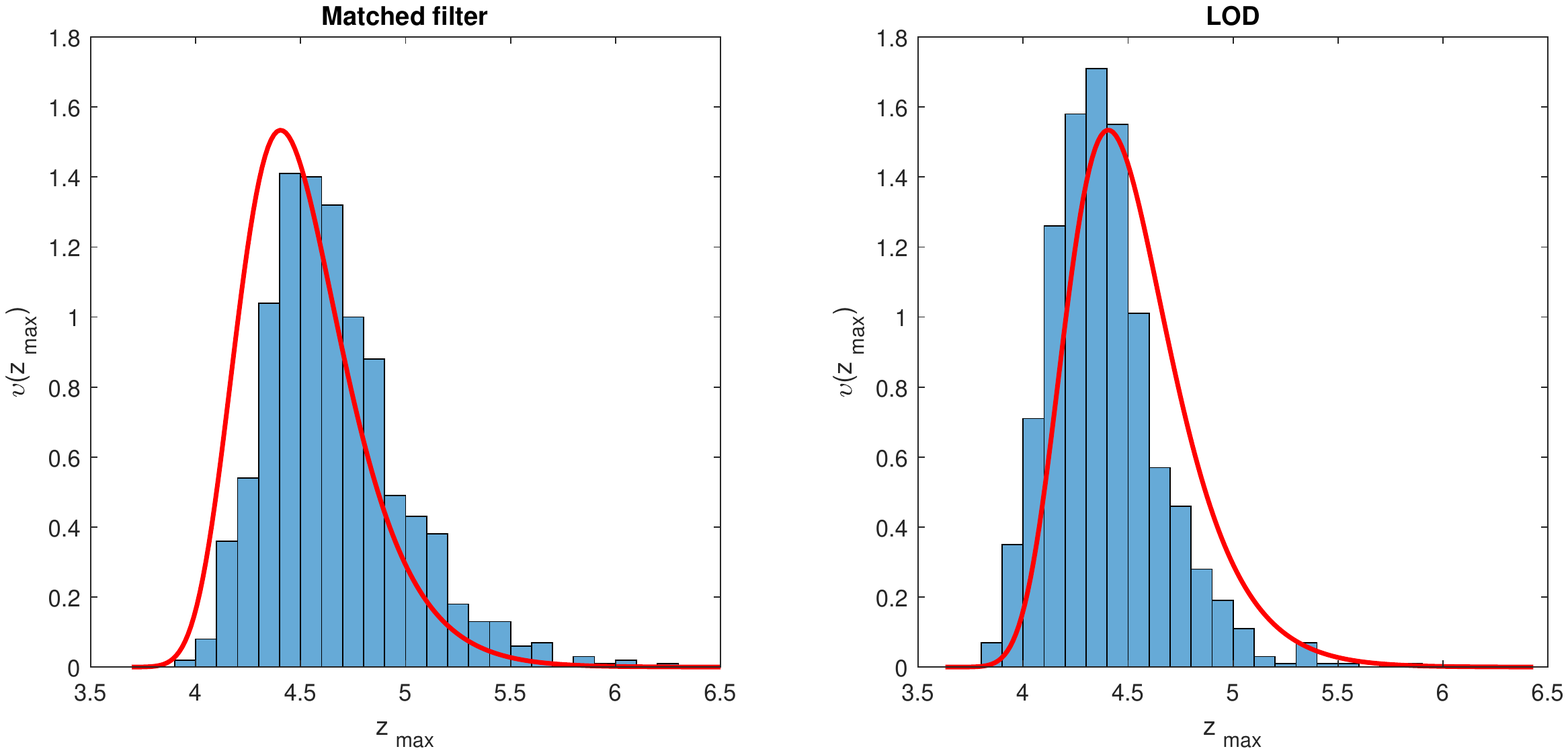}}
	    \vskip -0.7cm
		\caption{Histogram of the amplitude of the highest peak,  after the application of the MF and the LOD, of a set of one thousand  
			$1000 \times 1000$ pixels maps each containing the realization of a white-noise process with zero-mean, unit-variance double exponential PDF. The template $\Gmatb$ is assumed to be a circular Gaussian with dispersion $\sigma_s=3$ pixels. The red lines provide the PDF $\upsilon(z_{\rm max})$ of the highest peak expected in the case the noise was truly of of Gaussian type.}
		\label{fig:fig_gen_gauss_extreme}
	\end{figure}

	\begin{figure}
		\resizebox{\hsize}{!}{\includegraphics{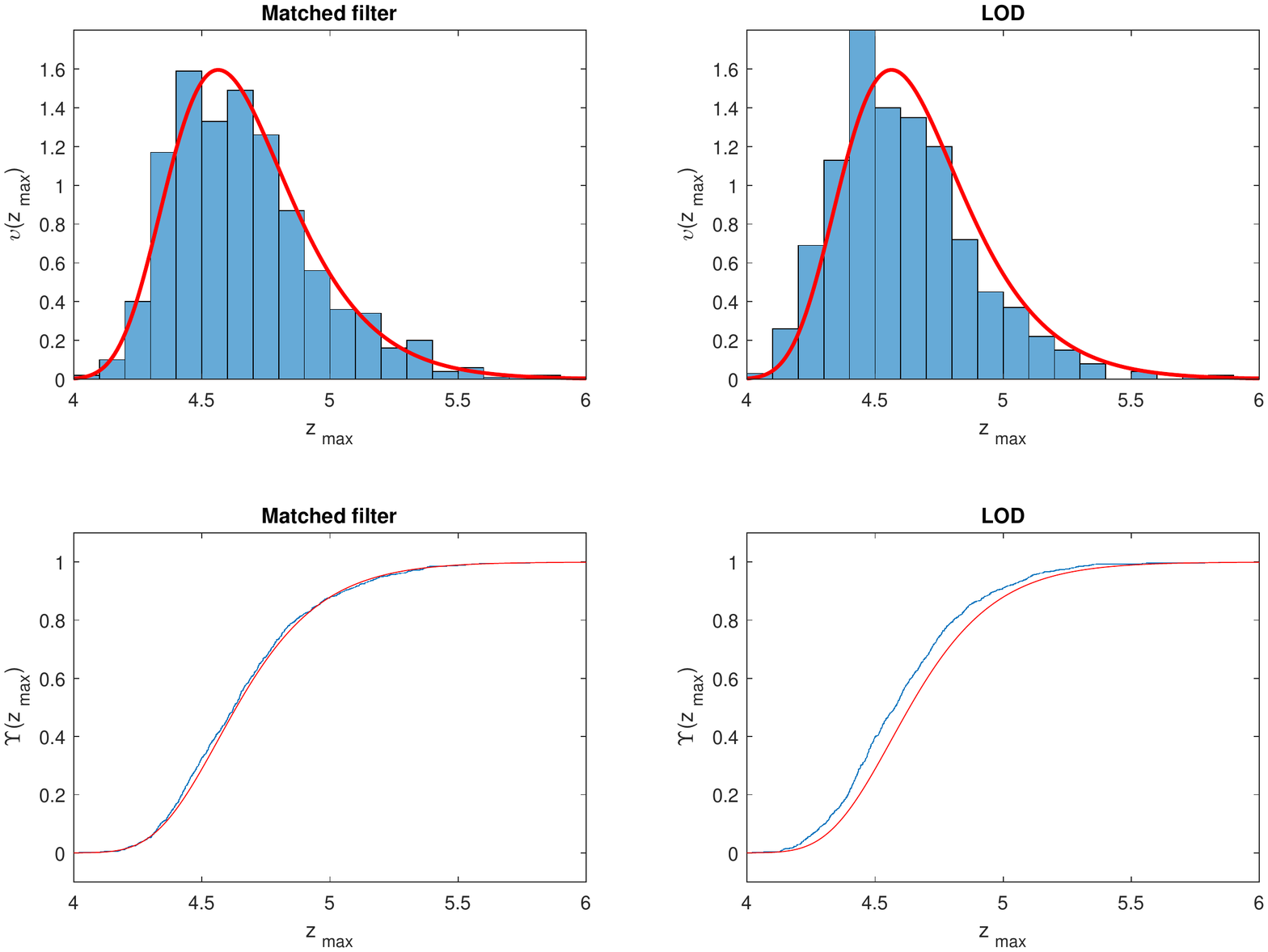}}
	     \vskip -0.7cm
		\caption{Top panels: histogram of the amplitude of the highest peak, after the application of the
			MF and the LOD, of a set of one thousand 
			$1000 \times 1000$ pixels maps each containing the realization of a white-noise process obtained form the mixture of a Gaussian and a double exponential PDFs both of zero-mean and unit-variance. Here, the mixing parameter is $\epsilon=0.05$
			(i.e. $95\%$ of the pixels have a Gaussian PDF). The template $\Gmatb$ is assumed to be a circular Gaussian with dispersion $\sigma_s=3$ pixels. The red lines provide the PDF $\upsilon(z_{\rm max})$ of the highest peak expected in the case the noise was truly of Gaussian type. Bottom panels: corresponding theoretical (red line) vs. the empirical (blue line) cumulative distribution functions.}
		\label{fig:fig_mix_gauss_dexp_extreme_005}
	\end{figure}

	\begin{figure}
		\resizebox{\hsize}{!}{\includegraphics{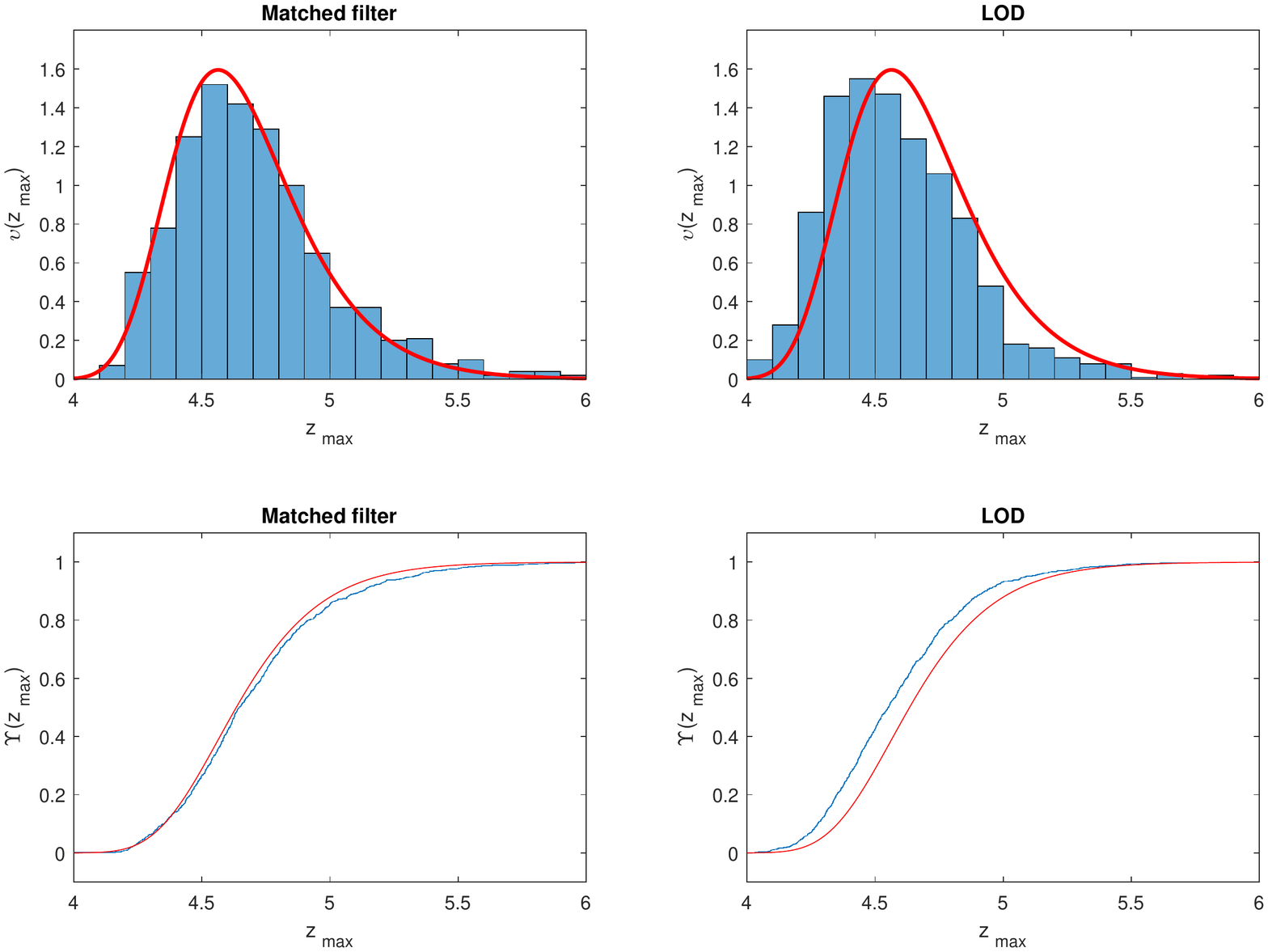}}
		\caption{As in Fig.~\ref{fig:fig_mix_gauss_dexp_extreme_005} but with $\epsilon=0.10$.}
		\label{fig:fig_mix_gauss_dexp_extreme_010}
	\end{figure}

\subsection{Optimal local detectors: the practice}

Testing the general performances of $\Tc(\xb)$ in the form~\eqref{eq:test3a} is not possible because the results critically depend on the kind of non-Gaussianity of the noise, the type of the data (one- or multi-dimensional) and on
the specific form of the searched signal $\ssb$. In any case, it is possible to get some useful indications by means of a few numerical experiments. 

\subsubsection{Double exponential noise} \label{sec:dexp}

In the first experiment, we check if, under the hypothesis $\Hc_0$, the condition~\eqref{eq:cond3} is really satisfied.  This is an important point since only in this case it is possible to fix a reliable detection threshold $\gamma$. 
To this aim, a $2000 \times 2000$ pixels map is simulated which contains a zero-mean white-noise with double exponential PDF
\begin{equation} \label{eq:dex}
	\varphi(\Xmatc; 0, \sigma_{\varphi})=\frac{1}{\sqrt{2} \sigma_{\varphi}} \exp{\left( - \frac{\sqrt{2}}{\sigma_{\varphi}} | \Xmatc | \right)}
\end{equation}
and standard deviation $\sigma_{\varphi}=1$.
It is also assumed that the template $\Gmatb$ is a two-dimensional circular symmetric Gaussian 
\begin{equation}
	\Gmat[i, j] = \exp{\left(- \frac{(i-N/2)^2+(j-N/2)^2}{ 2 \sigma_s^2}\right)}.
\end{equation}
Two values of the standard deviation $\sigma_s$ are considered, i.e. $\sigma_s=2$ and $3$. 
Figure~\ref{fig:fig_quantile_dexp_lo_mf} shows the theoretical quantiles 
\footnote{We recall that the quantiles related to a random variable $X$  with given PDF are obtained from the quantile function $Q(p)$ which assigns to each probability $p$ the value $x$ for which ${\rm Pr}(X \leq x)=p$.}
of the standard Gaussian distribution vs. the sample quantiles  of the detectors $\Tc(\Xmatb)$ given by Eq.~\eqref{eq:test1} (i.e. the classical MF) and Eq.~\eqref{eq:test2a}. In the first case, $\Tc(\Xmatb)$ is obtained for each pixel by cross-correlating the original map with the template $\Gmatb$ and then standardizing the resulting map to unit variance by means of the standard deviation $\sigma_{\Tc(\Xmatb)}$
expected for the matched filtered noise, $\sigma_{\Tc(\Xmatb)}=\sqrt{\sum_{i,j=0}^{N-1} \Gmat[i, j]}$. The same operation is carried out also for the second case with the difference that the map is pre-filtered  by means of $\flo(x)$.
Moreover, after the cross-correlation with $\Gmatb$, the  map is standardized to unit variance by means of  the expected standard deviation $\sigma_{\Tc(\Xmatb)}=\sqrt{2 \sum_{i,j=0}^{N-1} \Gmat[i, j]}$. From this figure it is visible that, especially for the greatest $\sigma_s$, the Gaussian assumption for the PDF of $\Tc(\Xmatb)$ is acceptable for both cases only in the central body of the PDFs. A certain discrepancy concerns the extreme tails. As it will be realized below, this last point may entail some important consequences. 

In order to compare the performances of the MF and of the LOD, ten thousands $61 \times 61$ pixels maps are simulated each containing the realization of a white-noise process again with a  zero-mean, unit-variance double exponential PDF. Each panel of Fig.~\ref{fig:fig_gen_gauss_detection_known} compares the statistical distribution of $\Tc(\Xmatb)$ when the maps contain only noise and when a circular Gaussian $\Smatb$ with amplitude $a=0.5$ and dispersion $\sigma_s=3$ pixels is added at the central pixel. The black dashed line marks the medians of the distribution of the resulting $\Tc(\Xmatb)$. The yellow line provides the expected Gaussian PDF of $\Tc(\Xmatb)$ under the hypothesis $\Hc_0$, whereas the red line provides the expected Gaussian PDF under the hypothesis $\Hc_1$. In this last case, the mean and the standard deviation the Gaussian PDF for the LOD are set to the corresponding sample quantities. This is due to the fact that, for the adopted value of the amplitude $a$, the theoretical values of the mean and standard deviation of $\Tc(\Xmatb)$ under the hypothesis $\Hc_1$ as given by Eq.~\eqref{eq:cond3} do not fully reproduce the corresponding sample values. From numerical experiments the theoretical mean value of $\Tc(\Xmatb)$ under the hypothesis $\Hc_1$ is generally greater than the sample value and it appears more sensitive to the magnitude of the amplitude $a$ than to the theoretical standard deviation.  This fact, however, has no consequences on the calculation of the detection threshold $\gamma$ and hence of the $\PFA$.
From this figure appears that the LOD is effectively characterized by a better detection capability. Indeed, with the MF, the median of the $\Tc(\Xmatb)$ under the hypothesis $\Hc_1$ has a $\PFA \approx 3.1 \times 10^{-2}$ whereas this value is $1.6 \times 10^{-2}$ for the LOD. 

This conclusion, however, is misleading. Again, the point is that $\Tc(\Xmatb)$ is computed assuming that the position of the signal $\Smatb$ is known. In practical application, often this is not the case and the above procedure is inapplicable \citep[see ][]{vio16, vio17, vio19}. As claimed in Sect.~\ref{sec:unknown}, the way out is to assume that, if present, the position of a signal coincides with one of the peaks in the data after the cross-correlation with $\Gmatb$ and a detection is claimed when the probability that a given peak is due only to the noise (i.e. the $\PFA$) is smaller than a prefixed threshold.
When the procedure illustrated in Sect.~\ref{sec:spurious} is applied to a set of one thousand of simulated 
$1000 \times 1000$ pixels maps each containing the realization of a white-noise process with a zero-mean, unit-variance double exponential PDF and assuming as signal template a circular Gaussian with $\sigma_s=3$ pixels,
Fig.~\ref{fig:fig_gen_gauss_extreme} compares the histogram of the amplitude of the greatest peak in each map with the theoretical PDF $\upsilon(z_{\rm max})$. The disagreement for both the MF and the LOD is evident. 
This is the consequence of the fact, as underlined above with respect to Fig.~\ref{fig:fig_quantile_dexp_lo_mf}, that the PDF of $\Tc(\Xmatb)$ is approximately Gaussian only if not too far from the mean. Hence, working with the extremes provides
poor results. The conclusion is that, in the case of noises with PDF somewhat different from the Gaussian and lacking the information on the position of $\Smatb$, the LOD is not able to improve the bad performances of the MF.

	\begin{figure}
		\resizebox{\hsize}{!}{\includegraphics{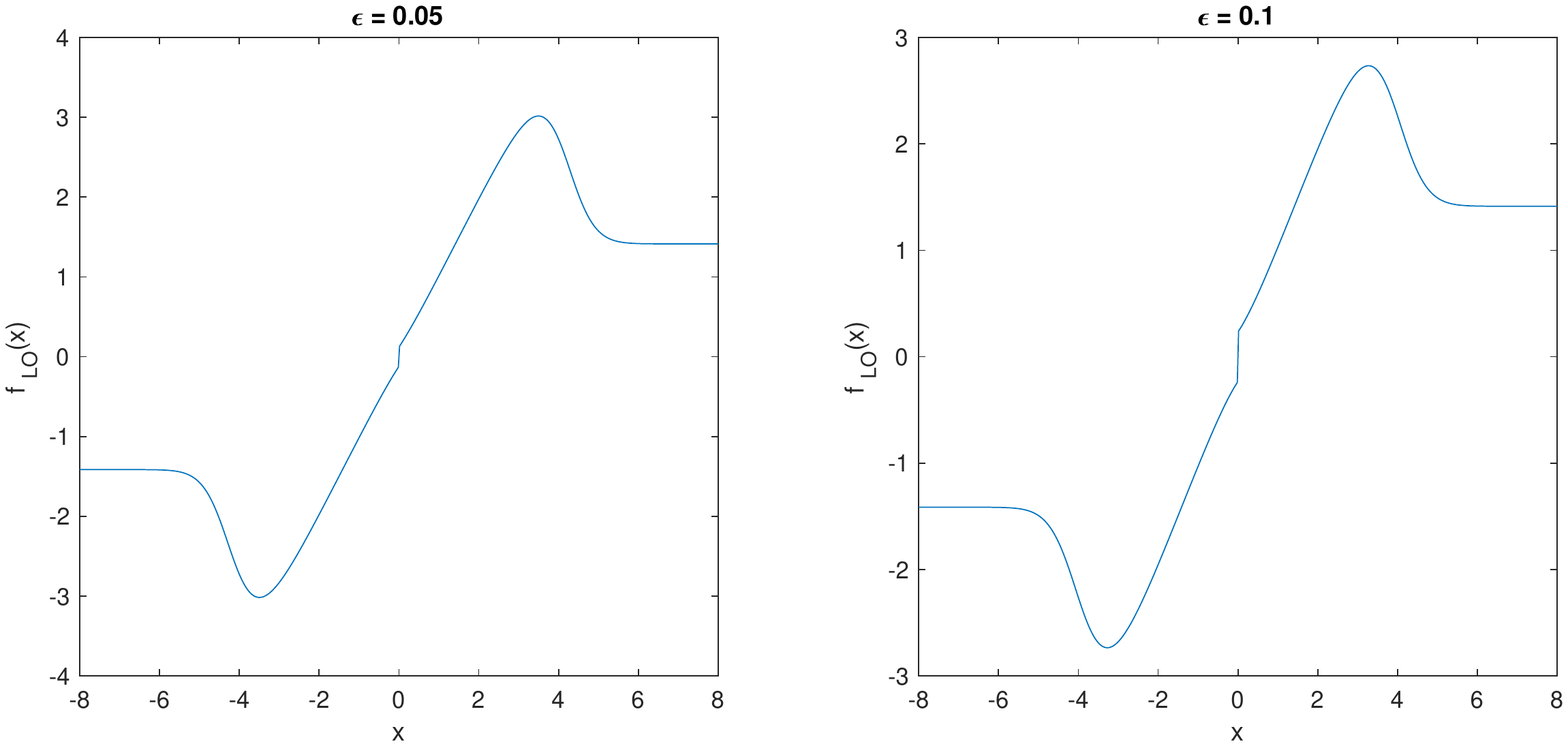}}
		 \vskip -1cm
		\caption{Filter $\flo(x)$ for the noise due to the mixture of a Gaussian and a double exponential distributions corresponding to the values $0.05$ and $0.10$ of the fraction $\epsilon$ of the double exponential 
			random noise component.}
		\label{fig:fig_LOs_mix_gauss_dexp}
	\end{figure}

\begin{figure}
	\resizebox{\hsize}{!}{\includegraphics{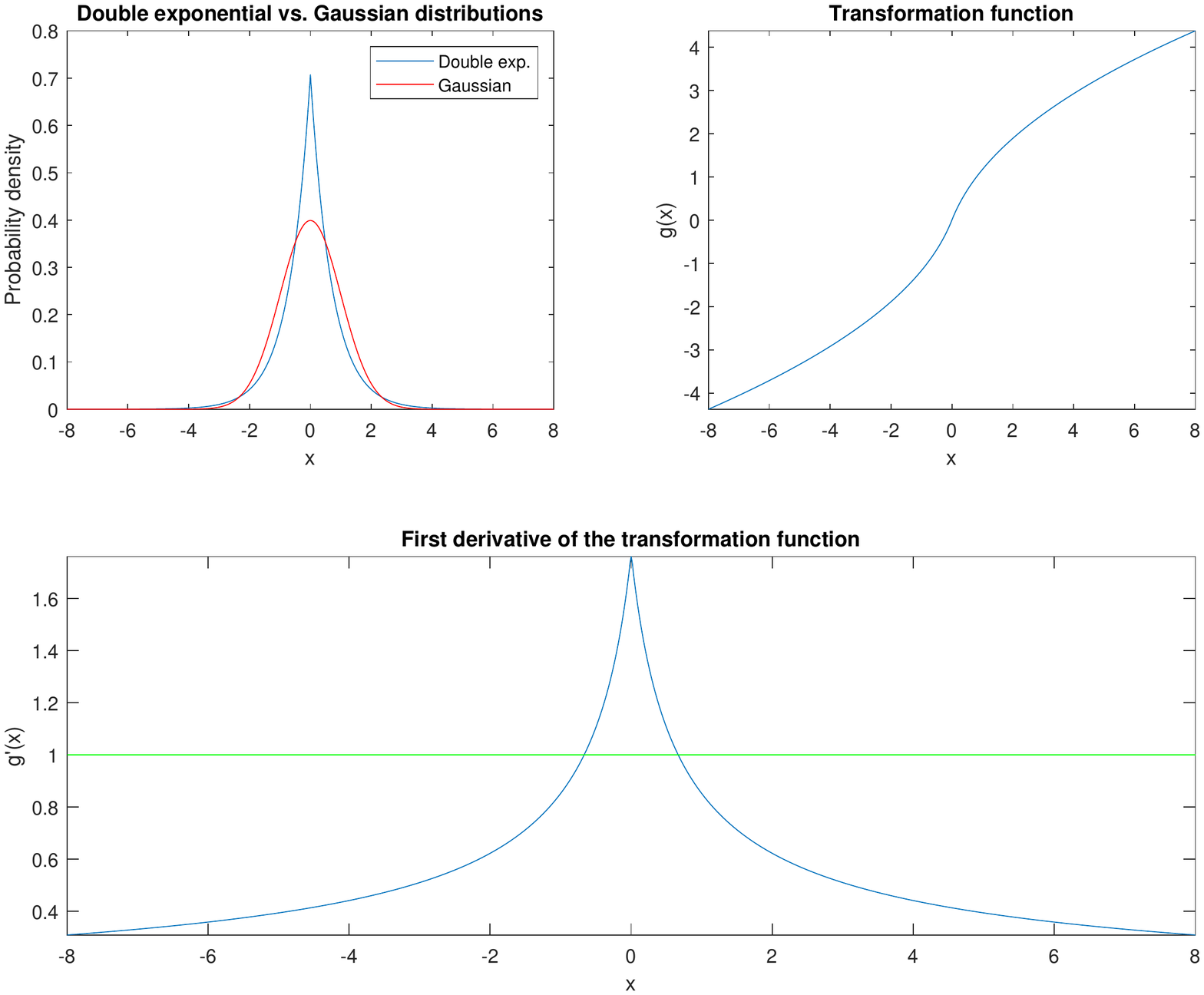}}
     \vskip -1cm
	\caption{Top left panel: comparison of the unit-variance PDFs corresponding to the double exponential distribution and the standard Gaussian PDF; Right panel: transformation function $\gc(x)$ from the corresponding PDF to the Gaussian PDF; Bottom panel: first derivative  $\gc'(x)$ of the function in the top right panel. The green line provides the level corresponding to $\gc'(x)=1$.}
	\label{fig:fig_test_source_taylor_dexp1}
\end{figure}

\begin{figure}
	\resizebox{\hsize}{!}{\includegraphics{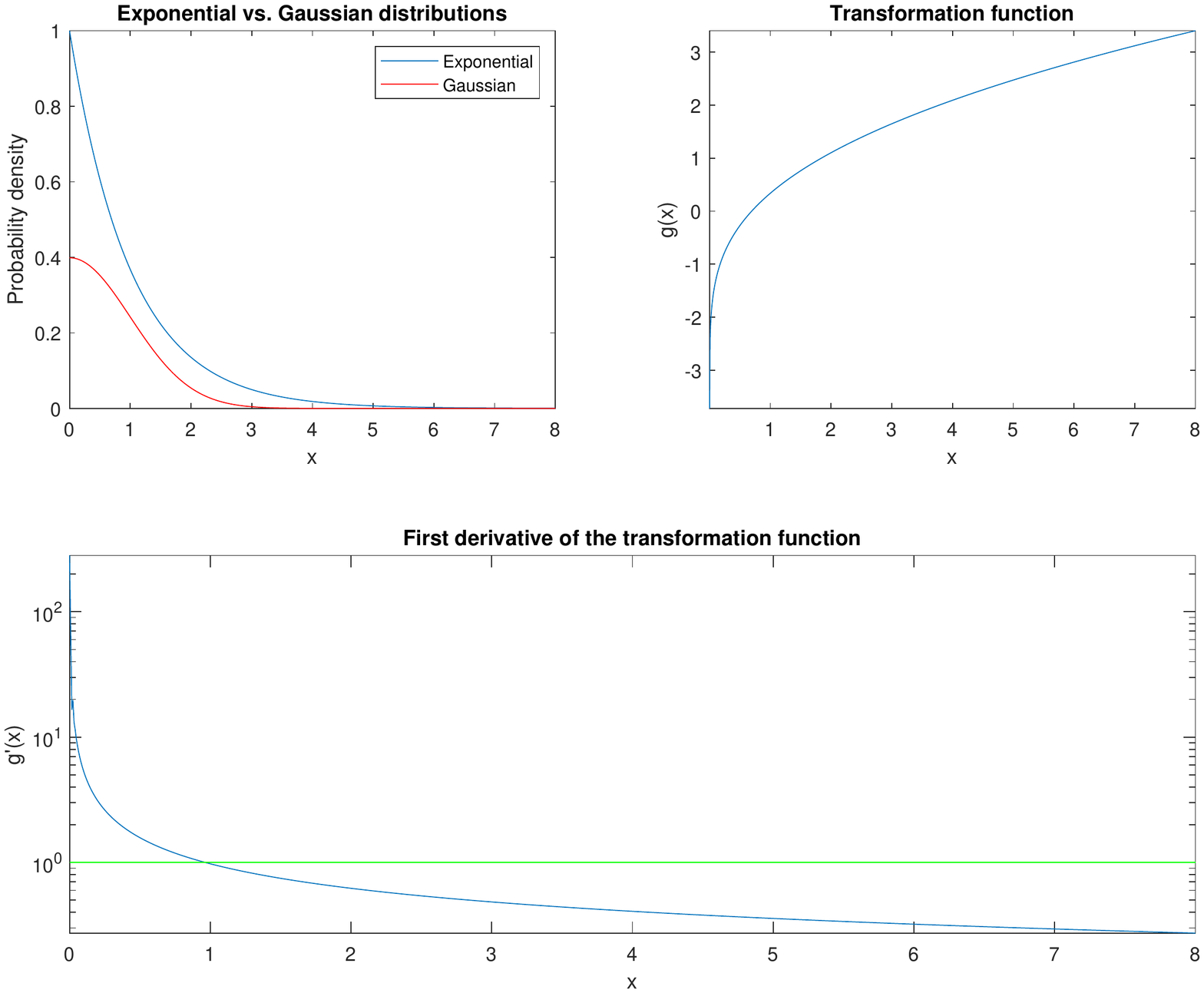}}
     \vskip -1cm
	\caption{Top left panel: comparison of the unit-variance PDFs corresponding to the exponential distribution and the standard Gaussian PDF; Right panel: transformation function $\gc(x)$ from the corresponding PDF to the Gaussian PDF; Bottom panel: first derivative $\gc'(x)$ of the function in the top right panel. The green line provides the level corresponding to $\gc'(x)=1$.}
	\label{fig:fig_test_source_taylor_exp1}
\end{figure}

\begin{figure}
	\resizebox{\hsize}{!}{\includegraphics{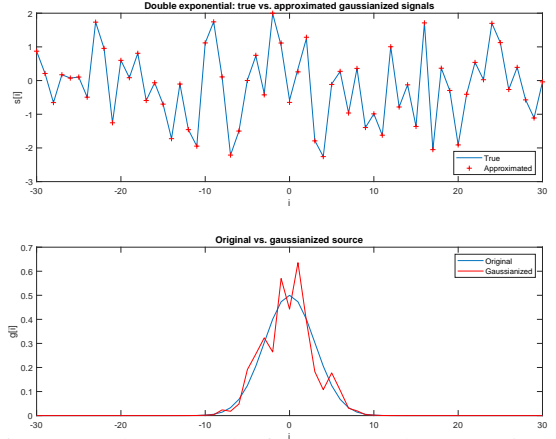}}
     \vskip -1cm
	\caption{Top panel: true vs. the first term Taylor approximation of a gaussianized one-dimensional signal containing a white-noise with unit-variance PDF corresponding to the double exponential distribution added to a Gaussian shaped signal with amplitude $a=0.5$ and dispersion $\sigma_s=3$ pixels. Bottom panel: original $\ssb$ vs. its gaussianized version.}
	\label{fig:fig_test_source_taylor_dexp2}
\end{figure}

\begin{figure}
	\resizebox{\hsize}{!}{\includegraphics{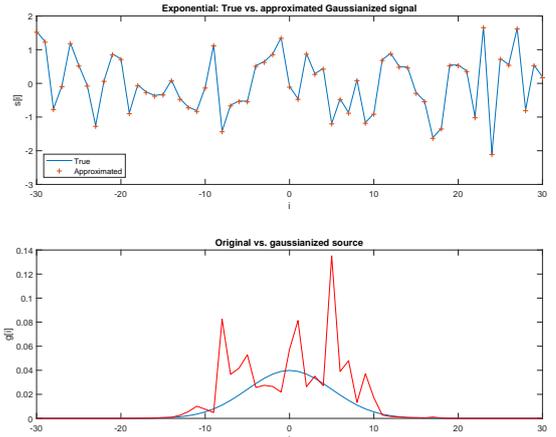}}
     \vskip -1cm
	\caption{Top panel: true vs. the first term Taylor approximation of a gaussianized one-dimensional signal containing a white-noise with unit-variance PDF corresponding to the exponential distribution added to a Gaussian shaped signal with amplitude $a=0.5$ and dispersion $\sigma_s=3$ pixels. In the case of the exponential noise, the signal is zero centered. Bottom panel: original $\ssb$ vs. its gaussianized version.}
	\label{fig:fig_test_source_taylor_exp2}
\end{figure}

\subsubsection{Mixture of a Gaussian and a double exponential noise}

An important experimental situation is when the noise is given by a mixture of two different PDFs. A typical case is when the zero-mean Gaussian PDF $\phi(x; 0, \sigma_{\phi})$ with standard deviation $\sigma_{\phi}$ is mixed with a zero mean PDF $\zeta(x)$ according to,
\begin{equation}
	p(x) = \phi(x) (1 -\epsilon) + \epsilon \zeta(x), \qquad 0 \leq \epsilon \leq 1.
\end{equation}
This model is typically used when part of the data are contaminated by ''spikes'' (a sudden change of the noise value). In this case, the PDF  $\zeta(x)$ is chosen with tails wider than that of the Gaussian PDF.

The filter $\flo(x)$ corresponding to this mixture is \citep{kas88}
\begin{equation} \label{eq:flogd}
	\flo(x) = - \frac{\zeta'(x)/\zeta(x) +[(1-\epsilon)/\epsilon] (x/\sigma_{\phi}^2) \phi(x)/\zeta(x)}{1 + [(1-\epsilon)/\epsilon] \phi(x)/\zeta(x)},
\end{equation}
where $\zeta'(x) = d \zeta(x) / dx$.
If $\zeta(x)= \varphi(x; 0, \sigma_{\varphi})$, i.e. $\zeta(x)$ is given by the double exponential PDF~\eqref{eq:dex}, then Eq.~\eqref{eq:flogd} becomes
\begin{equation}
	\flo(x) = - \frac{x (1-\epsilon) \exp{\left[-x^2/(2 \sigma_{\phi}^2)\right]}}{\sqrt{2 \pi} \sigma_{\phi}^3} - \frac{\epsilon \exp{\left[-\sqrt{2} |x|/\sigma_{\varphi}\right]} {\rm sign}[x]}{\sigma_{\varphi}^2},
\end{equation}
where ${\rm sign}[x]= \pm 1$ according to the sign of $x$.

Figures~\ref{fig:fig_mix_gauss_dexp_extreme_005} and \ref{fig:fig_mix_gauss_dexp_extreme_010} compare the histograms of the extremes $z_{\rm max}$ of two experiments similar to that in Fig.~\ref{fig:fig_gen_gauss_extreme} but now 
with the noise given by the mixture, with $\epsilon=0.05$ and $0.10$ respectively, of a Gaussian and a double exponential PDFs with $\sigma_{\phi}=\sigma_{\varphi}=1$. Although both methods show a good agreement with
the theoretical PDF $\upsilon(z_{\rm max})$ and CDF $\Upsilon(z_{\rm max})$, the MF appears to outperform the
LOD. At first sight, this could seem a surprising result. However, the reason is again  due to the fact that the optimal properties of the MF and the LOD hold only when the position of the signal $\Smatb$ is known. If this information is not available and one
is forced to work with the extremes of the maps, the quality of the results strictly depends on how good is the Gaussian approximation for the PDF of $\Tc(\Xmatb)$. Since with the MF 
the statistic $\Tc(\Xmatb)$ is obtained from the weighted sum of Gaussian plus a small fraction of non-Gaussian random quantities, then the Gaussian approximation for  $\Tc(\Xmatb)$ results better than the LOD
with which the highest and the smallest values of $\Xmatb$ are set to a fixed value  (see Fig.~\ref{fig:fig_LOs_mix_gauss_dexp}) with a consequent slower convergence to a Gaussian PDF.

\begin{figure}
	\resizebox{\hsize}{!}{\includegraphics{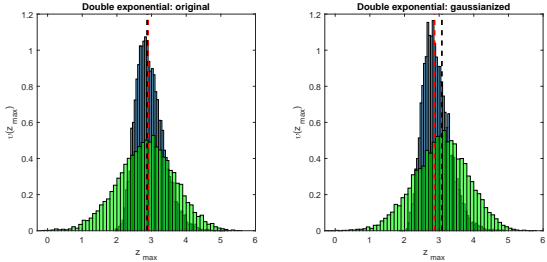}}
	\vskip -2cm
	\caption{Left panel: comparison of the histograms of the highest peak of a set of ten thousand $61 \times 61$ pixels maps after the MF filtering when they contain, respectively, only 
		a white-noise with unit-variance PDF corresponding to the double exponential distribution (blue bars)
		and when a circular Gaussian $\Smatb$ with amplitude $a=0.5$ and dispersion $\sigma_s=3$ pixels  is added at the central pixel (green bars). In the case of the maps containing the contribution of $\Smatb$, the highest peak is searched in a square area centered on the central pixel and with sides of seven pixels; Right panel: as the left panel but the maps that are gaussianized before the MF operation. The red and black dashed lines mark the median of the amplitude of the highest peak in the noise maps and in the noise maps plus $\Smatb$, respectively. }
	\label{fig:fig_test_gen_gauss_maxima_dexp}
	\end{figure}

	\begin{figure}
	\resizebox{\hsize}{!}{\includegraphics{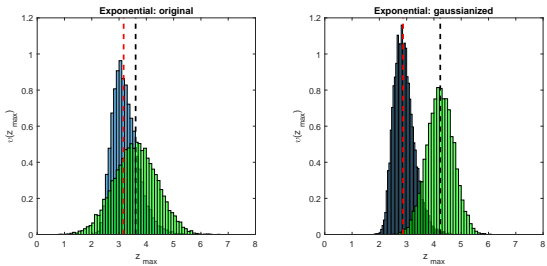}}
	\vskip -1cm
	\caption{Left panel: comparison of the histograms of the highest peak of a set of ten thousand $61 \times 61$ pixels maps after the MF filtering when they contain, respectively, only 
	a white-noise with zero-mean, unit-variance PDF given by the exponential distribution (blue bars)
	and when a circular Gaussian $\Smatb$ with amplitude $a=0.5$ and dispersion $\sigma_s=3$ pixels  is added at the central pixel (green bars). In the case of the maps containing the contribution of $\Smatb$, the highest peak is searched in a square area centered on the central pixel and with sides of seven pixels; Right panel: as the left panel but the maps that are gaussianized before the MF operation. The red and black dashed lines mark the median of the amplitude of the highest peak in the noise maps and in the noise maps plus $\Smatb$, respectively. }
			\label{fig:fig_test_gen_gauss_maxima_exp}
		\end{figure}

\subsection{An alternative technique}

\begin{figure}
	\resizebox{\hsize}{!}{\includegraphics{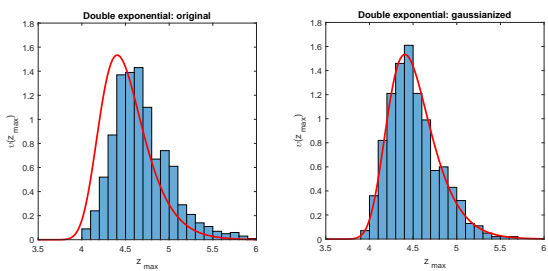}}
     \vskip -0.7cm
	\caption{Left panel: histogram of the amplitude of the highest peak in a set of $10^3$ MF filtered simulated maps containing, respectively, a white-noise with unit-variance PDF corresponding to the double exponential distribution. Here, the MF filter is assumed to be a circular Gaussian with dispersion $\sigma_s=3$ pixels; Right panel: the same as in the left panels but the maps are gaussianized before the MF operation. The red lines provides the theoretical PDF  $\upsilon(z_{\rm max})$ in the case of a zero-mean, unit-variance two-dimensional Gaussian noises.}
	\label{fig:fig_max_gaussianization_dexp}
\end{figure}

\begin{figure}
	\resizebox{\hsize}{!}{\includegraphics{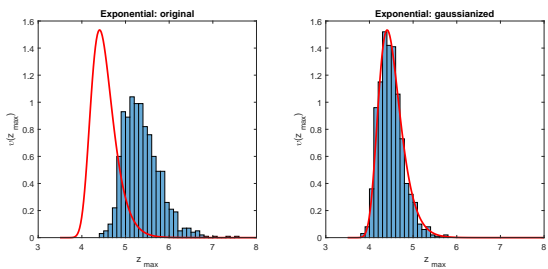}}
     \vskip -0.7cm
	\caption{Left panel: histogram of the amplitude of the highest peak in a set of $10^3$ MF filtered simulated maps containing, respectively, a white-noise with unit-variance PDF corresponding to the exponential distribution. Here, the MF filter is assumed to be a circular Gaussian with dispersion $\sigma_s=3$ pixels; Right panel: the same as in the left panels but the maps are gaussianized before the MF operation. The red lines provides the theoretical PDF  $\upsilon(z_{\rm max})$ in the case of a zero-mean, unit-variance two-dimensional Gaussian noises.}
	\label{fig:fig_max_gaussianization_exp}
\end{figure}

From the previous sections it is evident that, apart from some specific cases, when the noise is non-Gaussian both the MF and the LOD are not able to provide satisfactory results. The main problem is that 
the tails of the PDF of $\Tc(\xb)$ are not well approximated by the Gaussian distribution. Hence, 
an idea to overcome this problem is the gaussianization of the signal $\xb$ before the application of the MF, i.e. to work with a signal $\xhb$ whose entries are given by $\xhc[i] =\gc (x[i])$ where
\begin{equation}
	\gc(x[i])=\Phi^{-1}(\Theta(x[i]))
\end{equation}
is a monotonic increasing function with $\Theta(.)$ the CDF of the noise $\nb$ and $\Phi^{-1}(.)$ the inverse standard Gaussian CDF. Here, the problem is that $\gc(n[i]+s[i]) \neq \gc(n[i]) + \gc(s[i])$. Hence, in the case of white-noises, the MF is not given by the template $\gb$. However, if $a \approx 0$ the function $\gc(x)$ can be approximated by the first term of its Taylor expansion
\begin{align} \label{eq:appr}
	\gc(x) & \approx \gc(n) + \left. \frac{d \gc(x)}{d x} \right|_{s=0} (x-n), \\
	& \approx \gc(n) + \ssc.
\end{align}
with $\ssc = s \gc'(n)$. From this equation, it appears that $\boldsymbol{\ssc}$  has not a well defined shape since each entry $\ssc[i]$ is given by $s[i]$ multiplied by a positive random coefficient given by the derivative of the transformation function at the specific values of the noise.
This is visible in Figs.~\ref{fig:fig_test_source_taylor_dexp1}-\ref{fig:fig_test_source_taylor_exp2}. In particular, Figs.~\ref{fig:fig_test_source_taylor_dexp1} and ~\ref{fig:fig_test_source_taylor_exp1} show the transformation function $\gc(x)$ and its derivative $\gc'(x)$ for two different kinds of unit-variance noise, say the double exponential and the exponential PDFs. For such PDFs, Figs.~\ref{fig:fig_test_source_taylor_dexp2} and ~\ref{fig:fig_test_source_taylor_exp2} compare a one-dimensional simulation of a white-noise added to a Gaussian shaped signal with dispersion $\sigma_s = 3$ pixels and amplitude $a=0.5$ with the approximation given by Eq.~\eqref{eq:appr}. The goodness of this approximation is evident. In the same figure the original $\ssb$ is compared to $\boldsymbol{\ssc}$. As a consequence of the behavior of $\gc'(x)$, in both cases $\boldsymbol{\ssc}$ appears contaminated by a high frequency component that, especially for the exponential noise, make them  different from $\ssb$.  Here, however, two facts need to be considered. In particular, $\ssb$ and $\boldsymbol{\ssc}$ have the same support and, more importantly, $\gc'(x)$ is always positive and can assume very large values. {\it De facto}, this can drastically improve the SNR.
Indeed, the green line in the bottom panel of Figs.~\ref{fig:fig_test_source_taylor_dexp1} and ~\ref{fig:fig_test_source_taylor_exp1} provides the level corresponding to $\gc'(x) = 1$. From these figure, it can be
derived that the probability that $\gc'(x) \ge 1$ for a random number $x$ generated from both the PDFs is of order of $60\%$. This implies that $\ssc[i] \ge s[i]$ for most of the $\{ i \}$.
Moreover, it is necessary to keep into account that the functional form of the MF is robust with respect to the departure from the correct template $\gb$. 
As a consequence, although $\boldsymbol{\ssc}$ could be quite different from $\ssb$, as it happens for the exponential noise, the MF operation is still effective. This is confirmed by Figs.~\ref{fig:fig_test_gen_gauss_maxima_dexp} and \ref{fig:fig_test_gen_gauss_maxima_exp} that are obtained from the simulation of ten thousands $61 \times 61$ pixels maps each containing the realization of a white-noise given again by the two kind of processes used above. The left panel of each figure compares the sample distribution of the highest peak of each MF filtered map when they contain only noise and when a circular Gaussian $\Smatb$ with amplitude $a=0.5$ and 
dispersion $\sigma_s=3$ pixels is added to the central pixel. In the case of maps containing the contribution of $\Smatb$, the highest peak is searched in a square area centered on the central pixel and half-side set to $\sigma_s$.  The right panel of the same figures shows what happens when the maps  are gaussianized before the MF operation. The red and black dashed lines mark the median of the corresponding sample distribution. The benefit of the gaussianization operation is well visible.
Another benefit comes out from the fact that, after the gaussianization of the signal, it is possible to use the analytical forms of the PDF and the CDF of the extremum of a Gaussian noise that
are necessary in order to estimate the statistical reliability of the detection \citep{vio16, vio17, vio19}. This is not viable without the such operation (see Figs.~\ref{fig:fig_max_gaussianization_dexp}
and \ref{fig:fig_max_gaussianization_exp}).

\appendix
\section{The generalized Gaussian and Cauchy PDFs} \label{sec:appA}

In this appendix we illustrate two families of parametric continuous distributions: the generalized Gaussian and the generalized Cauchy densities,  which entail the majority of the zero-mean, symmetric PDFs commonly adopted to describe the noises in signal detection problems. One benefit of working with these families is that the corresponding LO filter $\flo(x)$ and the Fisher information $I(a)$ are available in an analytical form \citep{kas88}.

\subsection{The generalized Gaussian distribution}

The members of this family take the form:
\begin{equation} \label{eq:fg}
	p_k(x) = \frac{k}{2 A(k) \Gamma(1/k)} \exp{\left\{- \left( \frac{|x|}{A(k)} \right)^k \right\}},
\end{equation}
where
\begin{equation} \label{eq:ak}
	A(k) = \sqrt{\sigma^2 \frac{\Gamma(1/k)}{\Gamma(3/k)}}, 
\end{equation}
and $\Gamma(a)$ is the gamma function
\begin{equation}
	\Gamma(a)=\int_0^{\infty} x^{a-1} {\rm e}^{-x} dx.
\end{equation}
Parameter $\sigma$ provides the standard deviation of the distribution whereas $k$ is a real positive parameter which provides the exponential decay rate with which the tails decay.
For example, $k=2$ and $k=1$ corresponds to the Gaussian and the double-exponential distribution, respectively. More in general, the case $k < 2$ corresponds to distributions with slower
decay rate than the Gaussian, the reverse when $k > 2$. It is possible to show that the filter $\flo(x)$ and the Fisher information are given by \citep{kas88}:
\begin{equation}
	\flo(x)=\frac{k}{\left[ A(k) \right]^k} | x |^{k-1} {\rm sign}[x],
\end{equation}  
respectively,
\begin{equation}
	I(a) = \frac{k^2 \Gamma(3/k) \Gamma(2 -1/k)}{\sigma^2 \Gamma^2(1/k)}.
\end{equation}  

A further benefit of working with the PDFs belonging to this family is the easy generation of random numbers $x$ by means of the following procedure \citep{nar09}:
\begin{enumerate}
	\item Simulation of a a gamma random variable $z \sim \Gamma[k^{-1},A^k(k)]$; \\
	\item Application of the transformation $y = z^{1/k}$; \\
	\item Simulation of a random variable $r$ which can take with equiprobability only the values $-1$ and $+1$; \\
	\item Set $x= r y$.
\end{enumerate}

\subsection{The generalized Cauchy distribution}

This family contains distributions which have an algebraic rather than an exponential tail behavior. The corresponding PDFs  take the form
\begin{equation}
	p_{k,\nu}(x) = \frac{B(k,\nu)}{\left\{ 1 + \dfrac{1}{\nu} \left[ \dfrac{|x|}{A(k)} \right]^k\right\}^{\nu+1/k}},
\end{equation}
where
\begin{equation}
	B(k,\nu) = \frac{k \nu^{-1/k} \Gamma(\nu + 1/k)}{2 A(k) \Gamma(\nu) \Gamma(1/k)},
\end{equation}
with $A(k)$ given by Eq.~\eqref{eq:ak}.
To notice that here $\sigma$ it is only a scale parameter and it does not provide the standard deviation of the distributions which exists only when $\nu k > 2$ and it is given by 
$\sqrt{\sigma^2 \nu^{2/k} \Gamma(\nu - 2/k) / \Gamma(\nu)}$.
Important members of this family are obtained by  the values $k=2$ and $\nu=1/2$, which  provide the Cauchy distribution, and $k=2$ and $\nu=1$ which provide the Student distribution. 

It is possible to show that the filter $\flo(x)$ and the Fisher information are given by \citep{kas88}
\begin{equation}
	\flo(x) =\frac{\nu k +1}{\nu [A(k)]^k + |x|^k} |x|^{k-1} {\rm sign}[x],
\end{equation}
respectively
\begin{equation}
	I(a) = \frac{(\nu k +1)^2 \Gamma(3/k) \Gamma(\nu +1/k) \Gamma(\nu +2 /k) \Gamma(2-1/k)}{\sigma^2 \nu^{2/k} \Gamma^2(1/k) \Gamma(\nu) \Gamma(2 \nu + 1/k)}.
\end{equation}


\begin{thebibliography}{}
\bibitem[Baddeley et al. (2016)]{bad16} Baddeley, A., Rubak, E., \& Turner, R. 2016, Spatial Point Patterns (New York: CRC Press)
\bibitem[Barkat (2005)]{bar05} Barkat, M. 2005, Signal Detection and Estimation (London: Artech House)
\bibitem[Barreiro et al. (2003)]{bar03} Barreiro, R,B., Sanz, J.L., Herranz, D., \& Mart\'ines-G\'onzalez, E. 2003, MNRAS, 342, 119
\bibitem[Cheng \& Schwartzman (2015a)]{che15a} Cheng, D., \& Schwartzman, A. 2015a, Extremes, 18, 213
\bibitem[Cheng \& Schwartzman (2015b)]{che15b} Cheng, D., \& Schwartzman, A. 2015b, arXiv:1503.01328 [math.PR]
\bibitem[Chen (2016)]{che16} Cheng, D. 2016, {\it private communication}
\bibitem[Davis (1979)]{dav79} Davis, P. 1979, Circulant Matrices (New York: Wiley)
\bibitem[El-Samie et al. (2013)]{els13} El-Samie, F.E.A., Hadhoud, M.M., El-Khamy, S.E. 2013, Image Super-Resolution and Applications (London: CRC Press)
\bibitem[Galatsanos et al. (2005)]{gal05} Galatsanos, N.P., Wernick, M.N., Katsaggelos, A.K., \& Molina, R. 2005, in Handbook of Image \& Video processing, Editor Al Bovik, 203 (New York: Academic Press)
\bibitem[Goudail \& Refregier (2004)]{gou04} Goudail, F., \& Refregier, P. 2004, Statistical Image Processing for Noisy Images (New York: Kluwer Academic/Plenum Publishers)
\bibitem[Higham (2008)]{hig08} Higham, N.J. 2008, Functions of Matrices (Philadelphia: SIAM) 
\bibitem[Hippenstiel (2002)]{hip02} Hippenstiel, R.D. 2002, Detection Theory, (New York: CRC Press)
\bibitem[Hogg et al. (2013)]{hog13} Hogg, R.V., McKean, J.W., \& Craig, A.T. 2013, Introduction to Mathematical Statistics (New York: Pearson) 
\bibitem[Kassam (1988)]{kas88} Kassam, S.A. 1988, Signal Detection in Non-Gaussian Noise (New York: Springer-Verlag)
\bibitem[Kay (1998)]{kay98} Kay, S. M. 1998, Fundamentals of Statistical Signal Processing: Detection Theory (London: Prentice Hall)
\bibitem[Katsaggelos et al. (1993)]{kat93} Katsaggelos, A.K., Lay, K.T., Galatsanos, N.P. 1993, IEEE Transaction On Image Processing, 2, 417 
\bibitem[Jain (1989)]{jai89} Jain, A.K. 1989, Fundamentals of Digital Image Processing (London: Prentice Hall)
\bibitem[Lagendijk \& Biemond (1991)]{lag91} Lagendijk, R.L, \& Biemond, J. 1991, Iterative Identification and Restoration of Images (New York: Springer Science+ Business Media) 
\bibitem[Landau et al. (2007)]{lan07} Landau, R.H., Paez, M.J., \& Bordeianu, C.C. 2007, Computational Physics (Weiheim: Wiley-VCH)
\bibitem[Levy (2008)]{lev08} Levy, B.C. 2008, Principles of Signal Detection and Parameter Estimation  (New York: Springer)
\bibitem[L\'opez-Caniego et al. (2005)]{lop05} L\'opez-Caniego, M., Herranz, D., Barreiro, R.B., \& Sanz, J.L. 2005, MNRAS, 359, 993
\bibitem[Majumdar \& Comtet (2005)]{may05} Majumdar, S.N., \& Comtet, A. 2005, Journal of Statistical Physics, 119, 777
\bibitem[Macmillan \& Creelma (2005)]{mac05} Macmillan N.A., \& Creelman, C.D. 2005, Detection Theory (London: Lawrence Erlbaum Associates)
\bibitem[McNicol (2005)]{mcn05} McNicol, D. 2005, A Primer of Signal Detection Theory (London: Lawrence Erlbaum Associates)
\bibitem[Nardon \& Pianca (2009)]{nar09} Nardon, M., \& Pianca, P. 2009, Journal of Statistical Computation and Simulation, 79, 1317
\bibitem[Ofek \& Zakay (2018)]{ofe18} Ofek, E.O., \& Zakay, B. 2018, AJ, 155, 1690
\bibitem[Poor (1994)]{poo94} Poor, H.V. 1994, An Introduction to Signal Detection and Estimation (New York: Springer-Verlag)
\bibitem[Press et al. (2007)]{pre07} Press, W.H., Teukolsky, S.A., Vetterling, W.T., \& Flannery, B.P. 2007, Numerical Recipes (New York: Cambridge University Press)
\bibitem[Sanz, Herranz \& Martinez-Gonzales (2001)]{san01} Sanz, J.L, Herranz, D., \& Mart\'inez-G\'onzales, E. 2001, ApJ, 552, 484 
\bibitem[Tuzlukov (2001)]{tuz01} Tuzlukov, V.P. 2001, Signal Detection Theory (New York: Springer)
\bibitem[Vio, Tenorio \& Wamsteker (2002)]{vio02} Vio, R., Tenorio, L., \& Wamsteker, W. 2002, A\&A, 391, 789
\bibitem[Vio, Andreani \& Wamsteker (2004)]{vio04} Vio, R., Andreani, P., \& Wamsteker, W. 2004, A\&A, 414, 17
\bibitem[Vio \& Andreani (2016)]{vio16} Vio, R., \&  Andreani, P. 2016, A\&A, 589, A20
\bibitem[Vio et al. (2017)]{vio17} Vio, R., Verg\`es, C., \&  Andreani, P. 2018, A\&A, 604, A115
\bibitem[Vio \& Andreani (2018)]{vio18} Vio, R., \& Andreani, P. 2018, A\&A, 616, A25
\bibitem[Vio, Andreani \& Biggs (2019)]{vio19} Vio, R., Andreani, P., \& Biggs, A. 2019, A\&A, 627, A103
\bibitem[Vogel (2002)]{vog02} Vogel, C.R. 2002, Computational Methods for Inverse Problems (Philadelphia: SIAM)
\bibitem[Wickens (2002)]{wic02} Wickens, T.D. 2002, Elementary Signal Detection Theory (New York: Oxford University Press)
\end{thebibliography}
\end{document}